# CDIM

## Cosmic Dawn Intensity Mapper

A Probe Class Spectro-Imaging Astrophysics Mission for Reionization

# Table of Contents

## CDIM Mission Study Team

| Name | Institution/E-mail | Role |
|---|---|---|
| Asantha Cooray | University of California Irvine; acooray@uci.edu | Principal Investigator |
| Tzu-Ching Chang | Jet Propulsion Laboratory, California Institute of Technology; tzu-ching.chang@jpl.nasa.gov | Study Scientist |
| Stephen Unwin | Jet Propulsion Laboratory; stephen.c.unwin@jpl.nasa.gov | Study Manager |
| Michael Zemcov | Rochester Institute of Technology; zemcov@cfd.rit.edu | Instrument Scientist |
| Andrew Coffey | Jet Propulsion Laboratory; Andrew.S.Coffey@jpl.nasa.gov | Systems Engineer |
| Patrick Morrissey | Jet Propulsion Laboratory; Patrick.Morrissey@jpl.nasa.gov | Instrument Systems Engineer |
| Nasrat Raouf | Jet Propulsion Laboratory; Nasrat.A.Raouf@jpl.nasa.gov | Optics and filters |
| Sarah Lipscy | Ball Aerospace & Technologies Corp; slipscy@ball.com | Spacecraft Manager |
| Mark Shannon | Ball Aerospace & Technologies Corp | Spacecraft Engineer |
| Gordon Wu | Ball Aerospace & Technologies Corp | Spacecraft Engineer |
| **Science Team Co-Investigators** | | |
| Renyue Cen | Princeton University | Co-Lead, Active Galactic Nuclei/Quasars |
| Ranga Ram Chary | IPAC/Caltech | Co-Lead, Galaxy Formation |
| Olivier Doré | Jet Propulsion Laboratory California Institute of Technology | Intergalactic Medium |
| Xiaohui Fan | University of Arizona | Co-Lead, Active Galactic Nuclei/Quasars |
| Giovanni G. Fazio | Harvard Smithsonian Center for Astrophysics | Galaxy Formation |
| Steven L. Finkelstein | The University of Texas at Austin | Co-Lead, Galaxy Formation |
| Caroline Heneka | Scuola Normale Superiore, Pisa | Intensity Mapping Sciences |
| Bomee Lee | IPAC/Caltech | Galaxy Formation |
| Philip Linden | Rochester Institute of Technology | Intensity Mapping Science |
| Hooshang Nayyeri | University of California Irvine | Galaxy Formation, AGN/Quasars |
| Jason Rhodes | Jet Propulsion Laboratory California Institute of Technology | Intergalactic Medium |
| Raphael Sadoun | Osaka University | Lyman-$\alpha$ sources & Intergalactic Medium |
| Marta B. Silva | University of Oslo | Intensity Mapping Sciences |
| Hy Trac | Carnegie Mellon University | Lyman-$\alpha$ sources/Intensity Mapping Sciences |
| Hao-Yi Wu | The Ohio State University | Galaxy Formation |
| Zheng Zheng | University of Utah | Lyman-$\alpha$ sources & Intergalactic Medium |

## Acknowledgments

This research was funded by a NASA grant NNX17AJ80G to the University of California, Irvine. Part of this research was done by the Jet Propulsion Laboratory, California Institute of Technology, under a contract with NASA, and by Ball Aerospace & Technologies Corp. The CDIM team gratefully acknowledges the many members of the astronomical community who contributed their time and expertise in developing this concept and science case. The information presented about the CDIM mission concept is pre-decisional and is provided for planning and discussion purposes only. Cost information contained in this document is of a budgetary and planning nature and is intended for informational purposes only. It does not constitute a commitment on the part of JPL and/or Caltech.

# CDIM

**Cosmic Dawn Intensity Mapper**
A Probe Class Spectro-Imaging Astrophysics Mission for Reionization

CDIM will transform our understanding of the era of reionization when the universe formed first stars and galaxies, and UV photons ionized the neutral medium.

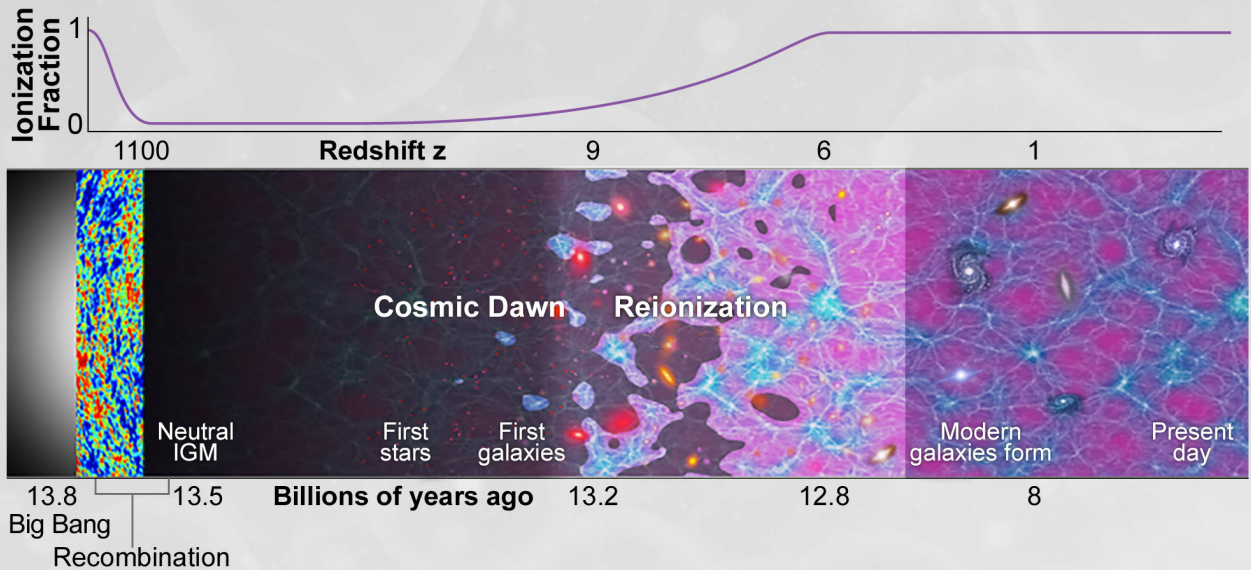

## CDIM FOCUSES ON THREE KEY SCIENCE GOALS

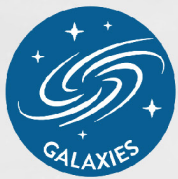 **Galaxies** — Measuring physical properties to z of 8

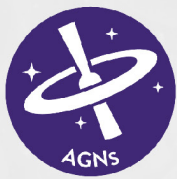 **AGNs** — Finding black holes to z of 8

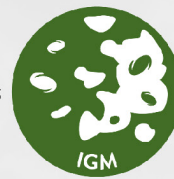 **IGM Tomography** — Mapping reionization topology & history from z of 5 to 10

## COMPLEMENTING FUTURE MISSIONS

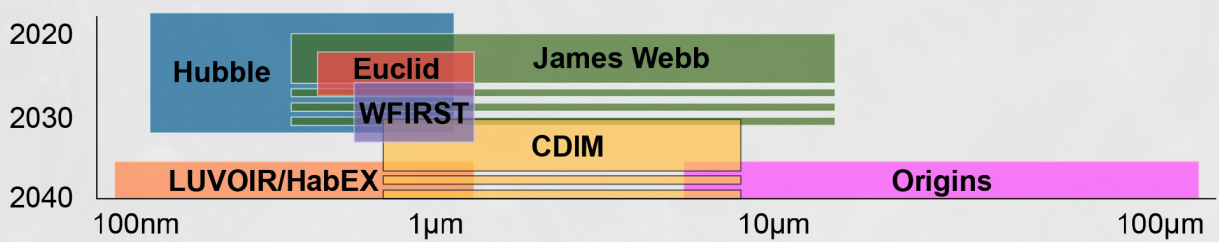

**Figure 1.** CDIM design and capabilities focus on the needs of detecting faint galaxies and quasars during reionization and intensity fluctuation measurements of key spectral lines, including Lyman-α and Hα radiation from first stars and galaxies. The design is low risk, carries significant science and engineering margins, and makes use of technologies with high technical readiness level for space observations. CDIM will fill the gap between LUVOIR/HabEX operating out to ~1–2 μm and Origins Space Telescope at above 5 μm.

## UNIQUE SENSITIVITY TO DETECT OPTICAL EMISSION LINES

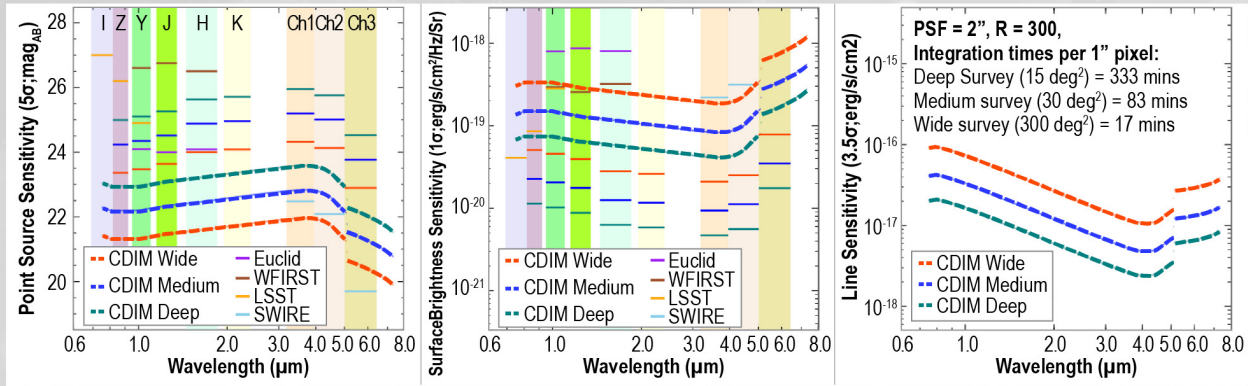

**Figure 2.** The three-tiered CDIM survey sensitivities in 840 bands (R = 300; dashed lines) and in broadband (R = 5; solid lines) are designed to match science requirements. CDIM is uniquely sensitive to detect rest-frame optical emission lines or their aggregate intensity over the entire cosmic history of galaxy formation and evolution, and out to the end of cosmic dawn (z ~ 10).

## LINEAR VARIABLE FILTER (LVF) OBSERVATIONS

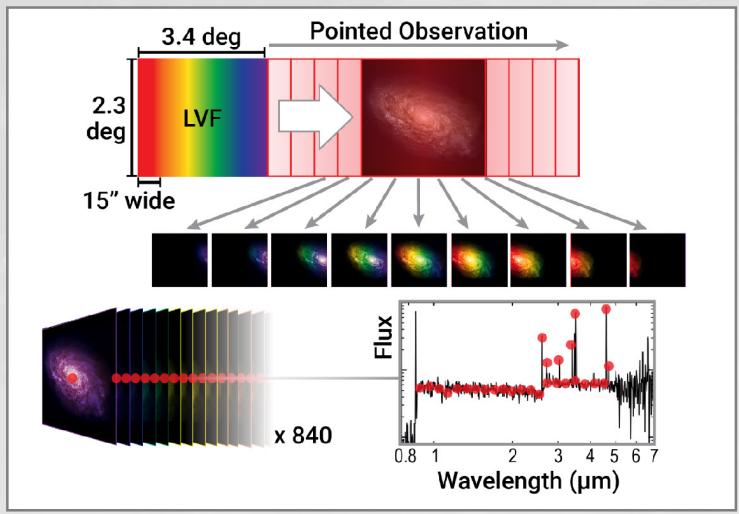

**Figure 3.** The full spectra are obtained by moving the telescope in small discrete steps corresponding to the CDIM spectral resolution across the dispersion direction of the LVF.

## INSTITUTIONS

JPL, University of California Irvine, RIT, Ball Aerospace, Princeton University, IPAC, Caltech, University of Arizona, Harvard, Scuola Normale Superiore Pisa, UT Austin, Osaka University, University of Oslo, Carnegie Mellon, Ohio State University, University of Utah

## CDIM FLIGHT SYSTEM

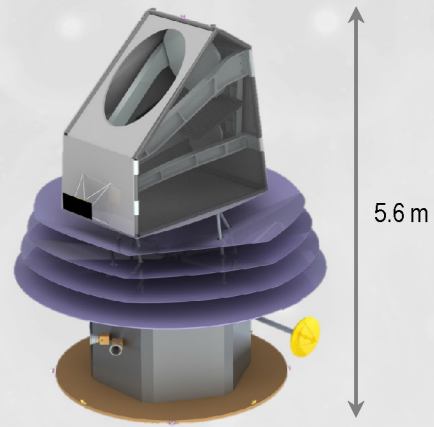

- 3-mirror all-reflective design with 0.83-m clear aperture
- Linear Variable Filters over 0.75–7.5 μm in 840 bands at R = 300
- V-groove radiators, passive cooling at T < 35 K in L2 halo orbit
- 4x6 H2RG detectors
- Data rate 413 Gbit/day; 1.5 hr/day downlink
- 4-year survey for key sciences; plus extended mission for community observations

## MISSION COST ESTIMATE

**Total cost estimated by Team X (including 30% reserves): $929M (FY18)**

## SCHEDULE

| | FY24 | FY25 | FY26 | FY27 | FY28 | FY29 | FY30 | FY31 | FY32 | FY33 |
|---|---|---|---|---|---|---|---|---|---|---|
| 2023 | 2024 | 2025 | 2026 | 2027 | 2028 | 2029 | 2030 | 2031 | 2032 | 2033 |

CDIM: Ph A (12m) | Ph B (12m) | PDR 10/2025 | CDR 7/2026 | ARR 7/2027 | Ph C/D (40m) | KSC | Phase E (49m) | PhF 4m

# 1 EXECUTIVE SUMMARY

The Cosmic Dawn Intensity Mapper (CDIM) will transform our understanding of the era of reionization when the Universe formed the first stars and galaxies, and UV photons ionized the neutral medium. CDIM goes beyond the capabilities of upcoming facilities by carrying out wide area spectro-imaging surveys, providing redshifts of galaxies and quasars during reionization as well as spectral lines that carry crucial information on their physical properties. CDIM will make use of unprecedented sensitivity to surface brightness to measure the intensity fluctuations of reionization on large-scales to provide a valuable and complementary dataset to 21-cm experiments. The baseline mission concept is an 83-cm infrared telescope equipped with a focal plane of 24 × $2048^2$ detectors capable of R = 300 spectro-imaging observations over the wavelength range of 0.75 to 7.5 μm using Linear Variable Filters (LVFs). CDIM provides a large field of view of 7.8 $deg^2$ allowing efficient wide area surveys, and instead of moving instrumental components, spectroscopic mapping is obtained through a shift-and-stare strategy through spacecraft operations. CDIM design and capabilities focus on the needs of detecting faint galaxies and quasars during reionization and intensity fluctuation measurements of key spectral lines, including Lyman-α and Hα radiation from the first stars and galaxies. The design is low risk, carries significant science and engineering margins, and makes use of technologies with high technical readiness level for space observations.

## 1.1 Scientific Objectives

The development of a comprehensive understanding of the physics that led to the formation of first stars and galaxies is challenging, but remains a fundamental goal of extragalactic astrophysics and cosmology. Over the next decade, multi-wavelength observations of the first galaxies and the early intergalactic medium will yield significant new clues in our quest to develop a fundamental understanding of the physics of the epoch of reionization (EoR). Though a great deal has been learned in the past two decades, and more information will be forthcoming with JWST and WFIRST over the coming decade, many key questions will remain unanswered: **(i) what is the history of stellar, dust, and metal build-up in early galaxies during reionization?; (ii) what is the contribution of quasars to the reionization history of the Universe?; and (iii) what is the exact reionization history of the Universe?**

One reason for this limitation in our science understanding of reionization is that JWST extragalactic surveys will be likely limited to a handful of deep fields, with a total area of several hundred $arcmin^2$. While WFIRST will be capable of wide area surveys, due to the lack of spectroscopic capability beyond 1.8 μm, spectroscopic studies involving Hα will be limited to $z < 2$; the selection of $z > 6$ galaxies will be limited to photometric data using the Lyman-break signature, leading to uncertainties on their exact redshifts. Thus, both JWST and WFIRST will not be definitive missions to complete our understanding of reionization – leaving open critical questions on the rate at which first galaxies formed their stars, dust and metals, the role of active galactic nuclei (AGN) in reionization, and the history and topology of reionization.

A Probe-class mission optimized for reionization studies will need to expand both imaging and spectroscopic capabilities over the expectations from JWST and WFIRST. The Cosmic Dawn Intensity Mapper (CDIM) is designed with an improved understanding of reionization as the primary science goal. The science program is directly linked to two of the NASA top-level goals in astrophysics (2018 NASA Strategic Plan and 2014 NASA Science Mission Directorate Plan): how does the Universe work? And how did we get here?

CDIM survey data will allow us to address the above three questions and many more, through the science program outlined in **Table 1-1**. In particular, CDIM spectro-imaging data with a spectral resolving power R=λ/Δλ of 300 can be used to: (i) determine the spectroscopic redshifts of WFIRST-detected Lyman-break galaxies (LBGs) out to $z \sim 8$–9; (ii) conduct a complete census of first-light galaxies in over 2–3 decades of stellar mass by establishing their mass, metal abundance, and dust content by spectrally separating [NII] from Hα, and by detecting both Hβ and OIII; (iii) establish the environmental dependence of star-formation during reionization through clustering measurements; (iv) detect Lyα emission from individual bright first-light galaxies and combine with Hα line detections of the same galaxies for studies

| Table 1-1. Science Program of the CDIM Probe Mission | | | |
|---|---|---|---|
| NASA Science Goals | How does the Universe work? How did we get here? (from *2014 NASA SMD Strategy Document*) | | |
| CDIM Science Themes | Galaxy Formation and Evolution at 5 < z < 8 | Active galactic nuclei at 5 < z < 8 | Reionization history through Lyman-α intensity |
| CDIM Science Goals | Trace the stellar mass buildup, dust production, and metal enrichment history during cosmic reionization. | Establish the role of active galactic nuclei (AGN) in cosmic reionization | Establish the progression and topology of reionization from cosmic dawn at z=10 to the end of reionization at z < 6. |
| CDIM Scientific Objectives | Determine if the rate of growth of metals and dust corresponds to the growth of stellar mass at 5 < z < 8. | Determine the fractional contribution of super-massive black hole/AGNs to reionization photon budget. | Determine the progress of reionization by measuring the ionization fraction in at least 10 redshift bins at 5 < z < 10, with accuracy better than 10%. |

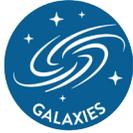

of variations of Lyα escape fraction with environment and other physical properties; and (v) use the redshift-evolution of spectral-line intensity fluctuations to determine the reionization history of the Universe.

While the primary focus is EoR, CDIM is also designed to study galaxy formation and evolution throughout cosmic history. It will map out, for example, Hα emission from $z = 0.2$ to 10 and will detect Lyα emission from galaxies present during reionization. Wide area spectral mapping with CDIM will allow searches for rare sources, such as bright quasars, active galactic nuclei (AGN), and galaxies that make up the bright-end of the luminosity functions. CDIM surveys will provide a three-dimensional view of the star-formation history, its environmental dependence, and clustering over 90% of the age of the Universe.

The three Level-1 scientific objectives that are summarized in **Table 1-1** flow to mission and instrument requirements. These in turn determine the scientific measurement capabilities of CDIM. For the first theme, using galaxy samples and their exquisite spectra from CDIM, we will address the rate of growth of metals and dust during reionization. The same CDIM data can also be used to address whether the initial mass function (IMF) of stars in reionizing galaxies at $z > 6$ is different from the IMF of stars in galaxies today. The difference could be attributed to differences in the physical properties of stars during reionization, which are likely to be metal poor and, on average, have masses that are higher than the stars in galaxies at low redshifts.

With AGN or quasar samples, our second theme seeks to address if they make a substantial contribution to the UV photon density budget during reionization. Current expectations are that galaxies are primarily responsible for reionization, but uncertainties remain on the role played by quasars and AGNs.

For the third theme, CDIM is capable of detecting Lyα emission from bright galaxies present throughout the full history of reionization, without the restrictions of ground-based narrow-band Lyα emitter surveys that are only limited to a handful of redshifts allowed by the atmospheric window. When combined with Hα and UV continuum detections of the same galaxies, CDIM will allow a complete statistical study of the escape fraction of ionizing radiation, relating galaxy properties to their environment. For the fainter population undetected as individual galaxies, CDIM is capable of sensitive surface brightness measurements leading to intensity mapping of spectral lines, including Lyα during reionization. These line intensity maps will be compared with 21-cm fluctuations that trace neutral hydrogen. Galaxies bright in Lyα emission are expected to have a relatively high ionizing photon emissivity, leading to a larger size for their surrounding ionized bubbles than the bubbles of Lyα-deficient galaxies. Therefore, galaxies bright in Lyα emission should be anti-correlated with the 21-cm emission from the intergalactic medium (IGM) on scales smaller than their bubbles. This anti-correlation provides ways to establish the bubble sizes during

| Table 1-2. CDIM Survey Requirements Summary | | | |
|---|---|---|---|
| **CDIM Science Themes** | Galaxy Formation and Evolution at $5 < z < 8$ 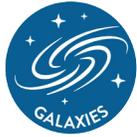 | Active galactic nuclei at $5 < z < 8$ 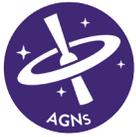 | Reionization history through Lyman-$\alpha$ intensity 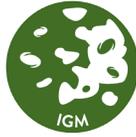 |
| **CDIM key Technical Requirement** | Wavelength range of 0.75–7.5 μm to detect Lyα and Hα. Spectral resolution R > 300 to separate [NII] from Hα for metallicity measurements. Line flux sensitivity to detect Hα below $L^*$ at $z = 6$ in a *deep survey* of 15 deg$^2$. | Large 300 deg$^2$ survey for determining the faint end of AGN luminosity function at $z = 6$. Point source sensitivity down to K band magnitude of 23.5 in a *wide survey* of 300 deg$^2$. | Large field of view to capture the large-scale reionization topology on >100 Mpc scales, with a survey area of 30 deg$^2$ to overlap with wide epoch of reionization 21-cm surveys. Surface surface brightness sensitivity to detect Lyα at a minimum signal-to-noise ratio of 500 in a *medium survey* of 30 deg$^2$. |

| Table 1-3. Proposed CDIM Surveys | | | | | | |
|---|---|---|---|---|---|---|
| **Survey** | **Area (deg$^2$)** | | **Sky location (§2.6)** | **Point source depth (K-band AB mag; 5σ)** | **Spectral line flux detection limit** (1 μm at R = 300; 3.5σ) [$10^{-17}$ erg/s/cm$^2$] | **Surface brightness sensitivity** (1.1μm at R = 300; 1σ) [$10^{-19}$ erg s$^{-1}$ cm$^{-2}$ Hz$^{-1}$ sr$^{-1}$] | **Scientific drivers** |
| | Requirement | Survey Design | | | | | |
| **Deep** | 15 | 15.6 | SEP or NEP | 25.7 | 1.5 | 0.70 | Galaxies – stellar mass, dust, and metallicity |
| **Medium** | 30 | 31.1 | ECDF-S | 25.0 | 3.2 | 1.5 | Deep Survey drivers, plus reionization history through Lyman-$\alpha$ intensity |
| **Wide** | 300 | 311 | SEP or NEP | 24.0 | 8.0 | 3.3 | Deep and medium Survey drivers, plus Quasars/AGNs |

reionization as a function of redshift. The two in combination also allow reionization history to be determined, improving over the current uncertainties by an order of magnitude or more, while not subject to systematics that can negatively impact 21-cm alone measurements.

## 1.2 Science Requirements

Currently prioritized science programs, conducted over four years, will be accomplished with a three-tiered wedding-cake survey with the shallowest spanning 300 deg$^2$ and the deepest tier of about 15 deg$^2$. **Table 1-2** summarizes survey requirements, while **Table 1-3** lists the surveys anticipated to meet the requirements. See §2.6 for the Science Traceability Matrix (STM). The primary science goal addressed by the deep survey (in the north or south ecliptic pole regions – NEP or SEP) is pioneering observations of the Lyα, Hα, and other spectral lines from individual galaxies throughout the cosmic history, but especially from the first generation of distant, faint galaxies when the Universe was less than 800 million years old. The deep survey requirement is a line sensitivity of better than $5\times10^{-18}$ erg s$^{-1}$ cm$^{-2}$ at 4.6 μm for Hα at $z = 6$, with spectral resolving power of 300 to separate Hα and [NII] for accurate metallicity measurements and galaxy/AGN separation. The medium survey aims for a three-dimensional tomographic view of EoR mapping Lyα emission from galaxies and the IGM, with a surface brightness sensitivity requirement of $1.3\times10^{-18}$ erg s$^{-1}$ cm$^{-2}$ Hz$^{-1}$ sr$^{-1}$ at 1.1 μm. The wide survey requirement will be K-band AB mag = 23.5 for detecting quasars at $z > 6$.

## 1.3 Science Implementation

The proposed NASA Probe Class Mission CDIM is an 83-cm effective aperture (1.1-m physical aperture), passively cooled telescope (down to 35 K), designed to meet NASA's Class B mission

CDIM: Cosmic Dawn Intensity Mapper Final Report    CDIMrequirements. CDIM is capable of three-dimensional spectro-imaging observations over the wavelength range of 0.75 to 7.5 µm, at a spectral resolving power $R=\lambda/\Delta\lambda$ of 300. The focal plane is made up of 24 × $2048^2$ detectors, leading to an instantaneous field-of-view (FoV) of 3.4 × 2.3 = 7.8 $\deg^2$ with 1″ pixels.

CDIM will make use of fixed linear variable filters (LVFs) to image the sky at narrow $\Delta\lambda$ wavelengths, rather than a using a dispersing element or a grism. This design has no moving parts, and is ideally suited to measuring spectra across a wide-field focal plane. An integral field spectrograph for such a large focal plane is not feasible on a Probe-mission budget.

CDIM will cover a spectral range between 0.75 and 7.5 µm with 840 independent spectral channels (**Fact Sheet Figure 2**). A fully Nyquist-sampled spectrum then requires 1680 individual pointings towards a given line of sight, with pointings offset by an angle that is $\delta\theta/2$ of the LVF width $\delta\theta$ on the sky. The LVF varies in wavelength along the 3.4° direction with a width of $\delta\theta$ at 15″. To ensure uniform spectral sampling, the step size must be accurate to about 0.5″. Even though a survey requires 1680 individual pointings in each sky position, CDIM is designed to be very efficient (§4.2); current best estimate for design and operations is 93%. The key science programs build on a survey strategy that requires 250 seconds of integration at each LVF position in the shallowest survey, while deeper surveys interleave multiple visits at the same integration time. The spacecraft requires less than 10 seconds to slew and settle.

CDIM builds on the existing heritage of using LVFs for spectral mapping on the sky. Past successes include WFPC2 on HST, OVIRS on OSIRIS-REx and LEISA on New Horizons. By taking repeated images, the spectrum is constructed in the data analysis process. CDIM is a natural follow-up mission to SPHEREx, an explorer class mission that utilizes R = 35–130 LVFs to map the whole sky between 0.75 and 5 µm using a 20-cm aperture telescope. While the focus on SPHEREx is all-sky cosmological measurements using shallow depths to cover the spatial distribution of galaxies out to *z* of 1.5, by going deeper on 15–300 $\deg^2$ patches on the sky, CDIM aims to focus on the era of reionization.

CDIM is low-risk. The two key technology development items are the production of LVFs to meet the requirements on filter shape and out-of-band leakage and the H2RG detectors that allow imaging out to the longest wavelengths covered by CDIM of 7.5 µm. The LVF technology has substantial development already by industry, while substantial work is underway to space qualify long-wavelength H2RGs as part of NASA concept missions such as NEOCAM. The CDIM telescope and other technical components are already commercially available and are already flight qualified or will be flight qualified. This includes components such as Teledyne SIDECAR ASICs that will be qualified as part of the Euclid mission to be launched in June 2022.

## 1.4 Science Operations

While the CDIM concept proposed here assumes four years of operations to meet the key scientific objectives, the mission lifetime is limited primarily by consumables. The design allows science operations that could last substantially longer than the four years necessary for the key science identified in the report.

CDIM is fully capable of functioning as a general-purpose observatory, enabling substantial community-led observing campaigns, to be selected via the usual peer-review process. These could be in the form of mapping the Galactic plane to targeted studies of nearby galaxies, or other deep cosmological fields, for example. Due to its large FoV and rapid mapping capability, CDIM can also function as a transient source identifier for time-domain science. Of interest in the 2030s will be identification of electromagnetic counterparts to gravitational-wave sources, especially super-massive black hole merger events detected by the LISA gravitational-wave space observatory. CDIM is capable of rapid scanning of LISA error boxes, expected to be on the order of 10 to 100 $\deg^2$.

## 2 SCIENCE INVESTIGATION

### 2.1 Science Landscape

**Cosmic Dawn and Reionization are ripe for discoveries.** Cosmic Dawn – when the first stars lit up the Universe, eventually coalescing into the first generations of galaxies – and the epoch of reionization that follows are among the most exciting frontiers in cosmology and astrophysics. Sometime between 200 and 800 Myr after the Big Bang ($z \sim 6$–$20$), the first collapsed objects formed in dark matter halos and eventually produced enough Lyman continuum photons to reionize all of the surrounding hydrogen gas in the IGM (Becker et al., 2001; Fan et al., 2006b). This epoch of reionization (EoR) marked the end of the dark ages, and is the first chapter in the history of galaxies and heavy elements (Barkana & Loeb, 2001). Large-angle CMB polarization suggests an optical depth to reionization of $0.054 \pm 0.007$ (Planck Collaboration et al., 2018), corresponding to an reionization redshift of $7.7 \pm 0.8$ (if modeled as instantaneous). It is likely that the reionization epoch is extended and inhomogeneous (Fan et al., 2006a).

Existing theoretical studies suggest that a combination of the first metal-free stars (Bromm & Larson, 2004), the subsequent generations of stars, and accretion onto remnant black holes (such as mini-quasars; Venkatesan et al., 2001) contributed enough UV photons to reionize the Universe. Given the existing estimates of the UV luminosity functions (LFs) at $z > 6$ (Bouwens et al., 2015; Finkelstein et al., 2015), reionization is clearly dominated by lower mass, lower luminosity systems (Salvaterra et al., 2011).

While galaxies are responsible for producing the bulk of ionizing photons, one major uncertainty is the *escape fraction*, the fraction of ionizing photons that are able to escape the galaxies to ionize the surrounding neutral intergalactic medium. Existing reionization models require escape fractions at the level of 20% (Robertson et al., 2015). Such a high escape fraction is inconsistent with direct measurements, which are at the level of a few % (Hayes et al., 2011; Vanzella et al., 2012). For this reason, quasars may also contribute a significant amount of the energy required to reionize the Universe (Madau & Haardt, 2015).

Although the basic framework is in place, we have significant deficiencies in our understanding of the physical processes at work during EoR. This is primarily due to the small number of detected galaxies and quasars at $z > 6$ in existing surveys, biases in current sample selection, and cosmic variance in pencil beam surveys. Furthermore, the ground-based telescopes are limited in sensitivity to $< 2.4$ μm allowing only strong UV lines like Lyα and HeII 1640 to be detected, which do not allow the tracing of many of the physical quantities in the early Universe such as the stellar mass distribution, metallicity, ionization parameter, and nature of stellar winds responsible for galaxy feedback. A necessary goal for reionization studies is therefore to overcome these small sample sizes and small survey volumes, and to extend in wavelength coverage to measure the rest-frame optical lines.

In addition, the favored cosmic downsizing scenario predicts that, as a result of the feedback mechanism, gravitational collapse and star-formation is more efficient in high-mass halos; this results in rare, massive galaxies forming more efficiently than lower-mass galaxies. Due to the steep shape in the dark matter halo mass function, detecting these galaxies at high redshift requires both deep and wide area surveys of many tens of deg². For instance, at $z\sim10$, there are 40 halos/deg² with masses $M_{halo} > 10^{10}$ $M_\odot$ while there are more than 4000 halos/deg² with $M_{halo} > 10^9$ $M_\odot$. Assuming that the earliest halos maintain the cosmological fraction of baryons to dark matter, this means that there are merely 40 galaxies per square degree more massive than $2\times10^9$ $M_\odot$ in baryons, with an even smaller fraction of stars. These massive sources would be easily missed in small, deep surveys due to their sparsity.

**Upcoming reionization missions.** JWST and the next generation of large, ground-based telescopes will be able to take deep spectra to characterize faint emission line galaxies in the epoch of reionization ($z > 6$). However, due to their relatively small (~10–50 arcmin²) fields of view, detecting the massive, rare sources that require deep, wide-area surveys of many tens of deg² will be difficult for JWST and TMT or ELT to undertake.

WFIRST and Euclid will increase the sample sizes of galaxies and quasars at $z > 6$. Due to the lack of wavelength coverage of the grism spectra beyond 1.8 μm, the identification of $z > 6$ galaxies will be limited to a photometric selection based on the Lyman dropout signature. Confirming and characterizing them requires deep, wide-field spectroscopy that extends to rest-frame optical emission lines. Thus, a dedicated mission with capability to cover wide areas like those of WFIRST, and to similar depths, but with wavelength coverage beyond 2 µm is essential in the post-WFIRST era to advance our knowledge of the distant Universe including reionization (**Fact Sheet Figure 1**).

**CDIM offers comprehensive and complementary probes.** CDIM therefore aims to provide a complete view of cosmic dawn and reionization processes by enabling wide-field, deep spectro-imaging surveys capable of capturing the rare, massive sources and the fainter population, and detecting the key diagnostic lines of Lyα, [OII], Hβ, [OIII], Hα and [NII], as well as the rest-frame UV and optical continuum from galaxies and quasars. With a three-tiered survey covering 0.75–7.5 µm, CDIM is uniquely sensitive to rest-frame optical emission lines over the entire cosmic timeline from the end of cosmic dawn ($z \sim 10$–15) through to the end of the reionization era ($z \sim 6$). CDIM will be able to extend the deep, small surveys by JWST to a wider area and complement data collected from WFIRST to longer wavelength (**Figure 2-1**), with depths sufficient to detect faint galaxies during reionization with stellar masses as low as $10^9$ M$_\odot$ (**Figure 2-5**). The expected CDIM point source depth, surface brightness sensitivity, and spectral line flux sensitivity are shown in **Fact Sheet Figure 2**. We now discuss each of our three key CDIM mission design science themes and their requirements.

## 2.2 Theme I: Galaxy Formation and Evolution during Reionization

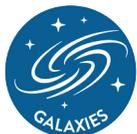

CDIM will access the Lyα and Hα emission from galaxies up to $z\sim10$, which will serve as tracers of EoR galaxy formation and evolution. They are sensitive to the instantaneous rate of star formation. These lines, combined with [OIII], Hβ, [NII] and the rest-frame UV and optical continuum, will provide critical clues to the stellar mass and metal content of first-light galaxies in the Universe. The CDIM *Galaxies* science theme aims to reveal the stellar mass and metallicity build-up of EoR galaxies, and determine their stellar initial mass function, which are critical for the understanding of galaxy formation and evolution at these early epochs.

### 2.2.1 CDIM measures EoR galaxy optical emission lines

**Simulated CDIM galaxy spectrum. Figure 2-1** shows a simulated CDIM spectrum of a massive galaxy ($10^{9.8}$ M$_\odot$) at $z = 6$, including realistic noise level from the CDIM deep survey sensitivity estimate shown in **Fact Sheet Figure 2**. CDIM can clearly detect strong emission lines such as Lyα, Hα, [OIII], Hβ, and [NII] over the rest-frame optical wavelength; the R=300 spectral resolution is sufficient to de-blend the Hα and [NII] emission lines, as shown in **Figure 2-2**, which is required for metallicity measurements. These are critical for the science cases discussed below. As shown in **Figure 2-3**, each of the CDIM three-tier surveys will be able to detect ~$10^4$ galaxies in multiple emission lines, including Hα, [OIII], [NII] and Hβ, at $z = 5$. The CDIM deep survey will detect greater than $10^3$ galaxies via [OIII] and hundreds of galaxies via [NII] at $z = 6$, and hundreds of galaxies via Hα, Hβ and/or [OIII] to $z$ of 8.

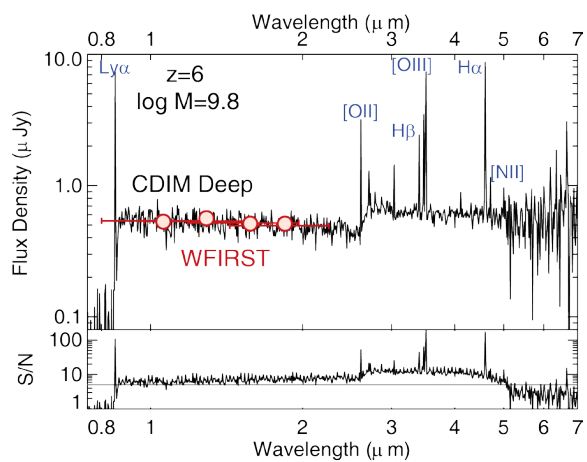

**Figure 2-1. Simulated CDIM spectra of a moderately massive galaxy at $z = 6$, including realistic noise from the CDIM deep survey.** While WFIRST (red) will probe only the rest-frame ultraviolet at R~5, the CDIM R~300 spectrum probes both the rest-frame UV and optical at S/N > 5 (see bottom panel, gray line denotes S/N = 5), allowing the detection of not only the metallicity and ionization-parameter sensitive emission lines, but also the stellar continuum. While JWST is also capable of observing spectra at this depth from 1–5 µm, JWST will only survey small areas and these galaxies will be rare, thus only the CDIM deep survey will build a statistically significant sample of spectra from massive galaxies.

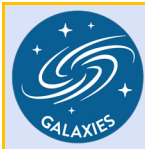

**CDIM spectral resolution requirement.** The nebular-line metallicity measurements require the ratio of [NII]/Hα; that is, the two lines [NII] and Hα need to be to spectrally resolved with adequate spectral resolution to distinguish Hα from [NII], in star-forming galaxies in the reionization epoch. This goal leads to the CDIM spectral resolution requirement of R ≥ 300. At R = 300 we are able to spectrally de-blend the 6549.86 and 6585.27 Å [NII] lines from the 6564.6 Å Hα line (**Figure 2-2**). Once resolved, the spectral line sensitivity is such that we are able to carry out metallicity measurements down to galaxy stellar masses of $10^{9.5}$ M$_\odot$ at $z > 6$ in the CDIM deep field.

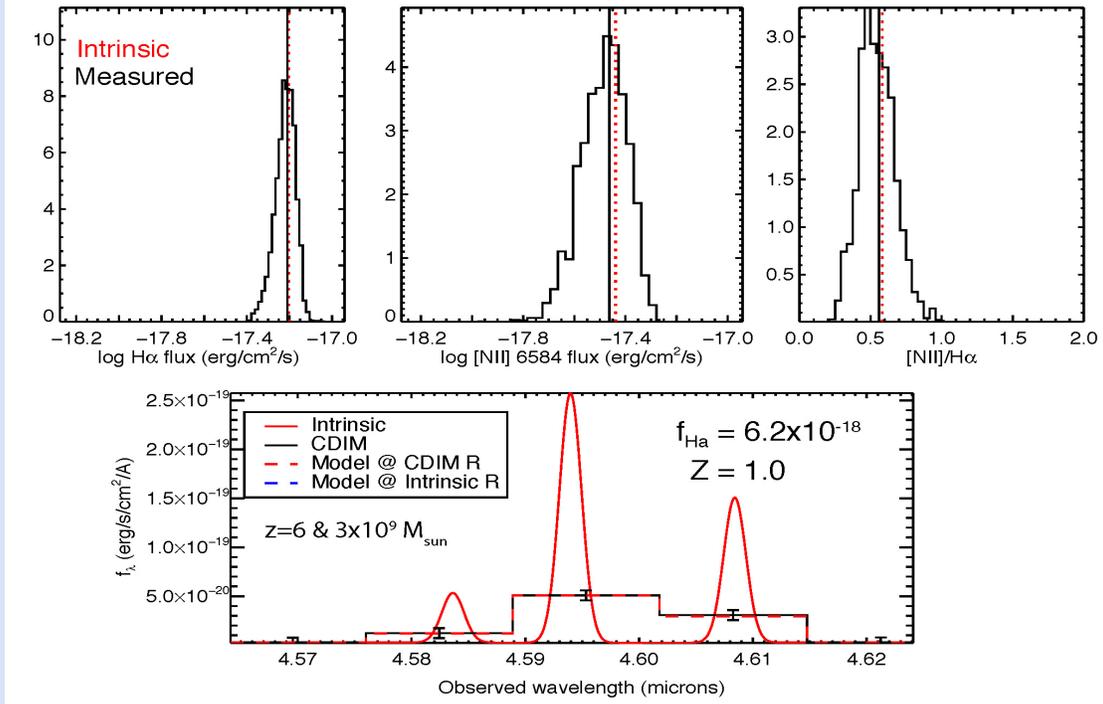

**Figure 2-2. Spectral de-blending of Hα and [NII] with CDIM.** The bottom panel shows a simulated spectrum for a galaxy with solar metallicity, with a Hα flux of 6.2 ×$10^{-18}$ erg s$^{-1}$ cm$^{-2}$. The narrow peaked curves show the input spectrum, while the histogram shows how this object looks sampled at CDIM spectral resolution. We then tested the ability of the CDIM deep survey to recover the ratio of [NII]/Hα with a Markov Chain Monte Carlo method. The top panels show the distribution of MCMC chains for the Hα flux, [NII] 6584 Å flux, and the ratio of [NII]/Hα. The vertical black and dotted red lines show the recovered median and intrinsic values, respectively. As can be seen, CDIM recovers this metallicity-dependent line ratio with high accuracy. In **Figure 2-6** we show the full distribution of the recovered [NII]/Hα ratio for a range of metallicities and Hα line fluxes.

### 2.2.2 Dust-corrected star formation history during reionization

**Hα galaxy detections.** Hα emission is a good tracer of star-formation. To estimate the expected Hα galaxy detections, we compile a set of models to first estimate the Hα luminosity functions, including models: (1) combining star-formation history with the Flexible Stellar Population Synthesis code (Conroy & Gunn, 2010); (2) using the Millennium simulations; (3) deriving from the UV luminosity function with a ratio between Hα and UV luminosity; (4) deriving Hα from the distribution from Stark et al. (2013). These models are broadly consistent with each other and encompass our current theoretical and observational uncertainties at high redshift.

Here we present an estimate of the expected Hα line count predictions based on the UV luminosity function and its evolution paired with the inferred Hα EW distributions at $z \sim 4$ (Stark et al., 2013; Smit et al., 2016), a nebular line model assuming 40% solar metallicity, and a moderate ionization parameter. The expected emission line counts are based on detections above the 3.5σ threshold for each survey configuration.

The CDIM wide survey will detect >$10^5$ galaxies via Hα emission lines at $z = 5$. **Figure 2-4** shows the anticipated depth in luminosity for Hα detections in the deep and wide surveys. Estimated Hα galaxy number counts and forecasts for the [OIII], [NII] and Hβ emission for the CDIM deep, medium, and wide surveys at $z \sim 4$–8, are shown in **Figure 2-3**.

**Cosmic star formation history.** The Hα emission line is a robust tracer of the global star-formation rate of galaxies (Kennicutt Jr, 1998). Given the expected large number of Hα detections, CDIM will provide an unprecedented opportunity to constrain the cosmic star-formation history from $z = 2$ to 8. In addition, the combination of Hα and Hβ measurements enabled by CDIM can be used to correct for dust extinction and improve the accuracy of star-formation rate estimation. **Figure 2-7** shows the expected constraints from CDIM on the global star-formation rate density between redshifts 4 and 8.

### 2.2.3 Stellar mass, metallicity, and dust build-up during reionization

**Stellar mass measurement.** Current state-of-the-art observations have discovered thousands of candidate galaxies as Lyman-break drop-outs at $z > 6$ (e.g., Song et al., 2016), including one at $z = 10$ (Oesch et al., 2016). However, apart from the brightest subset detected in Spitzer at 3.6 and 4.5 μm (e.g., Song et al., 2016), most of these galaxies lack reliable masses and other physical properties (Eyles et al., 2005; Yan et al., 2006). CDIM will increase the number of galaxies in the epoch of reionization with robust stellar masses by 3–4 orders of magnitude. It will measure the stellar mass function of galaxies from $z = 5$ to 8 and down to masses of $10^9$ M$_\odot$ much more robustly and accurately than present results.

The future large area of the WFIRST High Latitude Survey will discover tens-to-hundreds of thousands of galaxies in the epoch of reionization. The abundances of these galaxies trace galaxy evolution at this time in the early Universe (e.g., Behroozi et al., 2013; Finkelstein et al., 2015). CDIM will further provide the rest-frame optical spectrum of these galaxies to measure dust-corrected star-formation rate, determine accurate stellar masses, and allow the discovery of galaxies with little star formation (UV-faint) during EoR.

The left panel of **Figure 2-5** highlights the measurements allowed by CDIM, showing model spectra for both a $z = 5$ and a $z = 8$ galaxy, with stellar masses of $M/M_\odot = 10^{9.3}$ and $10^{8.9}$, respectively. Compared to the CDIM flux limits of deep and wide surveys, shown in red and green horizontal bars respectively, this highlights that galaxies at this modestly large stellar mass (for that epoch) will be detectable at $z = 5$ in the wide survey, and $z = 8$ in the deep survey.

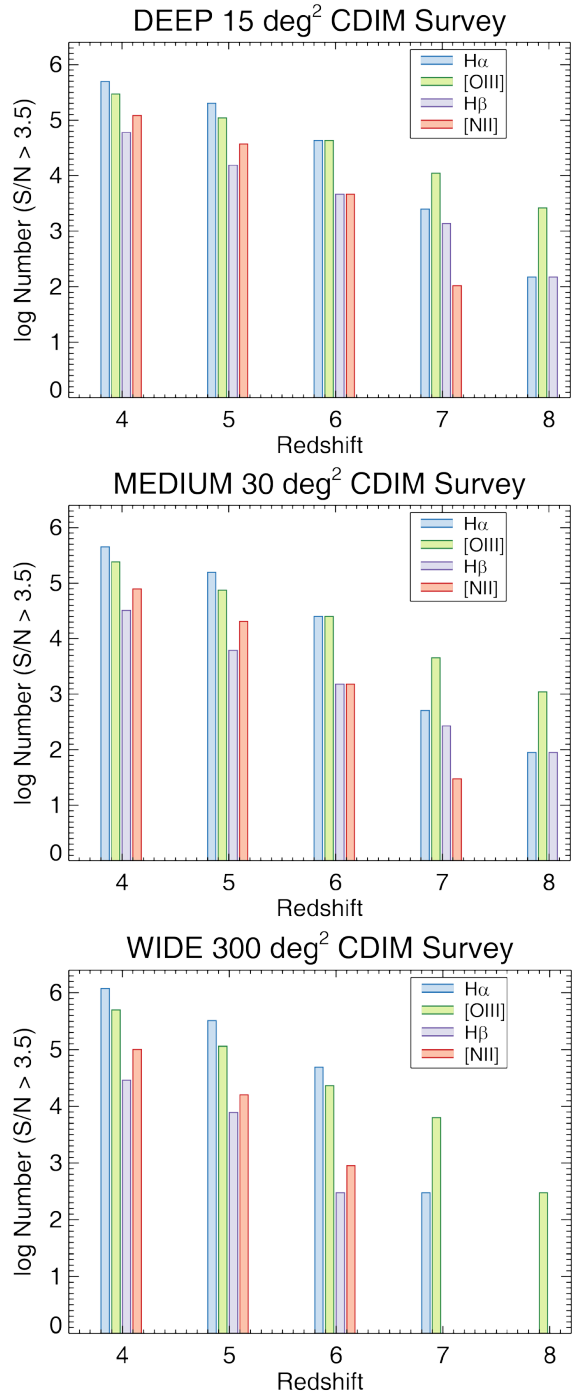

**Figure 2-3.** Predictions for the number of emission lines of Hα, [OIII], Hβ, and [NII] detected in the deep (*top*), medium (*middle*), and wide (*bottom*) surveys. These predictions were made by pairing the evolution of the galaxy luminosity function with inferred Hα EW distributions at $z\sim4$ (Stark et al., 2013; Smit et al., 2016), and a nebular line model assuming 40% solar metallicity, and a moderate ionization parameter.

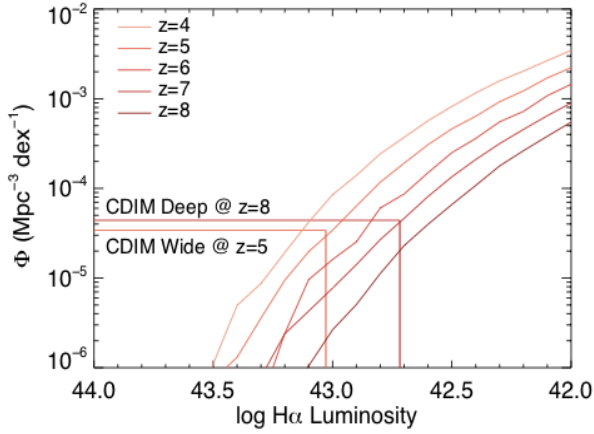

**Figure 2-4. Hα Luminosity function and CDIM luminosity depth.** Predicted Hα luminosity functions, created assuming the Hα equivalent width (EW) distribution from Stark et al. (2013), the rest-frame UV luminosity functions from Finkelstein et al. (2015), and a flat SED (consistent with a ~100–500 Myr continuously star-forming population). The vertical and horizontal lines show the expected luminosity limits for the CDIM Wide survey at $z = 5$, and CDIM Deep survey at $z = 8$.

The right panel of **Figure 2-5** shows the cumulative number of galaxies as a function of stellar mass expected per square degree. Down to our limit of $M/M_\odot = 10^{9.3}$ over the 300 deg$^2$ wide survey, we expect to detect ~100,000 $z > 6$ galaxies in their rest-frame optical emission (3–5 μm observed). In the deep 15 deg$^2$ survey, we will detect and characterize, through stellar mass and other SED-based measurements, at least 15,000 galaxies with $z > 8$.

**Metallicity measurements.** Metallicity can be directly inferred from the so-called $N_2$ index, the ratio of [NII] to Hα emission. Based on metallicity measurements of Lyman-break analogs in the local Universe (Shim & Chary, 2013), the estimated number of galaxies that will have both Hα and [NII] detected at each redshift with CDIM is highlighted in **Figure 2-3**. For instance, the CDIM deep survey can measure the metallicity of ~$10^4$ galaxies via detections of both Hα and [NII] emissions at $z = 5$, as well as the metallicity-sensitive [OIII] emission lines.

In **Figure 2-6**, we highlight the accuracy to which [NII]/Hα can be measured with CDIM spectra as a function of the stellar mass and assumed metallicity of the galaxies. These predictions make use of the existing $z \sim 5$ galaxy measurements to calibrate our simulations. Although the median value of [NII]/Hα is 0.3 in the present-day Universe, the expectation is that [NII]/Hα may be < 0.1 at $z > 4$ due to the limited cosmic time galaxies have had to build up their stellar mass and elemental abundance.

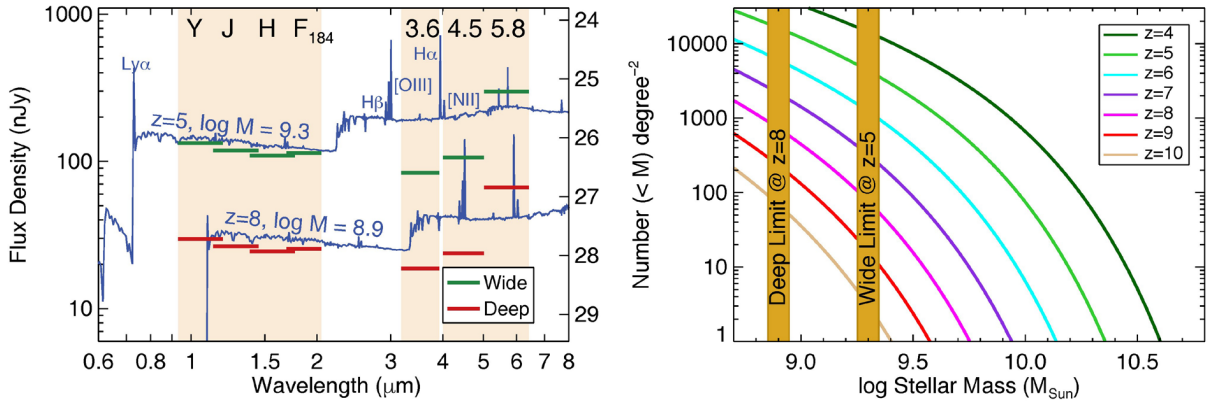

**Figure 2-5. CDIM detects faint galaxies at high redshifts.** *Left:* The two spectra shown are model star-forming galaxies at $z = 5$ and $z = 8$, with stellar masses of $\log(M/M_\odot) = 9.3$ (top line) and $\log(M/M_\odot) = 8.9$ (bottom line). These masses represent the detection threshold in notional wide and deep surveys of CDIM at $z = 5$ and $z = 8$, respectively, where the CDIM broadband (R~5) continuum sensitivities for the 300 deg$^2$ wide survey are indicated by the green horizontal bars and the 15 deg$^2$ deep surveys indicated by the red bars. *Right:* The expected cumulative number of galaxies per square degree as a function of stellar mass; the orange vertical bars denote the approximate stellar mass sensitivity of CDIM at $\log(M/M_\odot) = 9.3$ and $\log(M/M_\odot) = 8.9$ from the left panel. We expect clear detections leading to physical properties for at least $10^5$ $z > 6$ galaxies with CDIM, with detailed characterization of at least $10^4$ galaxies with $z > 8$ from the surveys.

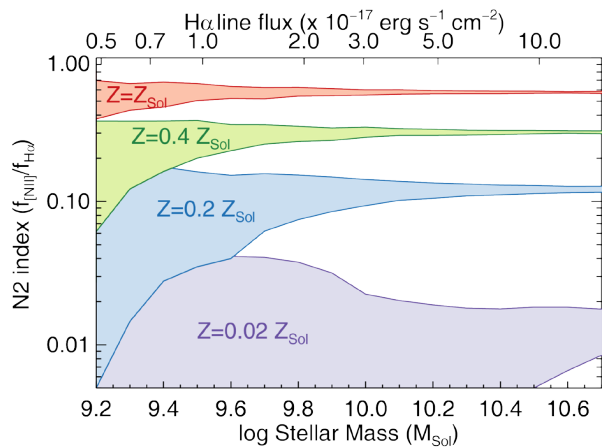

**Figure 2-6. The ratio of the [NII] flux to the Hα flux, which defines the metallicity-sensitive N2 index, versus stellar mass, for simulated galaxies in the CDIM deep survey.** The shaded regions denote the 68% confidence for the recovered N2 index as a function of simulated metallicity. For log (M/M$_\odot$) > 10, the CDIM deep survey will robustly measure the metallicities of galaxies, easily distinguishing between 0.02, 0.2, 0.4, and solar metallicity. At the lowest masses CDIM will probe, CDIM will still distinguish high (~solar) from low (~few percent solar) metallicities.

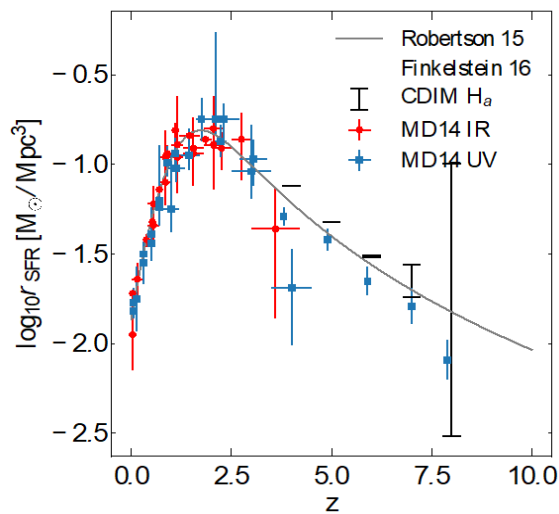

**Figure 2-7. The cosmic star formation history inferred from Hα galaxies detected by CDIM.** CDIM will be able to constrain the star formation history at redshift of 4 to 8, inferred from the measured Hα luminosity function; dust extinction can be corrected for from the Hβ measurements. Over-plotted are IR and UV based data points in Madau & Dickinson (2014), constraints from Finkelstein (2016) and a model from Robertson et al. (2015).

Due to uncertainties on the mass-metallicity relation, we make our predictions for a range of metallicities from solar abundance to 2% solar abundance. While for galaxy stellar masses above $10^{10}$ M$_\odot$, reliable (<5% error uncertainty) N2 index measurements are feasible, for low-mass galaxies the uncertainties are larger, especially when the metallicities are lower than 20% solar abundance.

The numbers, particularly the lower mass limit, are somewhat uncertain, due to the ignorance in the nature of the mass-metallicity relationship at these redshifts. However, by using the Hα derived star-formation rate observed in $z$~5 galaxies, and using the observed evolution of the specific star-formation rate with redshift (Shim et al., 2011), we can derive an estimate of the stellar mass and thereby their metallicity.

**Dust in First-Light Galaxies.** Finally, CDIM is sensitive to the continuum emission of stars as well as the nebular emission lines. The combination allows the evolution of dust extinction curve to be fit, as a function of age, stellar mass, and metallicity. **Figure 2-9** highlights a measurement of the dust extinction law at $z$ of 5, which will be extended to $z$ = 6–8 with CDIM data to address the formation and growth of dust in first galaxies. Using a combination of measurements, CDIM provides a unique window into the growth of all the key components of galaxies—stars, dust, and metals—at the epochs at which the first galaxies were being formed.

## 2.2.4 Separating star-forming galaxies from active galactic nuclei (AGN)

The [NII]/Hα measurements, when combined with [OIII]/Hβ have another application. Extending the Baldwin et al. (1981) BPT diagrams that have been so far constructed out to $z$ of 2 (Coil et al., 2015; Azadi et al., 2017), we can construct a BPT diagram during reionization to study the relative role of AGNs during this epoch. **Figure 2-10** shows results from a simulation matched to CDIM survey sensitivities. Such a construction can confirm if the high ionization parameters – the ratio of ionizing photon density to hydrogen density – expected of these galaxies is due to a top-heavy stellar IMF or due to activity from central massive black holes in them. This can be further confirmed when [NII]/Hα measurements of metallicity are combined with R23 (the ratio of [OII] and [OIII]) lines. This combination directly constrains the ionization parameter (Shim et al., 2011; Kewley et al., 2013), and the stellar initial mass function (Chary, 2008).

## 2.2.5 Stellar initial mass function

The deep CDIM surveys detect the key diagnostic lines allowing simultaneous spectroscopic confirmation and characterization of the physical properties therein. Measurement of [OIII] and Hα at $z \sim 8$ provides a direct measurement of the ionization parameter and the stellar initial mass function in the first stars. Based on deep Spitzer observations and UV spectral properties of galaxies at $z \sim 5$–$6$, there are already hints that the ionization parameter in EoR galaxies is much higher (**Figure 2-11**), although the reason for this is unclear.

It has been argued that this could be due to a top-heavy IMF ($d$N/$d$M$\sim$M$^{-1.7}$) (Chary, 2008), which in turn results in a population of super-stellar mass black holes, a hypothesis which is supported by the enhanced gamma ray burst rates at these redshifts (e.g., Chary et al., 2016).

Alternately, these galaxies could be harboring massive seed black holes that in turn would be bathing the ISM in these galaxies with a strong UV photon field and potentially enhancing the CIV and NIV emission (e.g., Fosbury et al., 2003). Distinguishing between these and other hypotheses, such as a higher binary fraction of stars and an elevated angular momentum distribution (e.g., Levesque et al., 2012; Dorn-Wallenstein & Levesque, 2018), would require using BPT type relationships (**Figure 2-10**) at $z > 6$ which is only possible with the multi-line detection capabilities of CDIM.

## 2.3 Active Galactic Nuclei (AGN) and Quasars during Reionization

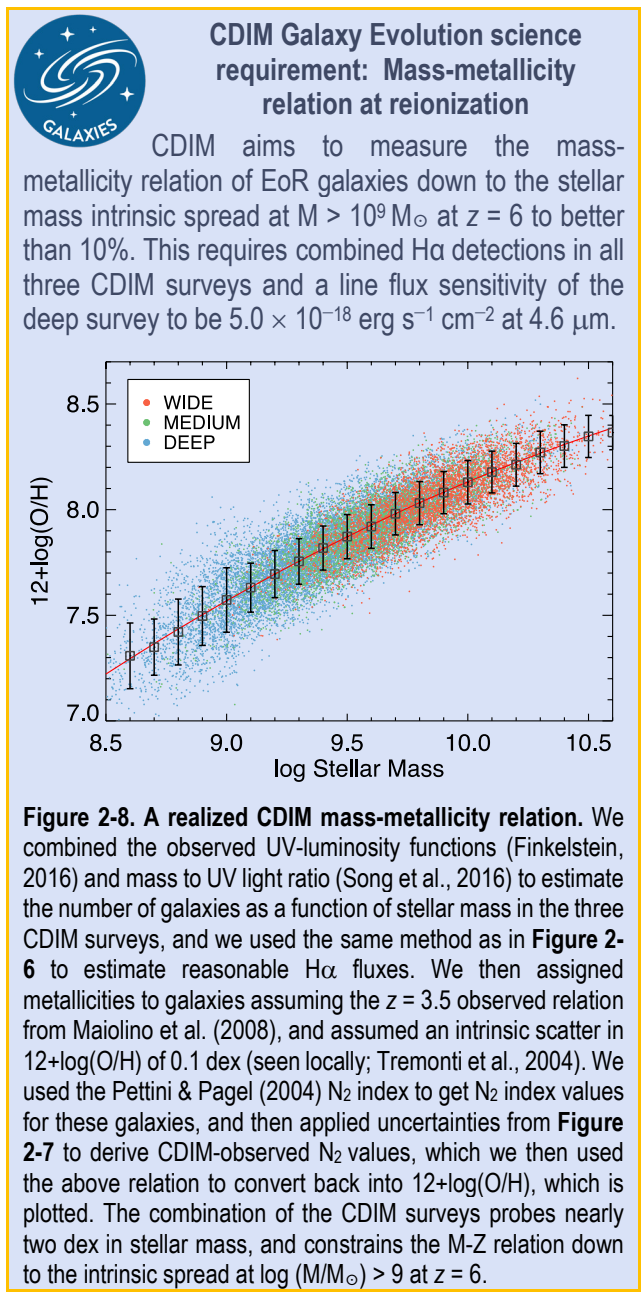

**CDIM Galaxy Evolution science requirement: Mass-metallicity relation at reionization**

CDIM aims to measure the mass-metallicity relation of EoR galaxies down to the stellar mass intrinsic spread at M > $10^9$ M$_\odot$ at $z = 6$ to better than 10%. This requires combined Hα detections in all three CDIM surveys and a line flux sensitivity of the deep survey to be $5.0 \times 10^{-18}$ erg s$^{-1}$ cm$^{-2}$ at 4.6 μm.

**Figure 2-8. A realized CDIM mass-metallicity relation.** We combined the observed UV-luminosity functions (Finkelstein, 2016) and mass to UV light ratio (Song et al., 2016) to estimate the number of galaxies as a function of stellar mass in the three CDIM surveys, and we used the same method as in **Figure 2-6** to estimate reasonable Hα fluxes. We then assigned metallicities to galaxies assuming the $z = 3.5$ observed relation from Maiolino et al. (2008), and assumed an intrinsic scatter in 12+log(O/H) of 0.1 dex (seen locally; Tremonti et al., 2004). We used the Pettini & Pagel (2004) N$_2$ index to get N$_2$ index values for these galaxies, and then applied uncertainties from **Figure 2-7** to derive CDIM-observed N$_2$ values, which we then used the above relation to convert back into 12+log(O/H), which is plotted. The combination of the CDIM surveys probes nearly two dex in stellar mass, and constrains the M-Z relation down to the intrinsic spread at log (M/M$_\odot$) > 9 at $z = 6$.

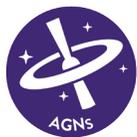

CDIM will provide an unbiased high-redshift AGN sample covering both their rest-frame UV and optical spectra. Quasars can be selected and identified directly from the R=300 spectroscopy, which fully resolves their broad and strong emissions, in particular Lyα, MgII, and CIV, based on which black hole masses can be estimated. CDIM's *AGNs* science theme will measure the anchor luminosity important for determining the faint end of the AGN luminosity function at $z = 6$, and constrain their ionizing photon contribution to reionization.

Understanding the formation and growth of supermassive black holes (SMBHs) is fundamental to two key frontier science issues at the epoch of reionization (EoR). First, how was the Universe reionized? And second, what allowed supermassive black holes to form at high redshift? We discuss these in turn.

Current observations indicate a rather high hydrogen ionizing photon escape fraction (~20%) under the stellar reionization framework, in order for the observed galaxies to have the ability to maintain the state of ionization at $z \sim 6$ under the already somewhat generous assumption with respect to the clumpy factor of the intergalactic medium (e.g., Robertson et al., 2015). An alternative significant source of ionizing photons may be provided by active galactic nuclei (AGN). How much contribution that AGN make to the ionization depends critically on the faint end slope of their luminosity function.

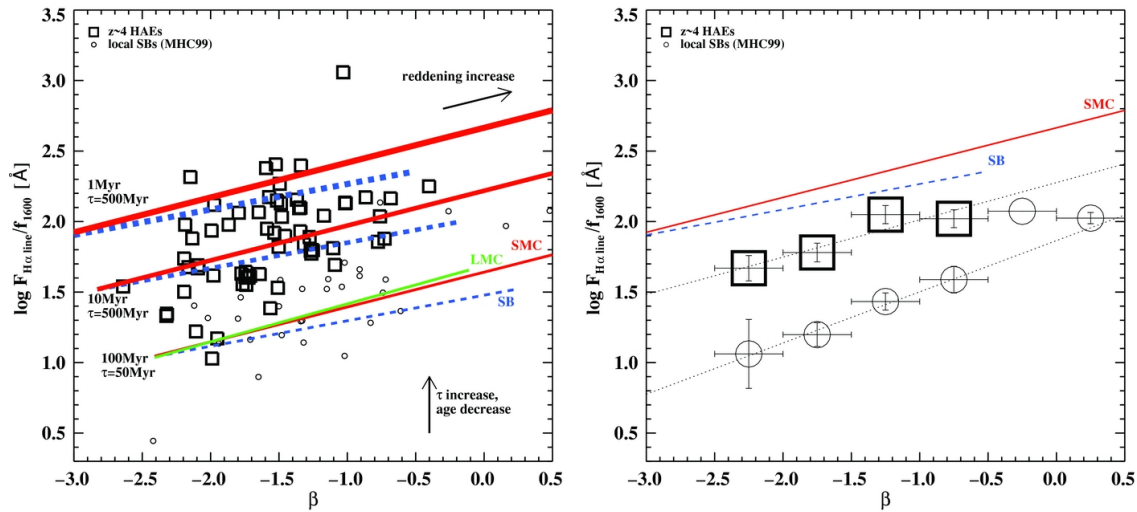

**Figure 2-9. CDIM is capable of studying the growth of dust during reionization.** The detection of optical Balmer lines, particularly Hα to $z \sim 8$ combined with the UV continuum emission, enables a unique characterization of the growth in early galaxies as shown in this figure at $z \sim 5$ from Shim et al. (2011). The left panel shows the Hα/1500 Å UV continuum ratio relative to the UV continuum slope, β, as measured for Hα emitters (HAEs) at $z \sim 4$ compared to the same for local Universe dusty starbursts (SBs). The Hα/1500 Å UV continuum is a proxy for the reddening over a factor of 4 in wavelength while the UV continuum slope β measures the reddening slope over a short wavelength range. The combination of the two is sensitive to both the type of dust and the amount of dust In galaxies. The expected differences are captured by the lines of different star-formation histories and different extinction laws shown in the plot for SMC-type, LMC-type, and SB-type dust. The current data at $z$ are more consistent with SMC/LMC-type extinction law with more small dust grains (slope captured by the red line) dominated by a very young (~10 Myr old) stellar population, with a total lifetime since the initial star-formation $\tau$ of 500 Myr. This observation implies that small grains may have formed first. The right panel shows the same data points, binned by UV slope for clarity. CDIM measurements will extend these studies to the epoch of reionization at $z > 6$ and statistics will allow measurements for UV slope $\beta > -0.5$ to improve the existing determinations that are currently somewhere uncertain even at $z \sim 4$. CDIM will directly address whether SMC type dust continues to appear in the galaxies at $z > 4$ and out to $z \sim 8$. This is particularly important since ALMA has been rather unsuccessful thus far in detecting the FIR continuum from dust emission at $z > 6$ for all but the most extreme massive galaxies.

**AGN luminosity function.** The exact slope of the AGN luminosity function is hotly debated, in large part due to limited statistics of faint high-redshift AGN. Even though the rapid decline of quasar number density at $z > 5$ (e.g., Wang et al., 2018) suggests an AGN contribution on the level of 10% or less (Onoue et al., 2017; Parsa et al., 2017), other deep AGN surveys at high redshift suggest that AGN could be a primary contributor to the ionizing photon budget at $z \sim 6$ (Giallongo et al., 2015). The faint-end slope of the AGN luminosity function is currently below the flux limit of any ground-based AGN survey; even surveys such as Euclid will not be able to reach the AGN population that contributes most to the photo budget, especially considering the limitation of follow-up spectroscopy.

For predictions related to CDIM and to establish survey and mission requirements related to quasar science, we estimate the quasar number counts in the CDIM wide survey based on two quasar luminosity functions: Model A adopts a luminosity function estimated at $z \leq 6$ from Kulkarni et al. (2018), with redshift evolution for $z > 6$ from Wang et al. (2018). Model B is based on the quasar luminosity function from Willott et al. (2015). The key differences are: (1) Kulkarni et al. has a steeper faint end slope and a very bright break magnitude; (2) Wang et al. shows that the evolution at $z > 6$ is steeper than at $z \sim 3$–6, so the number at very high redshift is smaller compared to Model B.

Recent new measurement from a larger sample of faint quasars discovered in the Subaru HSC survey (Matsuoka et al., 2018) is largely consistent with the Willott et al. (2015) result. However, quasar luminosity function at slightly lower redshift ($z \sim 4$–$5$) in deep field with X-ray data showed a much higher density (Giallongo et al., 2015), suggesting that surveys based on optical (rest-frame UV) colors such as SDSS and HSC could have missed a significant fraction of quasars due to dust reddening and extinction. A key advantage of the CDIM quasar survey is that it is based on low-dispersion spectroscopic selection, therefore it is not strongly affected by either broad-band colors or dust extinction. It is expected to be highly complete and able to resolve this controversy and yield a definitive measurement of the faint-end quasar luminosity function and the AGN contribution to reionization.

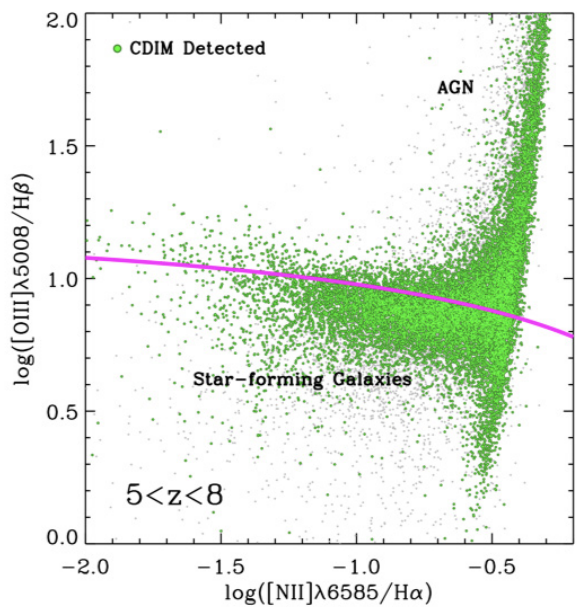

**Figure 2-10. The BPT diagram of [NII]/Hα vs. [OIII]/Hβ, separating the AGN population from star-forming galaxies.** The red line shows the line separating AGN dominated galaxies (above line) from star-formation dominated galaxies (below the line). EoR galaxies are thought to have high ionization parameters which arise from either a top-heavy stellar IMF or a population of massive black holes in them. Optical line diagnostics with CDIM will be able to distinguish between the two scenarios.

**Constraining the AGN ionizing photon contribution.** Based on the above estimates, CDIM is able to provide a reasonable sample at the location of the AGN anchor luminosity ($M_{1450} = -23$ to $-22$), below which the current controversy with respect to the faint slope arises, and sample the bulk of ionizing photon contributions from AGN. Thus, CDIM wide field data on quasar counts will significantly remove the statistical uncertainty that is currently present at this important anchor point of the AGN luminosity function at the faint end. Moreover, additional techniques, such as stacking, to statistically measure the AGN abundance at the luminosity fainter than this anchor point may allow us to significantly firm up the relative contribution of AGN to cosmological reionization. CDIM may be able to constrain the AGN contribution at about 10% at 1σ confidence level.

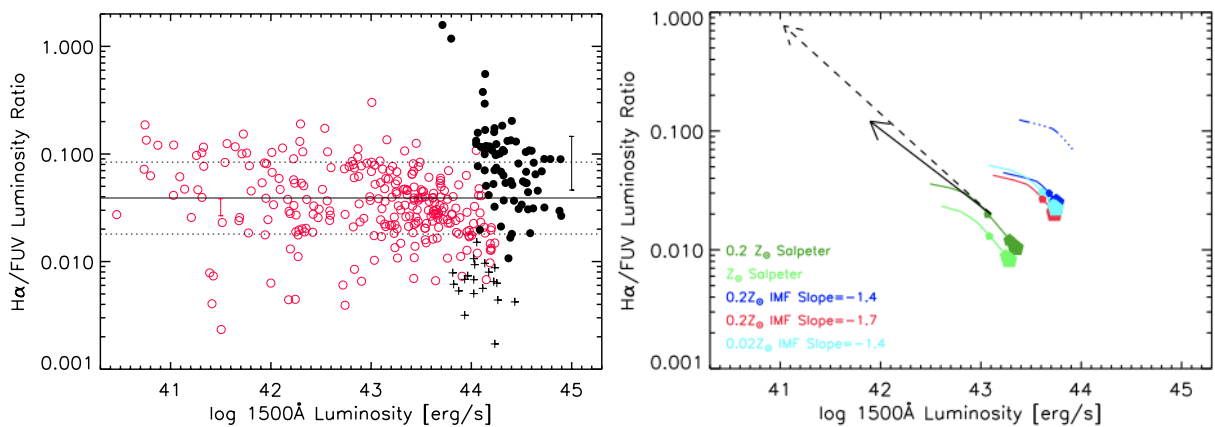

**Figure 2-11. CDIM infers the slope of stellar initial mass function (IMF) of early galaxies.** *Left:* The Hα/UV ratios of local dwarf starburst galaxies (red points) and $z \sim 5$ galaxies (black points) from Shim et al. (2011) and Shim & Chary (2013). Typical $z \sim 2$ galaxies are also shown as the plus symbols. The order-of-magnitude elevated Hα emission from high-$z$ galaxies and dwarf starbursts strongly points to a higher ionization parameter. Attempts to fit this elevated Hα emission (*right*) with models would naively assume dust as the primary cause. However, the non-detection of these galaxies in deep ALMA observations argues for a top-heavy IMF (Chary, 2008) and/or stars with high angular momentum values as argued in Levesque et al. (2012). CDIM will measure both Hα and UV continuum in massive halos to $z \sim 8$, and enabling the measurement of the ratio and thereby the evolution of the stellar IMF as a function of halo mass and redshift.

**Unbiased AGN sample.** The power of the CDIM high-redshift AGN survey relies on the fact that it will provide sensitive low-resolution spectroscopy. Quasars can be selected and identified directly from the low dispersion spectroscopy with their broad and strong emission lines, in particular Lyα, MgII, and CIV, and therefore not affected by the selection incompleteness from traditional color selection. The typical line width of quasar broad lines is ~5000 km s$^{-1}$, fully resolved with R ~300 CDIM spectra. Because the detection is based on the near-IR band, it is also less affected by dust extinction. At R ~300, the CDIM spectra will not only confirm and measure the redshift of AGNs, but also deliver key measurements of the AGN characteristics. In particular, for the redshift of interest, the MgII emission line will be in the sensitive part of CDIM wavelength coverage, and yield a reliable estimate of the central black hole mass measurements. This will produce a complete census of the active BH mass function, and the distribution of BH accretion rate in these systems, putting the strongest constraint yet on the growth rate of supermassive black holes (SMBHs), key to the second fundamental AGN science goal discussed below. In addition, the CDIM spectra will sample key emission lines, such as CIV and FeII. The relative line ratios of these transitions are sensitive to the chemical abundance and enrichment history in the quasar environment.

> **CDIM AGN science requirement: Ionizing photon contribution to reionization**
>
> CDIM aims to quantify the faint-end slope of AGN luminosity function (LF) at $z = 6$ by measuring down to the anchor luminosity of $M_{1450}$ = −23 to −22 and constrain its LF amplitude to better than 10%, which can be used to infer the total AGN ionizing photon contribution to reionization. This requires a detection of 100 AGNs at $z = 6$, independent of model assumption, at the corresponding flux limit. The expected number of quasars during reionization as a function of the K-band magnitude is summarized in **Figure 2-12**. CDIM thus requires a K-band magnitude limit of 23.5 (AB; 5σ) and a survey size of 300 deg$^2$. This is achieved with the CDIM wide survey and with sufficient margin in that the current design leads to a survey with 5-σ depth of 24 mag (AB).

**Black hole mass estimate.** The second question pertains to how the supermassive black holes at high redshift arose. The discovery of dozens of bright quasars with luminosities > $10^{47}$ erg s$^{-1}$ at redshift $z = 6-7$ indicates that at least some galaxies have been able to grow their SMBHs as massive as up to $10^{10}$ solar masses when the Universe was less than 1 Gyr old (Wu et al., 2015). Although these high-redshift quasars are relatively rare (with a space density of order 1 Gpc$^{-3}$), found in large volume surveys, their mere existence poses a severe challenge to the traditional model of SMBH growth via gas accretion, simply due to a finite number of e-folding (i.e., Salpeter) times available.

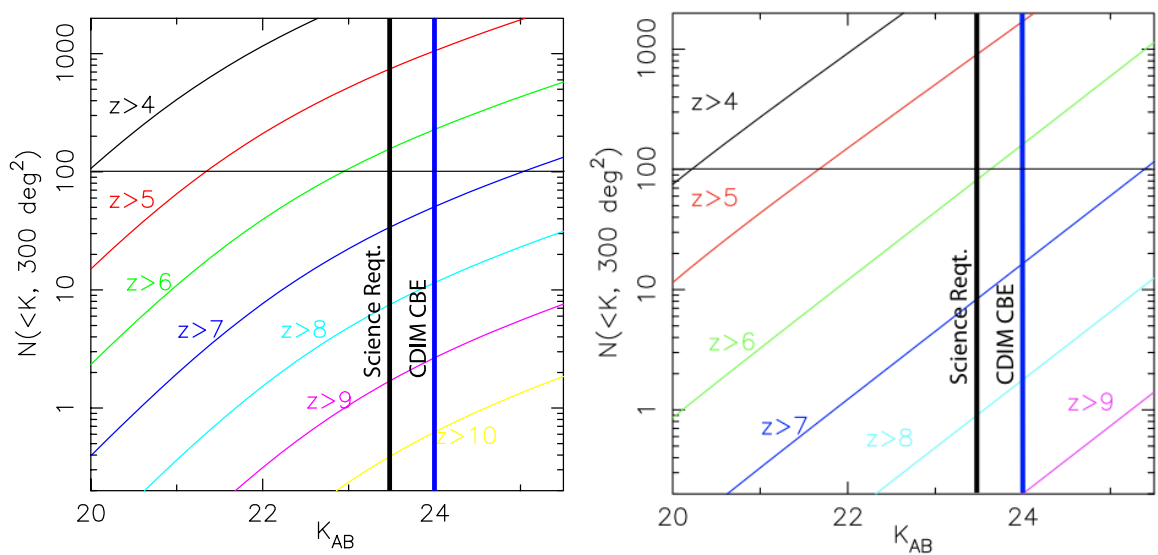

**Figure 2-12. Quasar number counts predictions for the CDIM wide survey as a function of K-band magnitude, based on two models in the literature.** *Left:* Based on the luminosity function estimated at $z < 6$ from Kulkarni et al. (2018), with redshift evolution for $z > 6$ from Wang et al. (2018). *Right:* Based on the quasar luminosity function from Willott et al. (2015). Our science requirement of 100 quasars at $z > 6$ requires a survey depth of about 23.5 AB K-band magnitude at 5σ. The CDIM design achieves a magnitude depth of 24.0 AB K-band (5σ).

The fact that accretion onto quasars shining at the peak of their activities at $z = 2-3$ appears to be able to account for the local SMBH mass density (e.g., Yu & Tremaine, 2002) and the fact that these observed high-redshift quasars look similarly mature compared to their counterparts at moderate redshifts (e.g., Fan et al., 2003) would be consistent with (though not necessarily required by) their SMBHs having also grown primarily via gas accretion.

In this picture, there is simply not enough time to grow stellar mass black hole seeds to the observed SMBH masses without invoking significantly super-Eddington accretion rates over a large fraction of the available time, regardless of the initial starting redshift. On the other hand, SMBH seeds of mass in the range of $10^5-10^6$ $M_\odot$, if they exist, would alleviate the time bottleneck to be able to grow to the SMBH mass of the observed quasars without having to resort to super-Eddington accretion. Although the physical origin of these proposed massive seeds remains an open question, several proposals have been put forward, including the channel of direct collapse of protogalactic massive clouds (e.g., Oh & Haiman, 2002; Latif & Ferrara, 2016).

Separately, despite observational challenges, extant observational evidence indicates a rapid decrease of the abundance of massive black holes below about $10^6$ $M_\odot$ in the local Universe (e.g., Greene & Ho, 2007; Goulding & Alexander, 2009), an intriguing hint that $10^6$ $M_\odot$ could be the mass of seed black holes. Despite the lack of direct sensitivity, due to the large volume covered by CDIM surveys, a combination of statistical probes (e.g., stacking) could allow us to probe down to this seed black hole mass level. This will allow us to determine the abundance of SMBHs at the lowest expected mass range leading to unprecedented quantitative insight on the formation mechanism and growth of SMBHs during reionization.

CDIM will be able to address these twin fundamental questions at the high-redshift frontier for the first time. The critical requirements for both hinge on our ability to detect an order-of-magnitude larger sample of AGNs with SMBH masses at around $10^7$ $M_\odot$ and, through innovative analysis technique, a statistical determination of the abundance of fainter AGNs below $10^7$ $M_\odot$ by a factor of 5–10.

## 2.4 Tomography and Reionization History of the Intergalactic Medium (IGM)

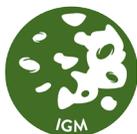

CDIM will provide tomographic measurements of the intergalactic medium (IGM) during reionization. CDIM will detect Lyα emitting galaxies and infer the Lyα escape fraction and neutral fraction of the surrounding intergalactic medium. In addition, CDIM will measure the intensity fluctuations of Lyα and Hα emission via the intensity-mapping technique, and construct a large-scale three-dimensional map of the cosmic dawn and reionization processes that is comparable in scale and redshift range and synergistic to 21-cm EoR experiments. The CDIM IGM science theme aims to reveal the history and topology of reionization and constrain the ionization fraction of the Universe across the key redshift range $z = 6–10$.

### 2.4.1 Lyman-α emitting galaxies

Lyα emission from massive stars is a key feature of star-forming galaxies. Lyα emitting galaxies (or Lyα emitters, LAEs) have become an essential probe of the high-redshift Universe, and studying such a population of galaxies can advance our understanding of galaxy evolution during reionization. The Lyα emission from LAEs is highly sensitive to the presence of neutral hydrogen, and observations of LAEs and their clustering, and Lyα intensity distribution can be used to constrain the ionization state of the IGM and thus learn about the reionization process.

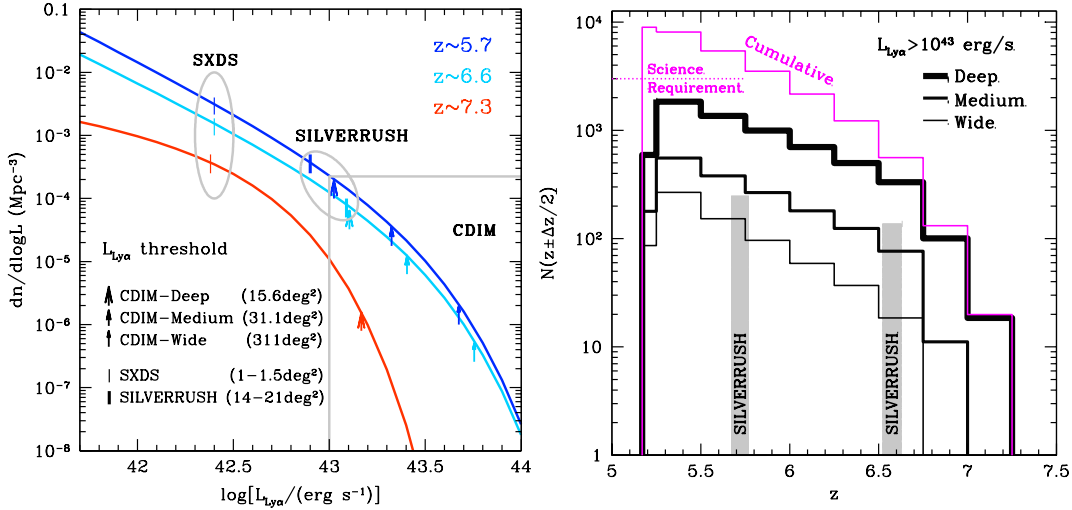

**Figure 2-13. Detecting Lyα emitters (LAEs) with CDIM.** *Left:* LAE luminosity function (LF) and CDIM detection thresholds. The three curves represent the best-fit luminosity functions of LAEs at three redshifts based on existing observations, with the short vertical lines indicating detecting thresholds and arrows indicating the detection thresholds of CDIM surveys. *Right:* Conservative forecast of the number of LAEs from CDIM. In total, about ~$10^4$ luminous LAEs (with Lyα luminosity above $10^{43}$ erg s$^{-1}$) may be discovered by CDIM in the redshift range of 5 < z < 7.5, well above the science requirement (dotted line). Compared to the SXDS and SILVERRUSH surveys with the Subaru Telescope, CDIM will provide significantly better constraints on the luminous end of the Lyα LF in a much wider redshift range.

With CDIM's spectro-imaging capability, we can directly identify individual LAEs from their Lyα emission. While such detections are expected to be limited to the most luminous galaxies, the Lyα emission from fainter LAEs can be studied with stacking techniques based on Hα observations.

**Lyα emitter (LAE) source counts.** First, we estimate the number of LAEs above $z \sim 5$ that can be discovered by CDIM with the expected sensitivity. At present, ground-based detections are focused on three redshifts determined by the atmosphere transmission windows. For the CDIM forecast, we adopt and interpolate/extrapolate the best-fit parameters of $z$ ~5.7, 6.6, and 7.3 Lyα LF of LAEs based on surveys with the Subaru Telescope (Konno et al., 2014; Konno et al., 2017). To be conservative, the calculation is done by assuming that images are constructed with two spectral resolution elements (to cover the broadest possible Lyα lines).

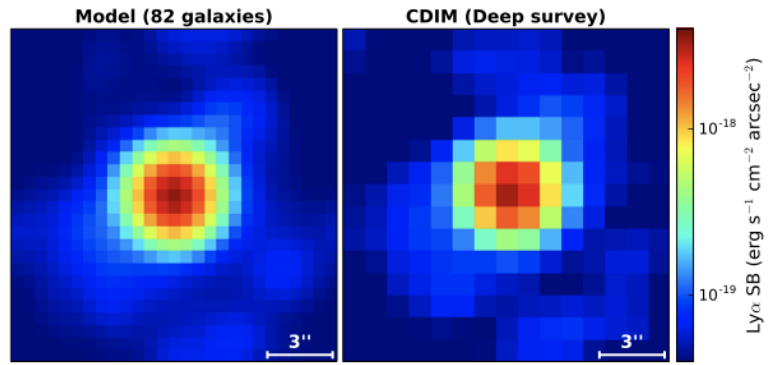

**Figure 2-14. A simulation of Lyα emission from galaxies at $z$~5.7 observed with CDIM.** *Left:* Map of Lyα emission around a stack of galaxies at $z$~5.7 predicted from Lyα Monte-Carlo radiative transfer calculations applied to a cosmological simulation of reionization. The map was obtained by stacking individual Lyα images around galaxies that will be detected in Hα with CDIM. Each image corresponds to a 15×15" region around a galaxy with a thickness of 1 CDIM spectral resolution element. *Right:* The same Lyα map observed with CDIM. The image has been rebinned to match the pixel size of CDIM (1") and smoothed with a 2D Gaussian filter with a FWHM of 2" to account for the CDIM PSF. Random white noise with a rms scatter corresponding to the CDIM line flux sensitivity (**Fact Sheet Figure 2**) has been added.

The results are shown in **Figure 2-13**. Compared to the Subaru SXDS and SILVERRUSH surveys, currently the largest, CDIM will provide significantly better constraints on the luminous end of the Lyα LF (*left*) in a much wider redshift range (*right*). Based on the conservative calculation, we expect to discover about $10^4$ LAEs at $5 < z < 7.5$ with Lyα luminosity above $10^{43}$ erg s$^{-1}$. In combination with the UV LF (e.g.,

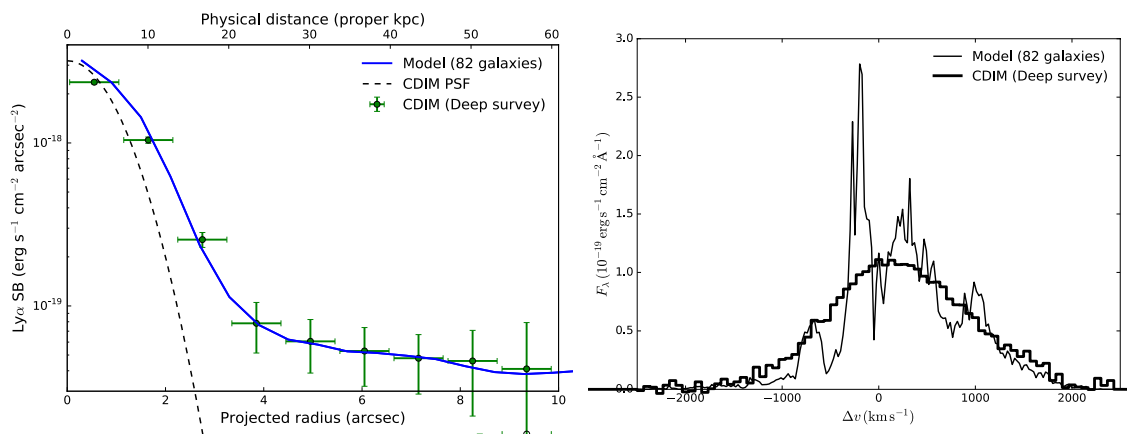

**Figure 2-15.** *Left:* **Measuring the Lyα surface brightness (SB) profile around galaxies at *z* ~5.7 with CDIM.** The profiles are calculated from the pixel values of the Lyα images shown in **Figure 2-14**. The blue line shows the profile corresponding to the Lyα map predicted from the simulations (**Figure 2-14**, *left*). Measurements of the Lyα SB profile obtained with CDIM are shown in green, with the vertical error bars corresponding to the CDIM SB sensitivity and the horizontal error bars showing the size of CDIM pixels. *Right:* **Stacking Lyα spectra of Hα detected galaxies at *z* ~5.7 with CDIM.** The spectra are calculated within an aperture of 1" projected radius around each individual galaxy. The blue line indicates the stacked spectrum obtained from Lyα radiative transfer calculations applied to a cosmological simulation of reionization. The green dotted line corresponds to the spectrum smoothed with a Gaussian filter of 1000 km s$^{-1}$ (R = 300). The green solid line shows the Lyα spectrum as observed by CDIM with a spectral resolution of R = 300.

Bouwens et al., 2015), we expect to improve the constraints on the evolution of the IGM neutral fraction (e.g., Konno et al., 2017).

At present, the luminous end of the $z \sim 7$ Lyα luminosity function is not well constrained, and there is a hint of substantial enhancement above $10^{43.2}$ erg s$^{-1}$ based on a handful of sources (Zheng et al., 2017). If this is true, it would be related to the reionization process in over-dense regions. However, given the limited volumes of current surveys (Konno et al., 2014; Zheng et al., 2017), the existence of the enhancement is still under debate. With more than 100 sources at this redshift, CDIM will settle the debate and provide the crucial data for investigating the implications for reionization.

**Lyα stacks of Hα sources.** Thanks to its large wavelength coverage, CDIM will be able to detect both Lyα and Hα emission lines from high-redshift galaxies. While CDIM will detect >$10^5$ Hα sources at $z > 6$, observation of Lyα emitting galaxies will be limited to the brightest objects (**Figure 2-13**). The resonant scattering of Lyα photons off neutral hydrogen in the circumgalactic medium (CGM) and intergalactic medium (IGM) makes them diffuse both spatially and in frequency (e.g., Zheng & Miralda-Escudé, 2002). A substantial fraction of Lyα emission from individual high-$z$ galaxies contributes to the diffuse halo component, making Lyα harder to detect than non-resonant lines such as Hα. In order to reveal this diffuse Lyα emission component, we will take advantage of the unique capabilities offered by CDIM by stacking images at the Lyα wavelength around positions of Hα detected galaxies with CDIM.

In **Figure 2-14**, we simulate a stacked Lyα image at $z \sim 5.7$ obtained from Lyα radiative transfer calculations applied to a 50 Mpc h$^{-1}$ (comoving) cosmological simulation of reionization (Trac et al., 2015). The left panel shows the Lyα image predicted from the simulation and the right panel is the image as observed with the CDIM deep survey. The corresponding Lyα surface brightness profiles are shown in

> 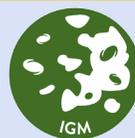 **CDIM Intergalactic Medium science requirement:**
> **Lyα escape fraction and IGM neutral fraction**
>
> A high precision measurement of the Lyα escape fraction from CDIM can provide valuable constraints on the IGM neutral fraction during reionization. We place the science requirement at 10% precision measurement of Lyα escape fraction at $z = 6.5$, as shown in **Figure 2-16**, which would constrain the IGM neutral fraction to be 0.22 ± 0.04 (Furlanetto et al., 2006; Dijkstra et al., 2007). This requires a CDIM line flux sensitivity of the deep survey to be 2.9 × 10$^{-17}$ erg s$^{-1}$ cm$^{-2}$ at 0.85 μm.

**Figure 2-15**, *left*. The two plots demonstrate the powerful capability of CDIM to detect the diffuse Lyα emission reliably from high-$z$ galaxies, which can be used to probe the CGM of high-redshift galaxies and constrain reionization (Momose et al., 2014, 2016).

**Lyα stacked profile.** In addition to the Lyα surface brightness profiles around galaxies, CDIM will also be able to measure the Lyα spectrum obtained from the stacking analysis of Hα detected galaxies (**Figure 2-15**, *right*). The advantage of CDIM to detect both Hα and Lyα is that Hα can be used to measure the systemic redshift of the galaxy. This is crucial in order to extract constraints from the Lyα spectrum as the Lyα line is often offset from the systemic redshift. CDIM will thus be able to measure this velocity offset from stacked Lyα spectra of Hα detected galaxies (green solid line in **Figure 2-15**, *right*), which can probe galactic wind and CGM properties for high-redshift galaxies.

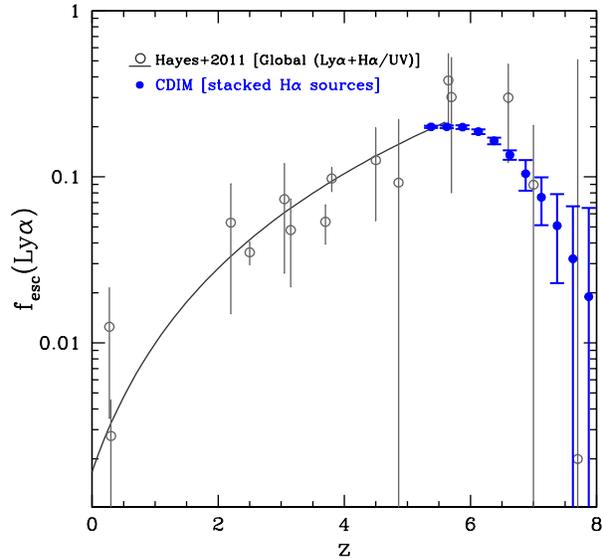

**Figure 2-16**. **Lyα escape fraction.** Lyα escape fraction with CDIM. Open circles and the corresponding fitting curve show the global Lyα escape fraction based on the ratio of Lyα and Hα/UV luminosity densities from the observed luminosity functions (Hayes et al., 2011). With stacked Lyα image of sources detected through Hα, CDIM will constrain the Lyα escape fraction at $z$ ~5.3–8 (filled circles). Besides the ISM effect, towards higher redshift, the observed Lyα escape fraction would be more and more affected by the IGM transmission (hence the drop), providing a sensitive probe of reionization (e.g., Konno et al., 2014; Konno et al., 2017).

**Lyα escape fraction and IGM neutral fraction.** Because of the radiative transfer effect, the observed Lyα luminosity is on average lower than the intrinsic one, usually characterized by the Lyα escape fraction $f_{esc}(Lyα)$. The global Lyα escape fraction can be estimated by the ratio of Lyα and Hα/UV luminosity densities (e.g., Hayes et al. (2011); open circles and the fitting curve in **Figure 2-16**). At low redshifts, the observed Lyα escape fraction is mainly contributed by the radiative transfer in the ISM and CGM (e.g., dust absorption and spatial diffusion). At high redshifts, IGM transmission can play a major role in shaping the Lyα escape fraction, which offers a sensitive probe of reionization.

For sources detected through Hα with CDIM, the intrinsic Lyα luminosity can be inferred. Together with the stacked Lyα image, we will be able to measure the mean Lyα escape fraction in the redshift range $z$ ~5.3 to ~8. The prediction is shown as filled circles in **Figure 2-16**, with error bars from both source counts and spectral uncertainties. The model assumes a mean Lyα escape fraction of 0.2 before IGM transmission (Hayes et al., 2011) and both interpolates and extrapolates the IGM transmission constraints at $z$ =5.7, 6.6, and 7.3 from the Subaru surveys (Konno et al., 2014; Konno et al., 2017). The distribution of Lyα escape fraction at fixed Hα luminosity is set to be log-normal with a scatter of 0.5 dex (Zheng et al., 2010). The Hα source counts are based on the redshift-dependent UV luminosity functions in Finkelstein (2016) and the CDIM sensitivity. CDIM can reliably constrain the redshift evolution of Lyα escape fraction at $z$ ~ 5–7, based on which valuable constraints on reionization, e.g., the characteristic size of ionized bubbles (Dijkstra et al., 2007) and the IGM neutral fraction (Furlanetto et al., 2006; Dijkstra et al., 2007; Konno et al., 2014; Konno et al., 2017) can be derived. For example, at $z$ ~ 6.7, a ~10% precision measurement of Lyα escape fraction would constrain the IGM neutral fraction to be 0.22 ± 0.04, based on the models shown in Figure 6 of Dijkstra et al. (2007) and Figure 1 of Furlanetto et al. (2006).

### 2.4.2 Tomography and History of Reionization

Intensity mapping of galactic emission lines at high redshifts provide a unique and new probe of reionization (Visbal & Loeb, 2010; Gong et al., 2011a; Gong et al., 2011b; Lidz et al., 2011). Intensity mapping refers to low-resolution mapping that measures the integrated emission from large volumes. It measures the *cumulative* emission of many faint galaxies and, by using spectral lines such as Lyα and Hα emission (Silva et al., 2013; Pullen et al., 2014; Heneka et al., 2017), the cosmological redshift allows us to map these sources in three dimensions. The measurement is sensitive to the integrated emission from all sources, including the faint dwarf galaxies that are thought to be responsible for reionization. It is thus complementary to both deep pencil-beam JWST surveys and planned WFIRST detections that individually resolve galaxies at higher luminosities (§2.2).

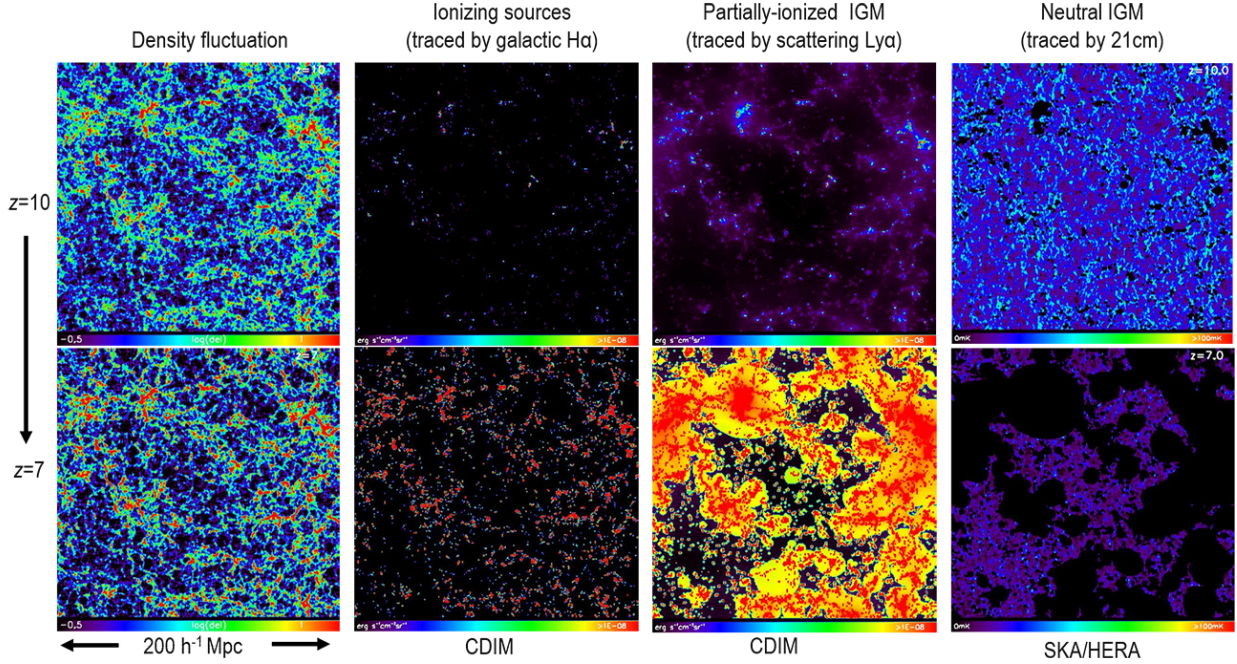

**Figure 2-17. Signals accessible by intensity mapping during reionization.** Simulations at *z*=7 and *z*=10 of density, Hα, Lyα, and 21-cm signals are from Heneka et al. (2017). The CDIM FoV extends beyond the simulation boxes of 200 h$^{-1}$ Mpc shown here. CDIM will map Lyα and Hα in the course of reionization, following the growth of regions ionized by galactic sources and enabling cross correlation and validation with the 21-cm signal as measured by SKA.

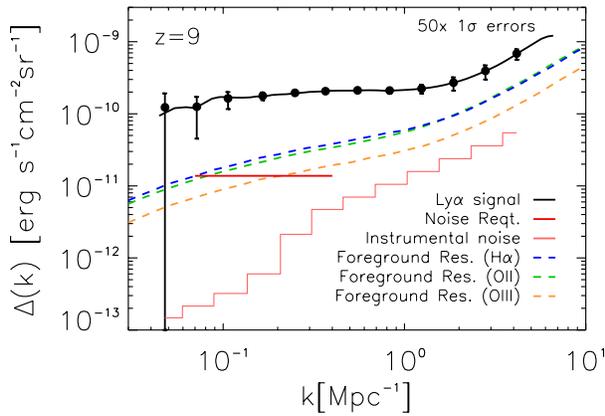

**Figure 2-18. Lyα intensity power spectrum during reionization at z = 9 in a redshift bin width of 1.0 (z = 8.5–9.5).** The data points in black are projected band-powers with the sensitivity specified by the CDIM experiment. The low-z Lyα foreground line emissions are shown in dashed lines for Hα (blue), OII (green) and OIII (orange). With the high flux masking of Hα, [OII], and [OIII] foreground sources, the total foreground residual shown in solid blue line becomes negligible to the targeted signal. CDIM detector noise is shown as a stair-step line, while the science requirement on the noise sensitivity to achieve at least a S/N ratio of 500 measurement in a Δz = 1.0 redshift bin during reionization (see Figure 2-19) is shown by a horizontal line.

**Intensity fluctuations of Lyα, Hα, and 21-cm.** Simulated Hα, Lyα, and 21-cm intensity maps, used as basis for our model predictions (Heneka et al., 2017), are shown in **Figure 2-17**, while the CDIM based Lyα power spectrum is shown in **Figure 2-18**. Hα and Lyα emission originating from star formation activities is expected to be associated with the location of galaxies, while Lyα photons can further resonant scatter in the IGM and appear spatially diffused. Combining Lyα and Hα tomographic maps can thus reveal the propagation of Lyα photons, inferring the physical state of the IGM through study of radiative transfer effects (e.g., Heneka et al., 2017).

**CDIM is complementary to 21-cm EoR experiments.** EoR tomography mapping of Lyα and Hα (Silva et al., 2013; Pullen et al., 2014; Comaschi

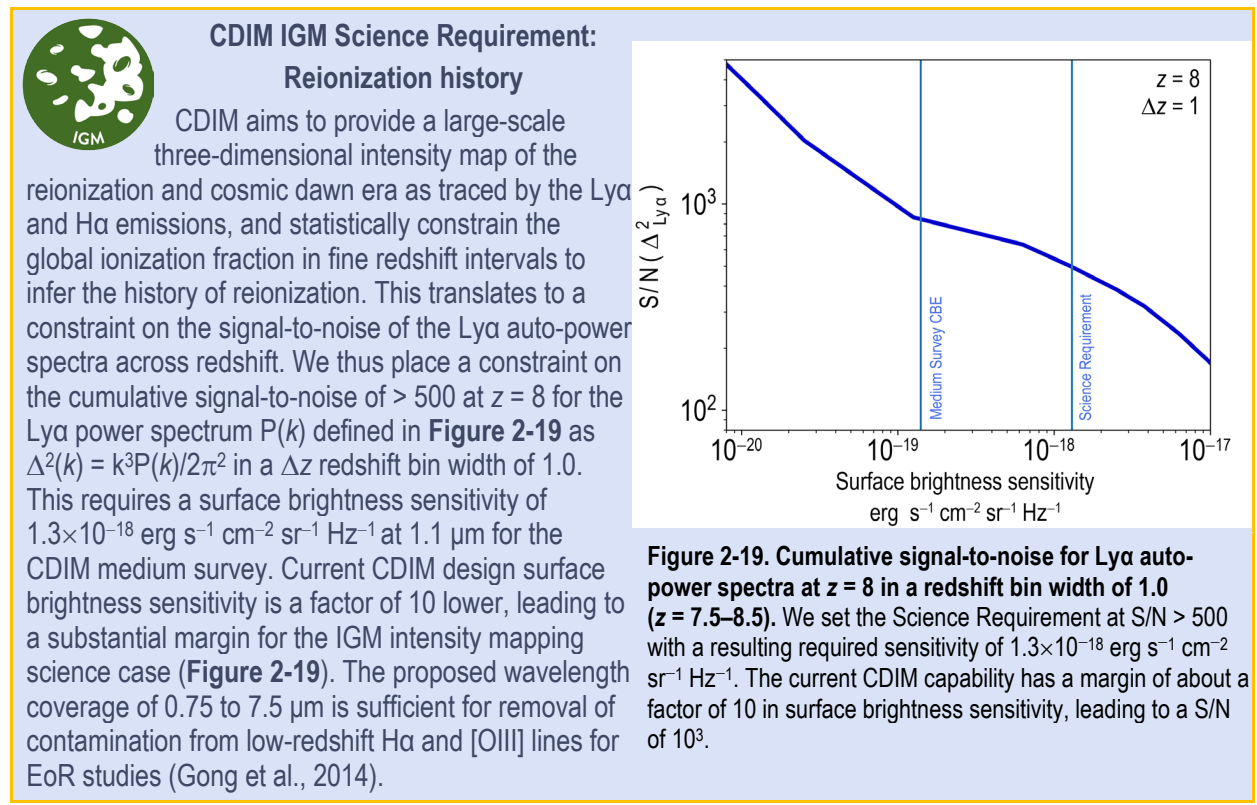

**CDIM IGM Science Requirement: Reionization history**

CDIM aims to provide a large-scale three-dimensional intensity map of the reionization and cosmic dawn era as traced by the Lyα and Hα emissions, and statistically constrain the global ionization fraction in fine redshift intervals to infer the history of reionization. This translates to a constraint on the signal-to-noise of the Lyα auto-power spectra across redshift. We thus place a constraint on the cumulative signal-to-noise of > 500 at $z = 8$ for the Lyα power spectrum $P(k)$ defined in **Figure 2-19** as $\Delta^2(k) = k^3 P(k)/2\pi^2$ in a $\Delta z$ redshift bin width of 1.0. This requires a surface brightness sensitivity of $1.3\times10^{-18}$ erg s$^{-1}$ cm$^{-2}$ sr$^{-1}$ Hz$^{-1}$ at 1.1 µm for the CDIM medium survey. Current CDIM design surface brightness sensitivity is a factor of 10 lower, leading to a substantial margin for the IGM intensity mapping science case (**Figure 2-19**). The proposed wavelength coverage of 0.75 to 7.5 µm is sufficient for removal of contamination from low-redshift Hα and [OIII] lines for EoR studies (Gong et al., 2014).

**Figure 2-19. Cumulative signal-to-noise for Lyα auto-power spectra at $z = 8$ in a redshift bin width of 1.0 ($z = 7.5$–8.5).** We set the Science Requirement at S/N > 500 with a resulting required sensitivity of $1.3\times10^{-18}$ erg s$^{-1}$ cm$^{-2}$ sr$^{-1}$ Hz$^{-1}$. The current CDIM capability has a margin of about a factor of 10 in surface brightness sensitivity, leading to a S/N of $10^3$.

& Ferrara, 2015) complement the planned 21-cm surveys from ground-based low-frequency radio interferometers (Chang et al., 2015; Feng et al., 2017; Heneka et al., 2017). Since galaxies are expected to trace the over-dense regions that are likely to be reionized first, and since the 21-cm emission traces the neutral IGM, line emissions from galaxies and the 21-cm fluctuations will anti-correlate on the size scale of the ionized regions (bubbles). Cross-correlations of Lyα or Hα tomographic maps with the 21-cm EoR surveys thus directly measure the characteristic bubble sizes during reionization (Silva et al., 2013; Chang et al., 2015).

The Square Kilometer Array Low Frequency Aperture Array (SKA1-LOW) construction is planned for 2019–2023, and its deep 21-cm EoR survey will directly detect and image the tomographic structure of neutral/ionized regions at $z > 6$ over 100–300 deg$^2$ (Koopmans et al., 2015). With a resolution of 5 arcmin and a FoV of 5° FWHM at $z = 6$, SKA1-LOW will reach a 1-sigma sensitivity to image 21-cm spatial structures on 5–300 arcmin scales at R > 500. CDIM is similar in scope to SKA1-LOW, and is thus an ideal probe of the EoR complementary to the large-scale 21-cm experiments. The CDIM medium survey is designed to optimize 21-cm cross-correlation tomography studies in terms of survey depth, size, and field location.

**Reionization topology: bubble size evolution.** The cross-correlation (shown in **Figure 2-20**, *left*) captures the fact that Lyα or Hα lines from galaxies are anti-correlated with 21-cm emission from neutral hydrogen in the IGM on a size scale proportional to the ionization bubbles, carved out of the neutral IGM by UV photons, a measure that is sensitive to the ionization history of the IGM (**Figure 2-20**, *right*). CDIM will not only establish that anti-correlation, but will also measure the average bubble sizes during EoR, establish the bubble size distribution function, and study the growth of ionization bubbles from $z = 8$ to 5. Cross-correlation techniques are robust against foreground contaminations, since continuum or line foregrounds in each of these maps are (to first order) not correlated. The anti-correlation can be reliably constrained even in the presence of residual foregrounds through cross-power spectrum measurements. Given that the 21-cm imaging with SKA1-LOW and other 21-cm experiments such as HERA (DeBoer et al., 2017) are likely to be foreground-limited, the combination with an external tracer, such as the three-dimensional Lyα and Hα intensity maps enabled by CDIM, will likely become crucial to fully extract information on EoR (Chang et al., 2015).

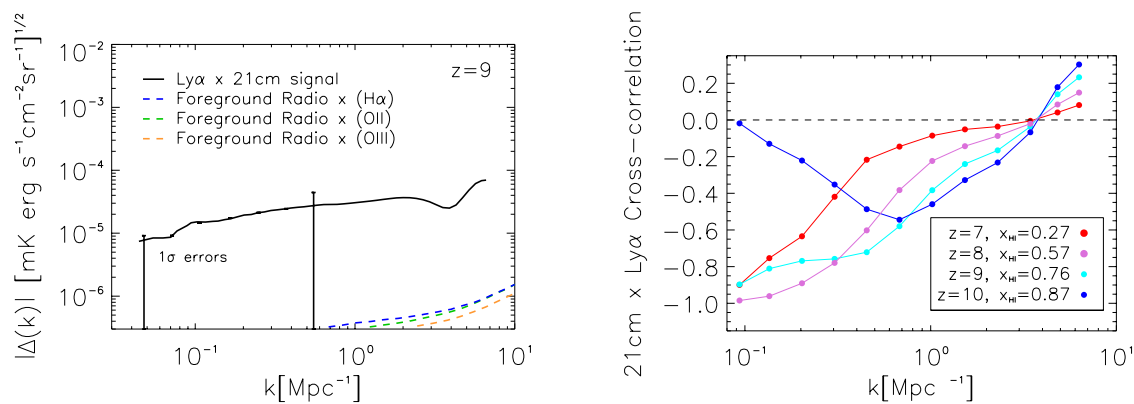

**Figure 2-20. The Lyα cross-correlation with 21-cm fluctuations.** *Left:* The theoretical 21-cm Lyα cross-power spectrum at $z = 9$ (black solid). The data points are projected band-powers with the sensitivity specified by the CDIM experiment. The low-$z$ Lyα foreground line emissions are correlated with radio galaxies detected in 21-cm and their cross correlations are shown in dashed lines for Hα (blue), OII (green), and OIII (orange), respectively. We have assumed here low-level mitigation strategies, such as high flux masking to reduce the impact of foregrounds in Lyα fluctuations; for 21-cm, the removal of the so-called foreground wedge was incorporated and SKA1-LOW sensitivities were assumed. Foreground residuals are negligible compared with the targeted signal, as is instrumental noise, for the angular scales probed by the CDIM. *Right:* Cross correlation coefficient (+1 for perfect correlation and −1 for perfect anti-correlation) of Lyα and 21-cm at $z = 7, 8, 9$, and 10 as a tracer of the growth of ionized regions during reionization. The cross-correlation coefficient directly captures the HII bubble growth during reionization as can be seen in this figure and the combination of 21-cm surveys and CDIM data will be a powerful probe of the topology of reionization.

**Reionization history.** CDIM will constrain the ionized hydrogen fraction in the IGM ($X_{HII}$) in at least four ways by (1) measuring the ionizing photon emissivity, (2) measuring the preferential escape mechanisms for Lyα and ionizing photons from galaxies, (3) probing the neutral content in the CGM, and (4) cross-correlating with 21-cm line maps to constrain the evolution and morphology of ionizing regions.

*Lyα and Hα from CDIM.* The Lyα and Hα intensity can constrain the rate of production of ionizing photons from galaxies. The ratio of Lyα to Hα intensity constrains the escape fraction of Lyα photons, which is expected to be higher than the ionizing photon escape fraction, and can help discriminate the mass range of galaxies where this escape fraction is higher. The spatial distribution of the Lyα signal will depend on the residual neutral hydrogen inside the ionized bubbles and thus the average size of these regions.

*CDIM cross-correlates with 21-cm.* The cross-correlation between 21-cm emission and Lyα or Hα from CDIM will be negative on sufficiently large scales, and the amplitude will be directly proportional to the mean neutral fraction, $X_{HI}$. 21-cm-Lyα cross-correlation therefore provides a powerful tool to measure the ionization history during EoR.

**Figure 2-21** shows two sets of constraints on $X_{HII}$ across $6 < z < 10$, obtained from CDIM data. The magenta dots indicate constraints on $X_{HII}$ using intensity maps of Hα, Hβ, and Lyα obtained by CDIM (using a combination of points (1) and (2)). These constraints are based on the ionizing photon budget derived from the amplitude of line fluctuations. The red dots show constraints obtained using simulated cross-power spectra of Lyα from CDIM and the 21-cm line from SKA1-LOW (using method (4) above) based on models in Heneka et al. 2017. These constraints are based on the anti-correlation scales discussed above. All error bars include instrument noise and cosmic variance.

## 2.5 Community-led General Observer Science

The key science we have discussed requires a 4-year science prime mission, but CDIM is expected to last substantially longer. Thus, we plan for CDIM to execute community-driven science programs through a Guest Observer (GO) campaign, selected from a standard peer-review process similar to those that are employed for observatory-class NASA missions like Hubble, Spitzer, and JWST.

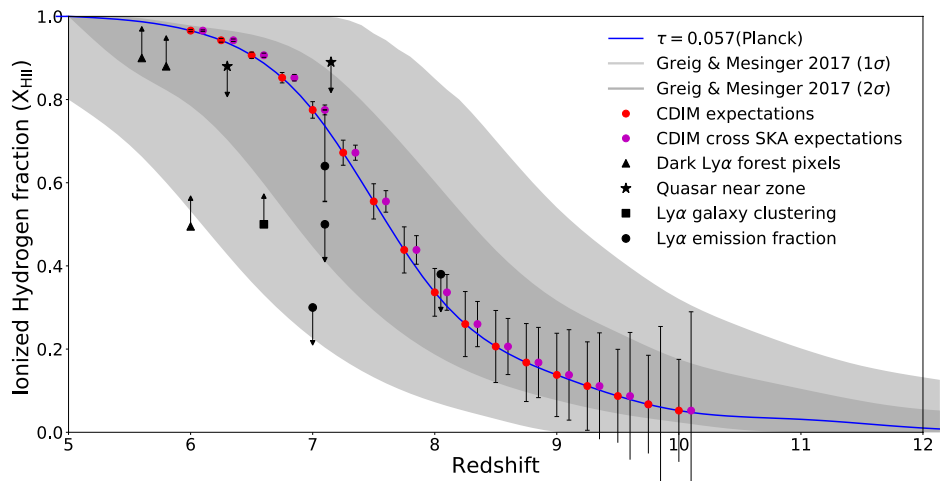

**Figure 2-21. Reionization History from CDIM**. CDIM constrains the IGM ionization fraction as a function of redshift. The magenta points mark the expected uncertainty on the ionization history derived from CDIM intensity maps of Hα, Hβ, and Lyα (shifted from the red points by a $\Delta z = 0.1$ for better visualization). The red points mark the same constraints derived from the cross-correlation of Lyα intensity maps from CDIM and a 21-cm EoR deep field from SKA1-LOW. Error bars account for the increasing uncertainty in the modeling towards $z = 10$. The assumed cosmology is from the Planck Collaboration et al. (2018) release. The black points show existing measurements and limits on the reionization history using Lyα emission fraction (Mesinger et al., 2014), Lyα clustering (Ouchi et al., 2010; Sobacchi & Mesinger, 2014), Quasar near zone (Greig et al., 2016), and dark Lyα forest pixels (McGreer et al., 2014). The 1σ and 2σ constraints on reionization history from Greig & Mesinger (2016) combine Planck and all existing data.

Community-led science is not baselined in this study since it does not drive any mission requirements. There are numerous GO observing programs that span a range of astrophysical topics, and that would benefit from the ~arcsecond spatial resolution and broad wavelength coverage of CDIM. For example:

1. SFR and metallicity measurements of galaxy properties in galaxy clusters which reveal the processes that quench star-formation in massive galaxies (e.g., Cerulo et al., 2014; Cerulo et al., 2016);
2. Spatially resolved measurements of local volume SFR, largely unaffected by dust attenuation, using the Paschen-alpha and Brackett-gamma line (e.g., Böker et al., 1999);
3. Measuring the strength of the 3.3 and 6.2-µm polycyclic aromatic hydrocarbon feature and ices as a function of dust mass and metallicity, which traces the abundances of organic compounds (e.g., HCOOH) and water in both circumstellar disks and star-forming regions (e.g., Pontoppidan et al., 2004);
4. Temporally-resolved, late-time, spectrophotometry of Advanced LIGO/LISA afterglows whereby the mid-infrared spectra reveal the production of light and heavy (A>140) r-process elements in the Universe (e.g., Kasliwal et al., 2018; Villar et al., 2018); and
5. Measuring the kinematics and mass of hot molecular gas in the vicinity of an AGN through the molecular hydrogen S3 (1.957 µm) and Q1 (2.406 µm) lines can place better constraints on the mass of nearby (z>0.3) supermassive black holes and the impact of feedback on the surrounding star-formation (e.g., Rodriguez-Ardila et al., 2004).

## 2.6 Science Requirements

**Table 3-1** shows the Science Traceability Matrix (STM) for CDIM. As part of this Probe study, lasting more than a year and half since the CDIM science team conceived this mission concept, we carried out a number of trades between science and mission/instrument requirements leading to the current point design (§3). The trades included wavelength range vs. sensitivity, depth vs. spectral resolution, and survey area vs. spatial resolution, among others.

Table 3-1. CDIM Science Traceability Matrix

| NASA Science Goals | CDIM Science Goals | CDIM Science Objectives | Science Requirements ||| Instrument Requirements ||| Driver | Mission Requirements |
| | | | Physical Parameters | Observables | Measurement Requirement | Instrument Parameter | Science Requirement | Capability | | Parameter |
|---|---|---|---|---|---|---|---|---|---|---|
| Explore the origin and evolution of the galaxies, stars and planets that make up our universe [NASA Science Plan]  How does the Universe work? How did we get here? [NASA 2014 Science Mission Directorate Strategy Document] | Trace the stellar mass buildup, dust production history, and metal enrichment history during cosmic reionization. | Determine if the rate of growth of metals and dust corresponds to the growth of stellar mass at $5 < z < 8$. | Metallicity of galaxies via the oxygen abundance, stellar mass, and dust attenuation (extinction rate, dust density) | [OIII], [OII], [NII]/H$\alpha$, H$\alpha$/H$\beta$ @$5 < z < 8$ | (i) Wavelength coverage to detect H$\alpha$ out to z of 10. (ii) Spectral resolving power to resolve [NII] and H$\alpha$. (iii) Sensitivity to detect galaxies $< 10^9$ M$_{sun}$ in a deep survey. | Wavelength range | $2.2 \leq \lambda \leq 6.0$ μm | $0.75 \leq \lambda \leq 7.5$ μm | Data Reliability and Systematic Error Control | Deep, medium and wide surveys each with $\geq 90\%$ voxel completeness for internal reliability.  Spatial resolution: Effective PSF FWHM $\leq 2"$ at 1 μm (from science requirements).  Stable cooling to $< 35$ K to control $> 5$ μm array dark current. |
| | | | | | | Spatial resolution (pixel scale) | $\Theta_{pix} = 1"$–$2"$ | $\Theta_{pix} = 1"$ | | |
| | | | | | | Spectral resolving power | $\lambda/\Delta\lambda \geq 300$ | $\lambda/\Delta\lambda = 300$ | | |
| | | | | | | Point source broadband photometric sensitivity (R=5, 5$\sigma$; deep survey) | 24.5 AB mag at J band | 25.2 AB mag at J band | | |
| | | | | | | Spectral line flux sensitivity (3.5$\sigma$; deep survey) | $5.0 \times 10^{-18}$ erg s$^{-1}$ cm$^{-2}$ at 4.6 μm | $2.7 \times 10^{-18}$ erg s$^{-1}$ cm$^{-2}$ at 4.6 μm | | |
| | Establish the role of active galactic nuclei (AGN) in cosmic reionization. | Determine the fractional contribution of super-massive black hole/AGNs to reionization photon budget. | Unbiased UV photon spectral density; black-hole masses via line widths of optical lines. | Rest-frame UV continuum @z = 5–8. [MgII] and other metal lines. | (i) Sensitivity to detect faint quasars in a wide survey. (ii) Spectral resolving power to detect equivalent width of broad metal lines. | Wavelength range | $2.9 \leq \lambda \leq 6.0$ μm | Same as above | Survey Strategy | Deep survey: 15 deg$^2$, imbedded in the Wide survey. |
| | | | | | | Spatial resolution (PSF; FWHM) | $\Theta_{FWHM} = 2"$ at K band | $\Theta_{FWHM} < 2"$ at K band | | |
| | | | | | | Spectral resolving power | $\lambda/\Delta\lambda \geq 300$ | Same as above | | |
| | | | | | | Point source broadband photometric sensitivity (R=5, 5$\sigma$; wide survey) | 23.5 AB mag at K band | 24.0 AB mag at K band | | |
| | Establish the progression and topology of reionization from cosmic dawn at z = 10 to the end of reionization at z < 6. | Determine the progress of reionization by measuring the ionization fraction in at least 10 redshift bins at $5 < z < 10$, with accuracy better than 10%. | Ly$\alpha$ luminosity function, escape fraction, and the spatial distribution. | Ly$\alpha$ | (i) Wavelength coverage to detect Ly$\alpha$ out to z of 10. (ii) Sensitivity to detect faint galaxies. | Wavelength range | $0.75 \leq \lambda_{Ly\alpha} \leq 0.98$ μm | $0.75 \leq \lambda \leq 7.5$ μm | | Medium survey: 30 deg$^2$, to overlap with 21-cm fields from HERA and SKA1-LOW. |
| | | | | | | Spectral resolving power | $\lambda/\Delta\lambda \geq 100$ | Same as above | | |
| | | | | | | Spectral line flux sensitivity (3.5$\sigma$; deep survey) | $2.9 \times 10^{-17}$ erg s$^{-1}$ cm$^{-2}$ at 0.85 μm | $2.0 \times 10^{-17}$ erg s$^{-1}$ cm$^{-2}$ at 0.85 μm | | |
| | | | Reionization history of the universe. | Ly$\alpha$ and H$\alpha$ | (i) Ability to perform cross-correlations, including Ly$\alpha$ and H$\alpha$, and 21-cm radio measurements. | Wavelength range | $0.75 \leq \lambda_{Ly\alpha} \leq 1.4$ μm  $3.9 \leq \lambda_{H\alpha} \leq 7.2$ μm | $0.75 \leq \lambda \leq 7.5$ μm | | Wide survey: 300 deg$^2$, driven by number of AGN detections. |
| | | | | | | Spectral resolving power | $\lambda/\Delta\lambda \geq 100$ | Same as above | | |
| | | | | | | Surface brightness sensitivity (1$\sigma$; medium survey) | $1.3 \times 10^{-18}$ erg s$^{-1}$ cm$^{-2}$ Hz$^{-1}$ sr$^{-1}$ at 1.1 μm | $1.5 \times 10^{-19}$ erg s$^{-1}$ cm$^{-2}$ Hz$^{-1}$ sr$^{-1}$ at 1.1 μm | | Read, reduce, and telemeter spectral imaging data. |

**Note:** Requirements derived from the three principal science themes are shown in colored boxes. White boxes denote non-driving requirements. Gray boxes show requirements common to all three themes.

To meet the science requirements as laid out in our STM, we envision a three-tiered survey taking a total of 4 years of observing time with CDIM (including a 10% overhead fraction for slews and data transfer). The sensitivity levels are summarized in **Fact Sheet Figure 2**. The survey areas are 15 deg$^2$, 30 deg$^2$, and 300 deg$^2$ for the deep, medium, and wide surveys, respectively. We plan to overlap the deep survey with one of the deep fields of WFIRST and/or Euclid surveys located in either the North Ecliptic Pole (NEP) or the South Ecliptic Pole (SEP) for synergistic science. The medium survey is chosen to overlap with one of the SKA1-LOW deep fields, likely the Extended Chandra Deep Field-South (ECDF-S), which will also overlap with the 21-cm EoR survey of the Hydrogen Epoch of Reionization Array (HERA). The wide survey is designed to surround the deep survey to optimize observing efficiency in either the NEP or SEP, which is visible year-round from L2 for ease of scheduling. The integration time per LVF step is 250 seconds, ensuring that photon-noise dominates, and the total integration time per spatial pixel per spectral resolution element are 333.3 minutes, 83.3 minutes, and 16.7 minutes each for deep, medium and wide surveys, respectively. In practice, the surveys will be completed with multiple surveys (visits) over the same areas, with a redundancy of 80, 20, and 4 visits for the deep, medium, and wide surveys, respectively. The redundancy is used for cross-checks on systematics and to minimize effects of Zodiacal light, among others, and allows Nyquist-sampling of the spectral resolution by a spatial offset of the pointing position per visit of half the LVF step size (§3.2). With time needed for calibrations as described in §3.11, key science objectives can be met within the 4-year prime mission.

## 3 SCIENCE IMPLEMENTATION

CDIM is an infrared survey mission that will provide spectra from 0.75 to 7.5 µm at R=300 for fields as wide as 300 deg$^2$. The instrument employs a 1.1-m cryogenically cooled telescope to image a 9 deg$^2$ instantaneous field of view. An array of Teledyne H2RG detectors provides sky-limited imaging while linear variable filters select the wavelengths with high efficiency. Spectra are constructed for each object by repointing the spacecraft to move the instantaneous field in a series of small steps across the detector array.

### 3.1 Survey Design

The CDIM survey is a three-tiered design conducted from L2 with overlapping wide, medium and deep target areas (300, 30, and 15 deg$^2$) sampled over 4 years that together meet the requirements of the Science Traceability Matrix (**Table 3-1**) including 10% overhead for slews and data transfer. Fields were chosen at high ecliptic latitudes to minimize background and maximize sensitivity (**Fact Sheet Figure 2**).

At the center, a deep survey will overlap either a WFIRST or Euclid set of fields to provide complementary spectral information. A medium survey surrounds the deep one and overlaps with one of the SKA1-LOW deep fields, likely the Extended Chandra Deep Field-South (ECDF-S), as well as the 21-cm survey of the Hydrogen Epoch of Reionization Array (HERA). The wide survey overlaps both of the others and is visible all year from L2.

The surveys will be completed with multiple visits to the same areas, with a redundancy of 4, 20, and 80 visits for the wide, medium and deep surveys, respectively. The redundancy is used for cross checks on systematics and to minimize effects of zodiacal light. With time needed for calibrations as described in §3.11, we expect the key science requirements can be met in < 4 years.

### 3.2 Spectroscopy Approach

The CDIM instrument images an instantaneous 9 deg$^2$ field of view (3.6° × 2.5°) onto a 6×4 element (spectral×spatial) array of Teledyne H2RG detectors (active detector area 3.4° × 2.3° or 7.8 deg$^2$) with 1″ pixels, each with its own proximity-mounted linear variable filter (LVF). The filters are designed to admit different wavelengths according to the lateral position of the object on the filter, with a spectral resolution of 300 at any given location. During the mission, a field is stepped 14.6″ at a time across the array in the spectral direction. These visits are then repeated with an offset of 7.3″. The resulting ~1680 samples provide Nyquist spectral sampling of each element in the field at R = 300 across the full 0.75–7.5 µm wavelength range of the spectrograph. The simple survey design requires no moving parts, and the 0.5″ RMS pointing stability of the spacecraft is well matched to the 1″ pixel (2″ FWHM PSF) of the CDIM detector array. Each individual sample will be sky-limited in 250 s. The total integration time per spatial pixel per spectral resolution element are 333.3 minutes, 83.3 minutes and 16.7 minutes each for deep, medium and wide surveys, respectively.

## 3.3 Instrumentation

CDIM is a simple and highly efficient instrument for multi-object spectroscopy from 0.75 to 7.5 μm. A block diagram of the concept is shown in **Figure 3-1**. It comprises a wide field, cryogenic telescope (§3.4) that forms an image of the sky on a 2-dimensional array of infrared detectors, each with an associated linear variable filter (LVF, §3.5). A planned mosaic of 24 Teledyne H2RG detectors (§3.6) has heritage from the JWST and Euclid missions. Some development is still needed to extend sensitivity to wavelengths longer than 5 μm and is already being successfully pursued by other mission studies. (McMurtry et al., 2016b). The associated LVFs provide high transmission, narrowband photometry for a wavelength that varies with position in one dimension as shown in **Figure 3-2**. Spectral response is adjusted simply by moving the image relative to the LVF. In CDIM, the set of high-throughput LVFs replaces the dispersing element of a conventional spectrograph, simplifying system design and minimizing weight. LVFs in the near infrared (up to R~200) have been flown in the New Frontiers New Horizons/LEISA and OSIRIS-Rex/ OVIRS instruments (Reuter et al., 2008; Reuter et al., 2018), and a study commissioned by our team for this paper has validated that the required set of filters can be fabricated with modest new development (Omega Optical Inc., 2018).

The instrument is passively cooled by a set of three V-groove radiators following the Planck thermal design (Leroy et al., 2006). A model of the anticipated thermal balance is shown in **Figure 3-3**. The telescope mirrors, baffles, and enclosure are cooled to below the 70-K requirement with margin to reduce thermal emission from the instrument, ensuring sky-limited measurements for five of the six

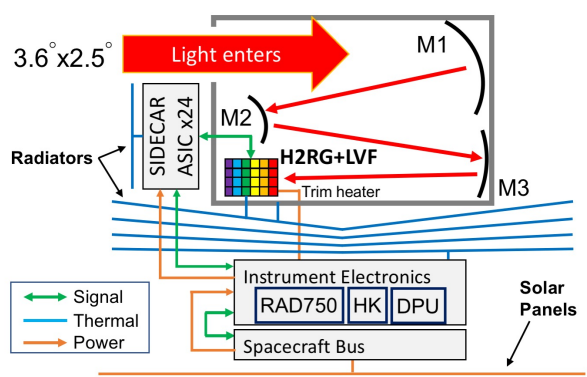

**Figure 3-1.** Block diagram of the CDIM instrument architecture showing signal, thermal, and power paths between subsystems

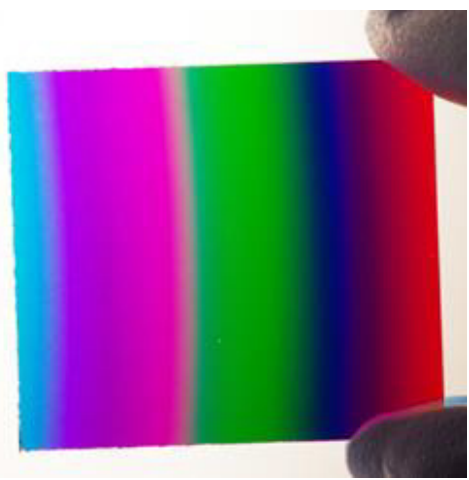

**Figure 3-2.** An example 40×40 mm LVF manufactured by Omega Optical (Brattleboro, VT) demonstrates existing manufacturing capabilities.

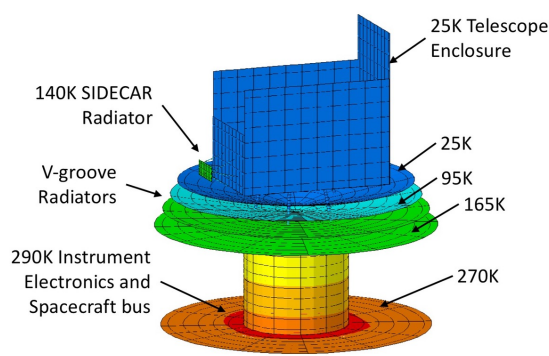

**Figure 3-3.** The CDIM thermal design has Planck heritage and meets requirements with margin to provide a low instrument background in a simple passive implementation.

CDIM bands. The detectors are cooled with margin to their 35-K requirement to minimize dark current, particularly in the long wavelength channel. An array of Teledyne SIDECAR ASIC-based detector controllers (Loose et al., 2006) is located near the H2RG detectors to minimize cable length; it has its own small radiator outside the telescope housing cooled to 140 K; there is substantial cooling margin as the radiator can be enlarged. Instrument control electronics are mounted in a compact PCI chassis below the radiators and operated close to room temperature.

All-beryllium light-weighted mirrors and mounting structures passively athermalize the system, simplifying alignment and test while a flexure mounted aluminum enclosure and baffling system controls stray light at the detector. A hexapod support structure kinematically attaches the instrument to the spacecraft bus. The side-looking telescope is protected from direct sunlight by the solar panel within a ±18° pointing range. A telescope aperture cover is the only deployable.

## 3.4 Telescope

The CDIM telescope is an unobscured, off-axis three-mirror anastigmat (TMA) design (cf Cook, 1979). Three beryllium elements leverage technology already developed for the James Webb Space Telescope (JWST) (Lightsey, 2007). A ray trace of the telescope is shown in **Figure 3-4** and a table of optical parameters is provided in **Table 3-2**. The entrance pupil is 83 cm in diameter, which underfills the 1.1-m primary mirror slightly for each field. Freeform Zernike polynomial surfaces enable a 3.6° field of view with < 0.6″ RMS diameter geometric image quality over a flat focal plane as shown in **Figure 3-5**. The as-built performance of the telescope is expected to be well matched to the size of the 1″ pixels, meeting the science requirement (<2″ pixel) with margin. The diffraction limit at 7.5 μm is 1.9″ FWHM, and performance at shorter wavelengths will likely be dominated by alignment and fabrication quality to a similar level. Gold coatings provide reflectivity greater than 95% across the desired 0.75–7.5 μm band. The telescope operates at f/4.5 and is nearly telecentric, providing uniform input to the bandpass selecting multilayer filters. Fields near the corners of the 233×162 mm image are tilted by less than 1° relative to the 6.4° half angle light cones, and have a maximum calibrated image distortion under 1.2 mm. Similar designs have flown on other instruments, for example on the NASA Advanced Land Imager flown in 2000 (Lencioni et al., 1999) while larger aspheric mirrors have already been fabricated for JWST.

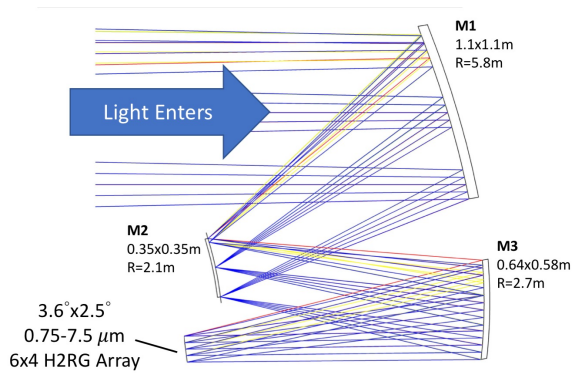

**Figure 3-4.** A ray trace of the CDIM telescope. Beryllium mirrors and structure provide passive athermalization in a lightweight design that simplifies integration and test. The 9 deg$^2$ field of view design has better than 0.6" RMS image quality, low distortion and a flat focal plane well matched to the 1" H2RG pixels.

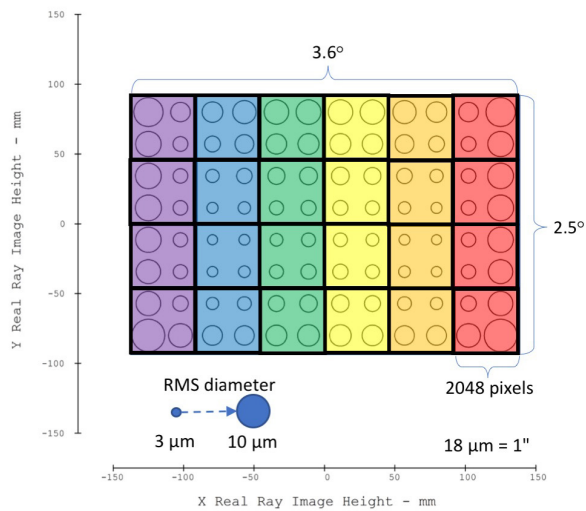

**Figure 3-5.** The distribution of RMS spot diameters predicted by the CDIM telescope optical design are 0.6" in the worst-case field corners. In this figure, colors represent the bandpass arrangement of the 6×4 LVF array. Our performance models allow margin for fabrication and alignment errors resulting in a design that is pixel limited (2 pixels full width at half maximum) across the full wavelength range.

| Table 3-1. Optical Parameters of the CDIM Design | |
|---|---|
| **Parameter** | **Value** |
| **Primary mirror** | Beryllium, Concave |
| Dimensions | 1083.4 × 1067.2 mm |
| Radius | 5818.5 mm |
| Primary–Secondary Distance | 1580.2 mm |
| **Secondary mirror** | Beryllium, Convex |
| Dimensions | 356.3 × 355.4 mm |
| Radius | 2054.2 mm |
| Secondary–Tertiary Distance | 1611.2 mm |
| **Tertiary mirror** | Beryllium, Concave |
| Dimensions | 644.0 × 582.1 mm |
| Radius | 2671.0 mm |
| Tertiary–Detector Distance | 1870.4 mm |
| System Focal length | 3738 mm |
| Entrance pupil diameter | 835.5 mm |
| Plate scale | 55.18"/mm (=0.993"/pix) |
| Focal ratio | 4.47 |
| RMS spot diameter (design) | <10 μm |
| FWHM (including alignment) | 2" across band |
| Field of View | 3.6° × 2.5° |

## 3.5  Spectrograph Design

Light from the telescope is separated into its component wavelengths by a 6×4 rectangular grid of 40 mm LVF elements placed immediately in front of the focal plane as shown in **Figure 3-8**. Wavelength varies continuously in the long direction of the grid, while the short direction provides identical parallel channels for more efficient sky sampling. In the CDIM design, a minimum spectral resolution of R > 300 is required in order to separate the H-alpha and [NII] spectral lines (2.1 nm at $z = 0$ and 23.1 nm at $z = 10$). We have designed the series of six filters with boundary conditions such that (1) the set of filters samples wavelengths from 0.75 to 7.5 µm; (2) the wavelength is continuous across the boundary from one filter to the next; and (3) the spectral range is proportional to the central wavelength of each filter.

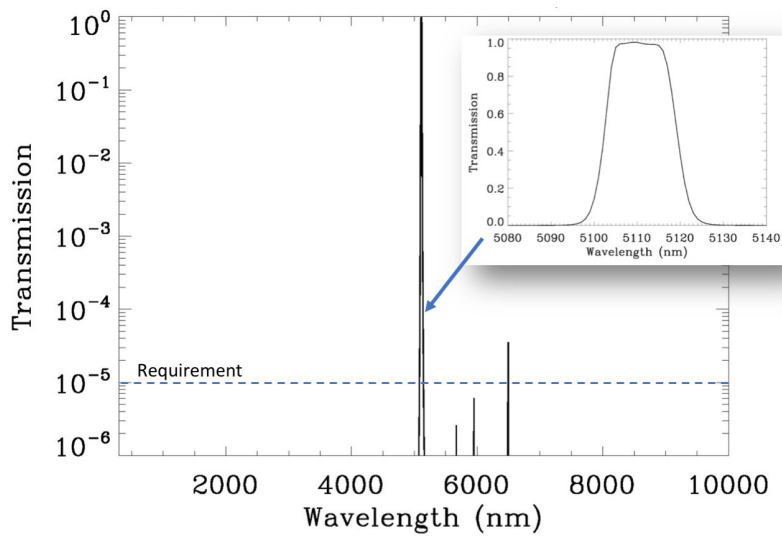

**Figure 3-6.** A representative bandpass plot for a location on the longest wavelength filter design corresponding to 5.1 µm. The transmission model is a composite of performance estimates for incidence angles from 0 to 7°, representative of the range of angles in an f/4.5 beam. Detector QE beyond 1.1 × the cutoff wavelength is modeled at <1% and assists with out-of-band suppression. The OD5 requirement is met with small excursions that will be addressed in a future technology development program.

A benefit of this design is that while each filter is linear, the rate of change of wavelength with position grows non-linearly across the set of filters as shown in **Table 3-3**. This design enables spectral sampling that is proportional to wavelength across the wide CDIM band simply by stepping the image a constant amount across the 6-filter width of the array. It consists of 4 identical 6-filter sets, allowing for increased sampling efficiency of the desired survey areas. During the mission, a field is stepped 14.6″ at a time across the array in the spectral direction. These visits are then repeated with an offset of 7.3″. The resulting ~1680 samples provide Nyquist spectral sampling of each element in the field at R = 300 across the full wavelength range of the spectrograph.

We commissioned a study by Omega Optical Inc. (Brattleboro, VT) to evaluate the feasibility of the filter set proposed in **Table 3-3**. The purpose of the study was to generate performance models of manufacturable filters in order to evaluate throughput and out-of-band blocking as a function of wavelength and telescope focal ratio (see §5.2). The filters were modeled as two elements optically in series on either side of the substrate. A fixed blocking filter was placed on the outside and designed to reject light outside the LVF pass band but within the sensitive range of the associated detector. The linear variable filter was placed on the inside surface to minimize the footprint of the f/4.5 beam on the filter surface. The filters will be placed within 100 µm of the detector surface to constrain the variation in the LVF passband across each beam footprint to less than 10% of the nominal filter width for all wavelengths.

**Table 3-2.** The CDIM LVF Filter Set

| Filter | Material | Pass Band | | Rejection Band | | Detector Cutoff µm |
|---|---|---|---|---|---|---|
| | | Begin | End | Short | Long | |
| A | B270 | 0.750 | 1.101 | 0.3–0.75 | 1.1–2.0 | 1.7 |
| B | Si | 1.101 | 1.615 | 0.3–1.10 | 1.6–2.5 | 1.7 |
| C | Si or InP | 1.615 | 2.371 | 0.3–1.61 | 2.4–3.5 | 2.5 |
| D | Ge | 2.371 | 3.479 | 0.3–2.37 | 3.5–6.0 | 5.3 |
| E | Ge | 3.479 | 5.106 | 0.3–3.47 | 5.1–6.0 | 5.3 |
| F | Ge | 5.106 | 7.439 | 0.3–5.10 | 7.5–10 | 7.5 |

A representative transmission model for the longest wavelength filter is shown in **Figure 3-6**. Peak throughput is similar for all six designs. Out-of-band blocking is required to meet OD5 ($10^{-5}$) to retain sky-limited performance as shown in **Figure 3-7**. The blocking band was specified on the blue side from 0.3 µm to the short wavelength end of the filter, and on the red side from the long wavelength end of the filter to the cutoff wavelength of the associated detector. Filter substrate materials were chosen where possible to improve filter out-of-band blocking performance and to simplify the blocking filter design. Since the cutoff wavelength of each detector is matched to the pass band, the filters do not generally need to strongly reject light all the way to the 7.5 µm red edge of the CDIM design. The combination of filter substrate, blocking filter and LVF together achieve an out-of-band rejection for each design of OD5 for the majority of each band, but with some excursions greater than OD4. We do not require perfectly uniform blocking at all wavelengths, but we do require that the effect of the blocking, integrated across the band, is at least equivalent to OD5. Small excursions are allowed as long as the total number of photons leaking into any particular band do not cause the measurement to become leakage- rather than sky-limited. One avenue for improved out-of-band performance would be to incorporate a linear variable rather than fixed blocker into each filter. Our plan for further development of these filter designs is outlined in §5.2.

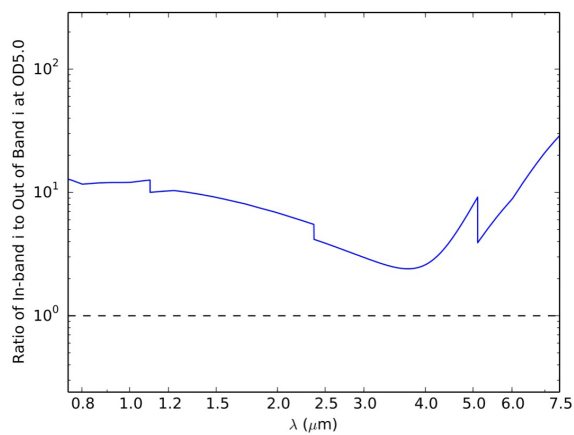

**Figure 3-7.** The out-of-band signal from filter leakage is easily controlled by a blocking requirement of OD5 (blue curve), which is sufficient to retain Poisson-limited performance (dashed line) from the combination of in-band sky signal and dark current (**Figure 3-10**) across the entire wavelength range.

## 3.6 Detectors

The instrument uses an array of twenty-four Teledyne 2k×2k H2RG detectors (Beletic et al., 2008) with 18 µm pixels (1″ pixel$^{-1}$). These detectors are hybrid sensors consisting of a photosensitive layer (either silicon or HgCdTe alloy) bump-bonded onto a Readout Integrated Circuit (ROIC). The H2RG detector is considered a TRL 9 product with examples having already flown on U.S. defense applications, and more planned for launch on JWST (Greene et al., 2017) and Euclid (Waczynski et al., 2016).

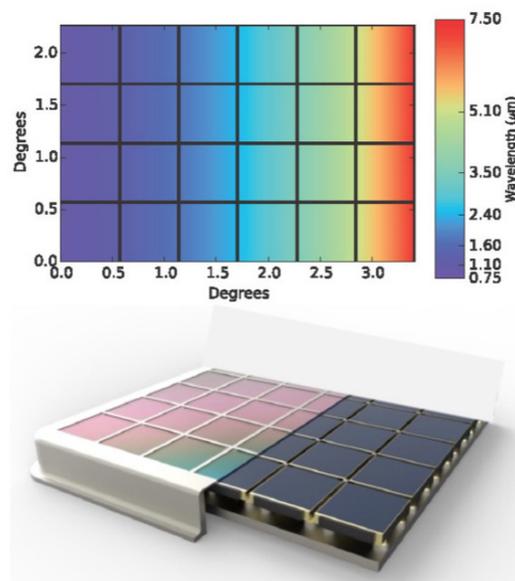

**Figure 3-8.** The CDIM focal plane and filter design. Detectors are Teledyne H2RG with cutoff wavelength matched to the band of interest. Infrared-sensitive HgCdTe with cutoff wavelengths of 1.7, 2.5, 5.3 µm are already available as standard products while the 7.5 µm cutoff devices are actively being developed for other missions (NEOCam).

The CDIM design employs HgCdTe alloy as the photosensitive material for all channels in order to promote data uniformity and continuous sensitivity across its wide wavelength range. A cutoff wavelength is selected appropriately for the required sensitive band of each detector by varying the ratio of Hg to Cd. Standard HgCdTe products are available with cutoff wavelengths of 1.75, 2.5, and 5.3 µm (Blank et al., 2012). Performance estimates are provided in **Table 3-4**. Long wavelength cutoff versions such as the required 7.5 µm detector (TRL ~4) are actively being developed by other groups (Dorn et al., 2018) and our planned development program is described in §5.1.

In the CDIM design, the detectors are mounted in a 6×4 grid in the telescope focal plane, each with an associated LVF element mounted within 0.1 mm of the focal plane to minimize any degradation of the spectral sampling across the beam footprint on the filter. Flight package options currently available from Teledyne are 3-side "close buttable" (Blank et al., 2012). An arrangement that minimizes gaps between the array elements would have approximately 2.1 mm (1.9′) gaps between active pixels in the spectral direction and 4.4 mm gaps (4′) in the spatial direction between the 6 element rows. The array of detectors is passively cooled to 35 K by the instrument V-groove radiator. A decontamination heater is provided to evolve ice and other volatile contaminants, as well as to provide some mitigation of radiation-induced hot pixels (McKelvey et al., 2004). At this low temperature, the instrument background will be dominated by zodiacal light from the sky, and therefore the thermal control requirements set by dark current, which is relatively stable below 35 K, are modest (± 0.5 K).

**Table 3-3.** H2RG Detector Performance Estimates

| Parameter | Value | | | |
|---|---|---|---|---|
| Type | Teledyne H2RG | | | |
| Format | 2048×2048 | | | |
| Pixel size | 18 µm (plate scale = 1") | | | |
| Operating Temperature | <35 K | | | |
| Operability | >95% | | | |
| Flatness | <20 µm | | | |
| QE | >70% | | | |
| Cutoff wavelength (µm) | 1.75 | 2.5 | 5.3 | 7.5 [i] |
| Required number (flight) | 8 | 4 | 8 | 4 |
| Read noise (electrons, single CDS) | 10 | 11 | 11 | 15 |
| Dark ($e^-$ $s^{-1}$) | 0.001 | 0.001 | 0.002 | 0.166 |
| Full well (electrons) | 80,000 | 65,000 | 65,000 | 60,000 |

Each CDIM detector will be read out through 32 channels in "slow mode" once every 2 s (<100 kHz pixel rate) for minimum read noise. The arrays are non-destructively sampled 125 times during each 250-s spectral sample time while the signal is integrating. This technique is referred to as "sampling up the ramp" and is a means of reducing the contribution of detector read noise by combining information from all of the samples to determine a best fit to the slope of the signal. It is estimated that the typical effective read noise for 125 CDIM samples will be <3.5 $e^-$ RMS using this technique (Benford et al., 2008).

The radiation environment performance of H2RG arrays has been studied extensively for JWST in the context of a 10 year mission life at L2 (McKelvey et al., 2004). The environment consists of a steady stream of ~GeV galactic cosmic ray protons (5 $s^{-1}cm^{-2}$) and a sporadic contribution of ~0.1 GeV (peak) solar protons that accounts for the majority of the integrated exposure. Damage can occur in the ROIC elements, where trapped charge (ionizing damage) causes flat band shifts required to properly operate the transistor elements. Damage can also occur in the thin photosensitive layer in the form of displacement damage that increases the dark rate of affected pixels. Since the H2RG does not transfer charge across pixels, it is not considered as sensitive to damage as a CCD in the same environment. The JWST studies indicate that approximately 10% of pixels would exceed the initial dark current by 6 standard deviations after an equivalent 10-year exposure, and therefore the effect would be expected to be of order 5% for a planned 4-year CDIM mission in the same environment.

High energy galactic cosmic rays (Girard et al., 2014) will penetrate the spacecraft shielding and are estimated to contaminate about 5% of pixels in each 250-s frame at the conservative L2 estimated rate of 5-$s^{-1}$ $cm^{-2}$ assuming 10 pixels per event. Since the ionization tracks of cosmic rays are erased by resetting the array, they are mainly considered nuisance artifact. The potential contribution of cosmic rays is one additional reason to choose the 1.75 µm cutoff HgCdTe array type over the silicon HyViSi for the shortest wavelength band, since the thicker silicon (250 µm vs. 10 µm) is much more sensitive to cosmic ray contamination.

JWST H2RG testing also showed no significant changes in read noise, amplifier linearity, or power dissipation after a 5 krad(Si) (10 year) dose. ROIC test measurements were made of the source-follower output voltage as a function of dose, up to levels as high as 50 krad(Si) in what was then a Rockwell Scientific device that showed essentially no radiation damage effects. Rockwell was purchased by Teledyne in 2006.

---

[i] The measured performance of existing 10 µm cutoff detectors (McMurty et al., 2013) provides the basis of estimate for the performance of the planned 7.5 µm cutoff sensors. Dark current can vary widely in the current generation of long wavelength test sensors, but these also achieve values similar to our baseline for the majority (about 80%) of pixels. It is reasonable to expect 7.5 µm variants to have relatively improved dark current performance.

Teledyne has made substantial progress on two known H2RG artifacts: inter-pixel capacitance (IPC) and image persistence (Blank et al., 2012). The first, IPC, results from capacitive coupling of pixels in the photosensitive layer, primarily. The effect is to broaden the PSF slightly and amounts to about 1% crosstalk between pairs of pixels, which is not expected to be a limiting effect given the modest spatial resolution requirements (<2″ pixel). The second effect, image persistence, is the retention of a residual image after the array is reset. It is an area of active development at Teledyne and substantial improvements have been made. The effect is expected to be of order 0.1% or less of the input signal, and therefore is only measurable for bright sources. A typical CDIM field will have sky signals of order 0.1–1 $e^-$ pixel$^{-1}$s$^{-1}$, and the resulting persistence will be much less than an electron and well below the read noise of the detector. Exceptions to this will come in the form of bright stars. We estimate that there will be of order 100 stars in each field near the ecliptic pole that will have significant persistence, producing an after-image that will last about 1000 s (Tulloch, 2017). These will be easy to track as the image is stepped across the field. Assuming the typical decay time for persistence from bright objects, then there will be ~four after-images at a time for each star in the field (since an exposure is typically 250 s). We therefore expect the quantity of flagged bright star persistence pixels to be much smaller than flagged cosmic ray pixels.

## 3.7 Electronics

The CDIM instrument electronics consists of a compact PCI chassis populated with a BAE RAD750 processor board, four Vertex V FPGA data processing units (DPU), and a housekeeping board with the capability to measure 128 temperatures and voltages at 1 Hz with 16-bit resolution. The design is based on one that has been qualified for Orbiting Carbon Observatory 3 (OCO3).

The RAD750-based command and data processor unit controls the detectors and moves data from the instrument to the spacecraft. The DPU reduces demand on the RAD750 by providing real time processing of survey data. Each detector is read out through 32 channels by a dedicated SIDECAR controller. The units are lightweight (0.12 kg) and low power (0.25 W). The 24-element SIDECAR array is mounted outside the telescope enclosure directly behind the detector array on a thermally-isolated sub-bench with a dedicated radiator that cools the 6 W load to below 140 K, simultaneously meeting the requirements for low thermal background and a maximum cable harness length of 70 cm. The twenty-four element H2RG array produces 201.3 MB of data per read every 2 s (0.8 Gb s$^{-1}$) that is stored in a real time buffer in instrument memory. This data is processed as it arrives by the set of four DPU boards, which adds it to an accumulation buffer where the slope for each pixel is determined and cosmic rays are flagged and corrected. The image resulting from the 250-s accumulation is transferred to the spacecraft during the estimated 20-sec slew to the next spectral sample along with raw engineering data from a 128×128 pixel section of each 2-s slice that will be used for data quality monitoring.

Less than 1 GB total instrument storage is required for this effort assuming the data is continuously off-loaded to the spacecraft. With modest 2:1 lossless compression, the data rate to the spacecraft is estimated to be 245 Gb day$^{-1}$ without raw engineering data and 413 Gb day$^{-1}$ with engineering data included, well within the ~1.5 terabit capacity of the proposed spacecraft bus.

## 3.8 Mass and Power

The instrument mass is 694 kg, dominated by the V-groove radiators, enclosure, and telescope. The instrument power is 50 W during normal observing, dominated by the instrument control electronics. A summary of the mass and power is provided in **Table 3-5**. Note: in the Team X Report (p. 77), the V-groove radiators, bipods and aperture cover (216.0 kg) are bookkept as 'Spacecraft levy' and included in the system MEL (p. 84) under Spacecraft. The instrument has four modes: standby, observe, decontamination and survival. The highest power mode is detector decontamination (125 W), which is available for contingency and would most likely run at most monthly with the instrument turned off. The majority of the time, power is consumed primarily by the instrument control electronics (43 W) and the array of SIDECAR ASICs (6

W). The power consumption of the 24 H2RG arrays is about 1 W, therefore the estimated power requirement of the instrument is 50 W while observing. The instrument is powered off during safe mode.

## 3.9 Performance

A performance model has been developed for the CDIM instrument with current best estimate assumptions for optical and detector performance. Given the detector performance parameters provided in **Table 3-4** coupled with an average gold coating reflectivity of 95% and an LVF throughput of 85%, we estimate the throughput of the CDIM instrument will be about 50% across the band as shown in **Figure 3-9**.

For faint light observations, a low instrument background is required to maximize signal-to-noise ratio. We estimate that the brightest instrumental signal will be the detector dark current for most of the band. By design, thermal radiation from the telescope enclosure contributes only longward of 7 µm. A comparison of instrumental and sky backgrounds is presented in **Figure 3-10**. Sky background is the dominant source shortward of 5 µm and it sets the limiting magnitude for the CDIM surveys. The minimum margin in sensitivity is 0.5 magnitude relative to the requirements (point source sensitivity). The longest wavelength channel is dominated by detector dark current (based on measurements of existing 10-µm cutoff test detectors being evaluated for the NEOCAM program (McMurtry et al., 2016b).

| Table 3-4. CDIM Instrument mass and power (CBE) | | |
|---|---|---|
| Item | Mass (kg) | Power (W) |
| Instrument control electronics | 10 | 43 |
| Detectors + LVF Filters + SIDECAR | 23 | 7 |
| Cables + Baffles | 40 | 0 |
| V-Groove Radiators | 168 | 0 |
| Aperture cover + bipods + thermal enclosure | 183 | 0 |
| Telescope | 270 | 0 |
| **Total** | **694** | **50** |

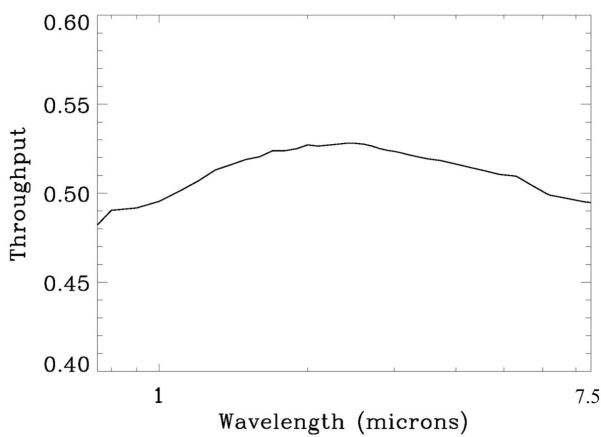

**Figure 3-9.** The CDIM design results in high throughput across the whole wavelength range. Projected performance includes three reflections from gold-coated mirrors, the transmissive LVF array, and QE of each detector.

## 3.10 Data Analysis

CDIM basic data products take the form of R = 300 narrow-band images of the sky. The data rate of about 413 Gb day$^{-1}$ (§3.7) is achieved with on-board processing and standard data compression techniques. The photo-current imaging data plus house-keeping data related to mission operations will form the basic Level-0 dataset. Our best estimates are that less than 0.3% of the pixels will be flagged for cosmic rays, pixel non-linearities, and bright source saturations. The Level-0 data will be transferred to the ground and will be housed at a central location, maintained by one of the standard astronomical datacenters such as IPAC to be used by the CDIM science team. The decompressed dataset that is properly matched

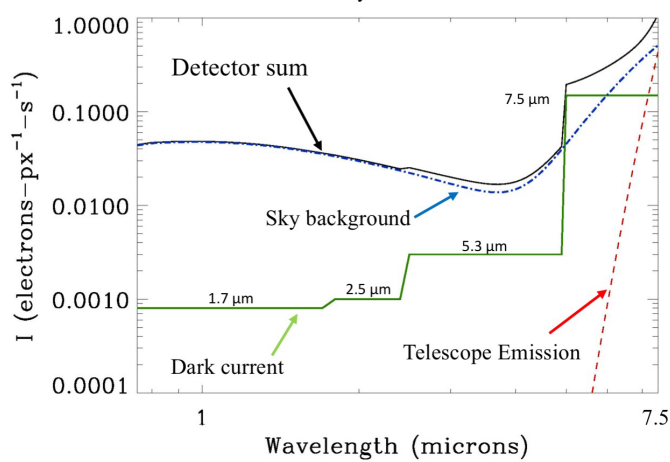

**Figure 3-10.** Total background seen by the detectors (Detector sum), comprising the primary instrumental and sky backgrounds for CDIM. Sky background in the form of zodiacal light is the dominant source of signal for all but the longest wavelengths, when detector dark current and telescope emission become significant.

with flight data forms the Level-2 dataset.

The calibrated LVF images, as described below, will form the Level-3 dataset and will be used to identify galaxies, quasars and to carry out intensity mapping and other science described in §§1 and 2. At the Level-3 step in analysis, the team will also produce multi-wavelength catalogs, with photometry performed at each narrow band image. For the intensity mapping sciences, LVF spectral images will be separated from each sky frame and will be mosaicked together to produce narrow-band images of the deep and medium-depth surveys. The large number of observations, spanning 4 years, allow us to produce high fidelity mosaics that are not impacted by the zodiacal light. These images will make use of the self-calibration algorithms that have been implemented by team members for fluctuation studies with Spitzer, CIBER and other datasets. Science data products will be at Level-4, in the form of public-release images and catalogs, as well as intensity mapping power spectra and associated tools and software codes.

The team will develop a data management plan and will pursue a suitable strategy—with the help of an established data center such as IPAC—for data analysis and archiving of astronomical community-led programs, especially if CDIM is to have a General Observing campaign in addition to the key science program that will be led by the CDIM Science Team.

## 3.11 Calibration

The detection of lines with LVF observations requires pixel intensity measurements that are stable at the 5% level and wavelength calibration that is stable at the 1% level. This tolerance is estimated based on the high equivalent width (rest-frame >500 Å) of nebular emission in high-redshift galaxies. Using either a calibration lamp and/or the zodiacal light, we expect to reach sub-percent accuracy in the flat fields.

The variation of the effective point spread function with wavelength due to telescope jitter is likely to play a bigger role in the uncertainty in spectral extraction. We will obtain spectra in dense stellar fields such as globular clusters to monitor the variation of the PSF with wavelength. We will also use interleave our science observations with a combination of planetary nebulae and spectrophotometric stellar standards. The strong lines from ionized gas of planetary nebulae will help with wavelength calibration to within the required tolerance. Observations of spectrophotometric standards will be undertaken at multiple points across the detector. Since the field of view of CDIM is large, we anticipate that standards will need to be observed in an array of 10×10 locations across the detector. Intermediate locations will then be interpolated from these measurements to obtain the desired threshold. By interspersing low cadence (once per month), large number of positions observations of standards with high cadence (one per week), few position observations, we anticipate being easily able to achieve the required characterization of the PSF and thereby the desired photometric accuracy.

The absolute calibration of CDIM will be based on spectral standard stars, of which a handful will fall in our wide field. We will also routinely measure a number of secondary standards and well-understood stars within our fields. Photometric calibration of CDIM will be better than 2% using these calibration sources, which exceeds the requirement on the accuracy required on both the absolute level of power spectra and the accuracy of measuring reionization source fluxes.

Another calibration requirement is set by the channel-to-channel gain calibration, which is important for reconstruction of both source spectra and for cross-correlations between wavelengths. The inter-channel calibration will combine information from ancillary catalogs, ground calibrations, and in-flight monitoring to produce stable and reliable gain calibration to better than 2% accuracy. This will permit source spectra and cross-power spectra to be limited by statistical rather than calibration uncertainty.

We will measure the point spread function (PSF) in each exposure using a stacking technique to reconstruct the optical transfer function of the end-to-end system (Zemcov et al., 2014). This point spread function will be measured to a level of at least $10^{-5}$ per exposure, and will be monitored for changes as a function of time. We can reach this level of accuracy by virtue of the number of stars in each field, which act as ideal point sources that each probe the end-to-end system PSF. This level of accuracy will allow us to perform precision photometry of point sources, as well as to reconstruct the beam response function to allow a precision measurement of the small-angle behavior of power spectra.

CDIM: Cosmic Dawn Intensity Mapper Final Report                                                                    CDIM## 3.12 CDIM Systematics Control

The CDIM survey design offers several levels of systematic error control and mitigation. Our survey strategy naturally provides several visits of the deep field in a given band per year, which allows us to measure and develop mitigation strategies using flight data. The redundancy of the data set is such that we can develop and constrain models for each of the effects contributing to our signal, and the statistical power of the data is such that each of these can be reconstructed to high fidelity. By using a combination of mitigations during the design phase, monitoring during observations, and statistical tests during data analysis, we will be able to measure and track these systematic errors over time. Below we list the main systematic errors and how they are controlled in the CDIM data.

**Survey Redundancy.** Each strip within a survey field will have some overlap with a previous strip. CDIM could, for instance, offset each strip by half the width of the detector array (i.e., offset perpendicular to the wavelength direction by 2 detector widths). CDIM does not have any constraint from Earth avoidance, so could observe a strip sampling every other point, and then double back and fill in the interleaved points. We could potentially extend that to stepping every third bin and taking additional passes to complete. If there are drifts in instrumental characteristics, this scheme exchanges spatial variations for spectral variations. Selection of an optimal strategy would require simulation of the data, including drifts.

**Point Spread Function.** CDIM's PSF doesn't need to be known particularly accurately for the spectral line intensity mapping science because the beam transfer function is unity at the angular scales that impact our science. The beam transfer function does matter for accurate reconstruction of the power spectrum for the smallest angular scales, but there is little science of interest there. A more powerful way of using the beam transfer function is as a check of the data analysis pipeline, by requiring that the output power spectra follow a shot-noise power spectrum at the smallest angular scales, as dictated by the statistical isotropy of galaxy point sources on the sky. This is a stringent test of other corrections in our analysis pipelines for instrumental and astrophysical effects that occur on fast time and small angular scales. Knowledge of the PSF is also important for masking out sources, but with 1″ pixels CDIM can afford to be conservative and remove a larger number of pixels in marginal cases. We will construct a source masking algorithm using simulations of the point spread function to determine the optimal masking procedure. This kind of procedure also gives detailed information of systematics in point source studies, including response variations, spatial behavior, and so on.

**Relative Pixel Response.** The relative gain response between pixels (sometimes called the "flat field") is an important instrumental quantity to measure and monitor as it can imprint large scale structure if improperly corrected. The relative pixel response will be calibrated on the ground and monitored using in-flight data. Specifically, we can control this systematic by noting we will have S/N>1 per pixel on zodiacal light per 250-s exposure. Zodiacal light does not exhibit fluctuation power on < 1° scales (Arendt et al., 2016), nor is it known to change color over time as the dust population producing it is well-mixed (Leinert et al., 1998). As a result, zodiacal light provides an excellent emission field with well-understood spatial and spectral properties from which to monitor the gain. The survey redundancy allows us to sample the sky more quickly than the temporal zodiacal light varies, so the largest effect will be a gradient term across the array. This can be modeled using the formalism of Fixsen et al. (2000), where the gain variation is modeled as part of a larger mosaicking algorithm.

**Pointing Jitter.** Short time scale motion in the attitude control system, or "jitter", in the telescope pointing will spread flux from discrete sources over a larger number of pixels than the ideal value. This has two connected effects: (1) to spread the flux from a given source into neighboring pixels; and (2) to broaden the PSF. The broadening of the PSF is not problematic, per se, since we will apply a conservative masking scheme to the data that is derived from the properties of the entire data set. Measurement of the PSF per exposure will allow us to determine the appropriate correction to apply to photometry to reconstruct accurate fluxes. We will also develop lookup tables of interpixel capacitance, electrical cross talk, and sub-pixel quantum efficiencies during ground testing, which could be applied to construct corrections for low-level detector effects.

**Dark Current and Out of Band Response.** Dark current and leakage of out-of-band photons through the optical filtering both yield a spurious photocurrent that is not related to the in-band astrophysical structure in an exposure. The dark current in these devices is known to be quite stable, and will be measured on the ground and after launch. The real-time dark current can be monitored using sacrificial pixels at the edges of the arrays, which will provide measurements of a representative subset of the array. It is possible that more aggressive schemes like a dark shutter could be used to provide occasional checks as well. For the out-of-band photocurrent, we will characterize the filters on the ground, and use the measured profiles to correct the data during data analysis. The filter coating is not expected to change or degrade over the course of the mission, so this will be a static correction. The extent to which it is not static can be traced using bright sources, which will allow us to trace dynamic leakage across the array.

**Stray Light.** Light from outside the direct optical path, whether a specular reflection or scattered, can contribute spurious signal that is related to the brightness of the astrophysical sky away from the field of view of the telescope. We will perform detailed simulations during the instrument design phase to minimize the stray light paths, and will include both coherent and diffuse paths. The telescope baffling will be designed to reduce the far-angle response to below the requirements set during a later design phase. We will then characterize the stray light performance in flight using scans over the Milky Way to place constraints on the maximum level at which it could be present in the data.

# 4 MISSION IMPLEMENTATION

CDIM's four-year astrophysics survey requires operation of a spacecraft at L2 to support a wide-field telescope and provide data downlink and telemetry. The Ball Configurable Platform is able to provide these capabilities without the need for technology development. Bus design and testing will follow NASA Class B requirements and as such CDIM is resilient to single fault failures.

## 4.1 Mission Architecture

CDIM observes the sky by performing repeated slew and observation sequences over its four-year mission lifetime. All sequences are pre-determined for the week and do not require updating from the Science Team over the course of the mission. Science observations consists of staring at a point in the sky then stepping half a spectral bin width and staring again. This sequence repeats for ~one week until a scheduled reaction wheel desaturation occurs. The only time a different sequence is used is when the current target has been fully sampled or when the spacecraft is approaching its sun-zenith constraint (**Figure 4-2**). The spacecraft downlinks science data once per day while simultaneously observing the sky.

The telescope is mounted on top of the spacecraft bus to provide unobstructed views of space and enable passive cooling via V-groove radiators as seen in **Figure 4-1**. The V-groove radiators provide 18° of rotational freedom between spacecraft Z-axis and solar zenith, as shown in the top right of **Figure 4-2**. Top-mounting the instrument also enables the spacecraft to be power-positive during all science portions of its mission while also allowing simultaneous science and telecom operations. The passively-cooled instrument thermal system is designed to maximize observation efficiency by providing the largest solar exclusion zone possible. Large V-groove radiators allow the telescope operational flexibility in mid latitudes and continuous viewing of fields above 72°N and below 72°S.

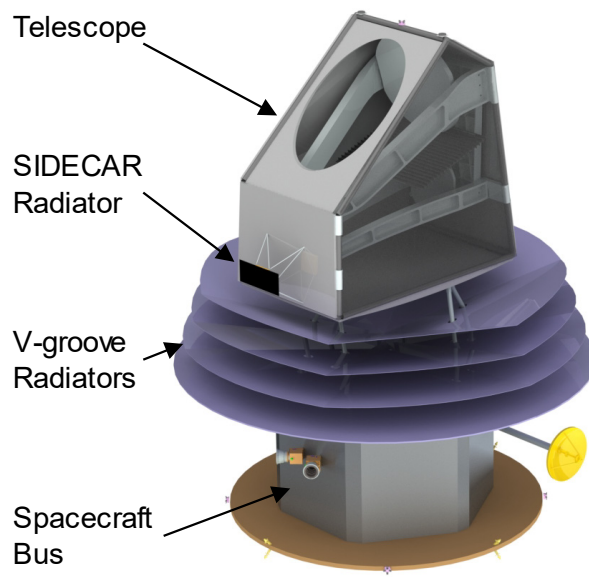

**Figure 4-1.** CDIM's telescope is mounted on V-groove radiators above the spacecraft bus to enable passive cooling and maximize observation time.

CDIM uses a dual string architecture to comply with NASA Class B mission requirements and ensure tolerance of single fault failures. The payload design is based on the successfully implemented Planck V-groove radiators, while the bus draws heritage from numerous successful Ball missions using the same platform (§6). Overall, the design is optimized to minimize operational risk by using an instrument with simple operating modes, a fault tolerant spacecraft bus that requires no deployments, as well as heritage components and operations (§6). Mission functional requirements are derived from science requirements shown in the Science Traceability Matrix, **Table 3-1**, and are shown in **Table 4-1**, the Mission Traceability Matrix.

## 4.2 Science Mission Profile

CDIM performs near continuous science operations at Earth-Sun L2 consisting of a series of 250-s images separated by half of a spectral bin. After each image the spacecraft steps 7.3″ in the spectral (3.4°) direction of the focal plane to obtain the next sample in a different bin. There are 1680 steps are taken to achieve Nyquist sampling over a single 3.4° × 2.3° FOV, requiring 58 hours of observations. While stepping between fields the High Gain Antenna can be pointed towards Earth to downlink science data once per day while continuing data collection.

**Table 4-1.** Mission Traceability Matrix. Colored dots indicate the flow of driving requirements. Colors originate in the "Mission Functional Requirements" column and appear next to the design requirements they impact.

| Mission Functional Requirements | Mission Design Requirements / Constraints | Spacecraft Requirements / Constraints | Ground System and Operations Requirements |
|---|---|---|---|
| Deep, Medium, and Wide surveys each with ≥ 90% voxel completeness for internal reliability | **Orbit Characteristics**<br>• L2 orbit with semi-major axis of 732,000 km<br><br>**Thermal Attitude Limits**<br>• S/C -Z axis ≤ 18° from local zenith<br><br>**Launch**<br>• Date: not constrained<br><br>**Mission Duration**<br>• 49 months: 1 mo. For cruise and decontamination; 48 mo. science operations | **Launch Vehicle Capability**<br>• 3453 kg (Falcon 9 to L2 capability)<br><br>**Orbit Average Power**<br>• 622 W required (MEV)<br><br>**Radiation Tolerance**<br>• 27 krad behind 100 mils Al<br><br>**Attitude Control**<br>• 0.5″ Pointing Knowledge<br>• < 0.5″/250-s RMS stability<br>• < 1″ control<br><br>**Downlink/Uplink**<br>• Science Data rate: Ka-band 150 Mbps<br>• TLM Data Rate: S-band 2 kbps<br>• Uplink Data Rate: S-band 2 kbps<br>• Contacts: 1/day<br><br>**On-board Data Storage**<br>• 48.8 GB per day | **Science Data Operations**<br>• DSN tracking and coverage (Ka-Band DL, S-band DL/UL)<br>• I/SOC @ JPL<br>• 48 mo. science operations<br><br>**Science Data Latency**<br>• 2 months<br><br>**Mission Operations**<br>• 49 mo. flight operations<br>• Schedule downlink contacts<br><br>**Data Downlink**<br>• Transmit 48.8 GB science / day<br><br>**Data Uplink**<br>• 1 S-band pass per day |
| Effective PSF FWHM ≤ 2″ at 1 μm | | | |
| Stable cooling to < 35 K to control > 5 μm array dark current | | | |
| Deep Survey: 15 deg², embedded in the Wide survey | | | |
| Medium Survey: 30 deg², to overlap with 21-cm fields from HERA and SKA1-LOW | | | |
| Wide Survey: 300 deg², driven by number of AGN detections | | | |
| Read, compress, and telemeter spectral imaging data | | | |

There are three primary fields: a deep (15 deg$^2$) and wide field (300 deg$^2$) both with observation centered around the SEP, and a medium survey (30 deg$^2$) to overlap with fields from HERA and SKA1-LOW. Each of these fields will be observed at least once per year for systematic control. Only one of CDIM's three proposed targets has viewing constraints during any part of the year, the medium field, due to solar avoidance criteria. It receives observation priority during all months when it is visible, approximately 84 days per year in two equally sized segments.

The most operationally-complex period for CDIM occurs four times per year when a change to or from the medium field is required. On these occasions the spacecraft will perform science observations, a large slew, and downlink science data, as shown in **Figure 4-2**.

## 4.3 Launch Segment

CDIM is compatible with 5-m EELV fairings and has a maximum diameter of 4.5 m. The largest element is the bottom-most V-groove radiator, which can be sized depending on the launch vehicle fairing to ease the Sun-zenith constraint (**Figure 4-3**).

CDIM's MEV dry mass is 1429.2 kg with an additional 24.5 kg for propellant and 25 kg for the launch vehicle adapter. The Falcon 9 ocean-recovery launch vehicle is capable of accommodating 3453 kg wet mass at launch to the desired L2 orbit, yielding a 136% launch mass margin. **Table 4-2** shows the subsystem current best estimate (CBE) mass breakdown as well as margins to the launch vehicle capability with a reduction for the launch vehicle adapter.

## 4.4 Flight Dynamics

The launch vehicle will insert the spacecraft into a Sun-Earth L2 halo orbit with a semi-major axis of 732,000 km (**Figure 4-2**). There are two small trajectory correction maneuvers (TCMs) that occur during cruise: a post-launch cleanup (10 m s$^{-1}$) and a pre-insertion cleanup (5–10 m s$^{-1}$). Station-keeping maneuvers occur twice per orbit, requiring 4 maneuvers per year (2 m s$^{-1}$ per year). Total mission delta-V is 28 m s$^{-1}$ for the 18 required maneuvers.

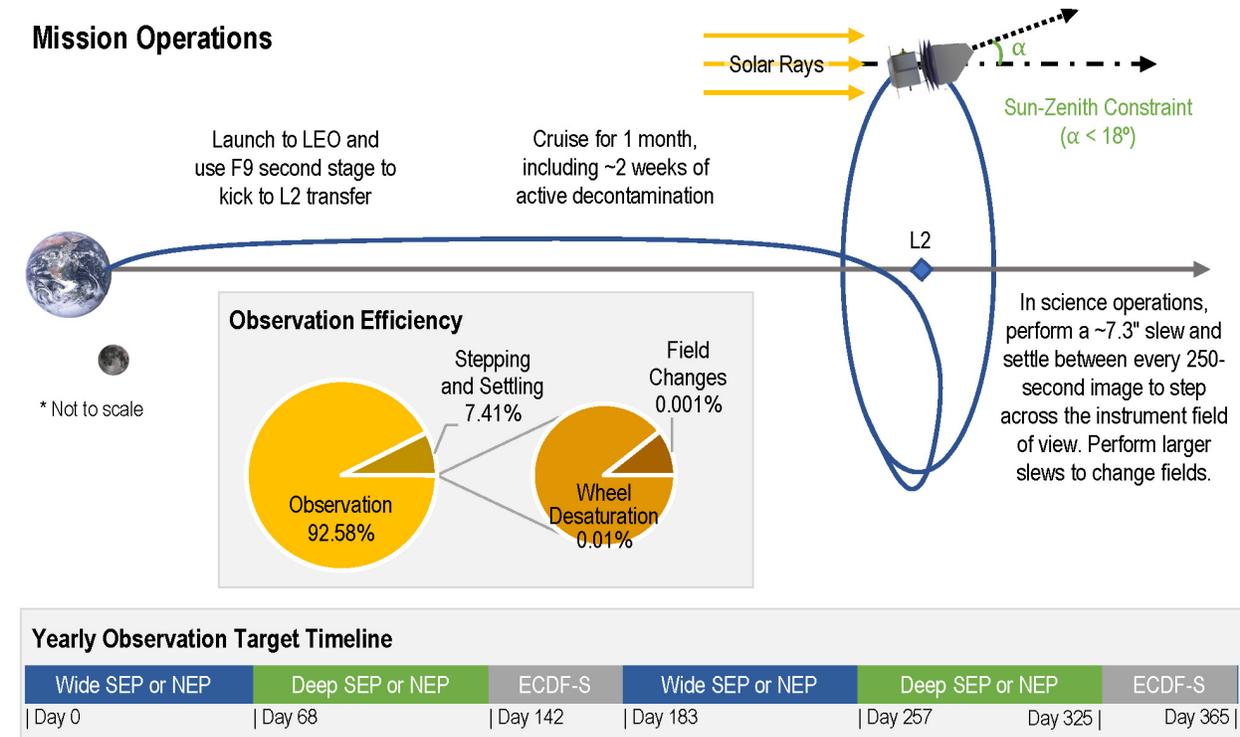

**Figure 4-2.** CDIM's operational scenario is designed to maximize the amount of time in science observation mode by performing as many continuous small step and stare activities, enabling coverage of each science region at least once per year.

## 4.5 Payload Accommodation

CDIM's payload consists of a 1.1 m primary aperture telescope with four V-groove radiators with a diameter set based on the launch vehicle fairing size (**Figure 4-3**). The telescope is rotated 90° with respect to the spacecraft's z-axis, shown in **Figure 4-1**, to enable unconstrained viewing of both NEP and SEP regions between 72° and 90°. The telescope sits atop three separate bipods that also support the four V-groove radiators. Large V-grooves enable the bus to orient ±18° around the sun zenith direction (**Figure 4-2**), providing a worst-case minimum observation time of 72 days per year for a point target at 0° latitude. Further payload thermal analysis may indicate that the number (and therefore height) of the V-groove radiators can be reduced while still maintaining thermal requirements (§8.3). This would allow the telescope to sit closer to the bus and relax the viewing angle constraint. The total payload mass is 694 kg CBE.

The payload's 24 SIDECAR detectors generate data at 5 Mbit s$^{-1}$ during observation mode, which will occur up to 92% of the mission (**Figure 4-2**). Daily data generation is conservatively set for 95% data collection or 413 Gbit/day. There are minimal data latency requirements imposed by the science team, permitting flexible science operations. During observation mode the telescope SIDECAR ASICs require a maximum power of 50-W CBE (**Table 4-2**). The only time power will exceed 50 W is during decontamination mode, which occurs during cruise to L2 and requires 125 W thermal power.

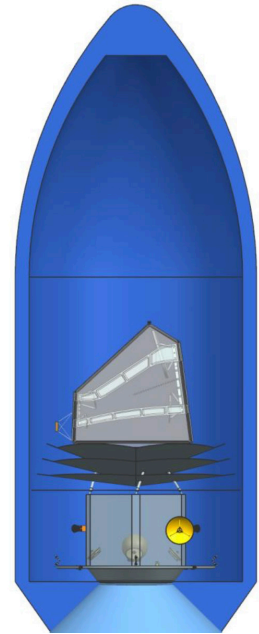

**Figure 4-3.** CDIM fits within EELV dynamic fairings and its bottom V-groove can be reduced in size to fit additional vehicles with minimal impact to instrument thermal performance.

## 4.6 Spacecraft Design and Subsystems

The spacecraft bus is designed to satisfy all science requirements (§2.6) while using heritage components and requirements within Ball Aerospace's current capability. CDIM only has one deployment, the telescope cover, and none are required for the spacecraft bus. Potential single fault failures are designed to be redundant according to Class B requirements. The spacecraft subsystem level mass and power breakdown is shown in **Table 4-2**. The Report from Team X – JPL's concurrent design facility – includes a full Master Equipment List (MEL) and component list. Spacecraft subsystems are shown in **Figure 4-4** and are described below.

**Attitude Determination and Control**. CDIM's most demanding Attitude Determination and Control Subsystem (ADCS) requirements are set to reduce PSF jitter and perform the small steps that enable LVF survey sampling with 7.3″ steps (**Table 4-1**). Pointing stability requirements derive from reducing

**Table 4-2.** CDIM's subsystems with Current Best Estimate values shown for mass and orbit average power (OAP) with healthy margins to MPV design parameters.

| Subsystem | Mass (kg) | OAP (W) |
|---|---|---|
| Attitude Control | 66.1 | 156 |
| Command & Data | 16.8 | 38 |
| Power | 46.4 | 60.5 |
| Propulsion | 13.1 | 8 |
| Propellant | 24.5 | 0 |
| Structures & Mechanisms | 224.2 | 0 |
| S/C-Side LV Adapter | 8.6 | 0 |
| Cabling | 48.5 | 0 |
| Telecom | 27.7 | 60 |
| Thermal | 22.3 | 61 |
| **Spacecraft Bus Total** | 489.6 | 383.5 |
| Instrument | 694.0 | 50.0 |
| Dry Total, CBE | 1183.6 | 433.5 |
| Dry Total, MEV | 1429.0 | |
| **Wet Total, MEV** | 1453.5 | 621.5 |
| **Maximum Possible Value** | 3428.5 | 820.0 |
| **Margin (%)** | 136% | 32% |

PSF jitter where pointing control requirements flow from the need to target spectral channels to a small fraction (0.15) of the channel width. Pointing knowledge flows from the control and jitter requirements.

The 3-axis stabilized attitude control system is designed to meet science requirements and maximize science observational efficiency by decreasing the amount of time spent slewing and settling between observations. Small 7.3″ steps are designed to take less than 20 seconds to slew and settle on reaction wheels, with desaturation events occurring ~once per week. These steps must be controlled to <1″ to achieve the required spectral sampling. The spacecraft does not need to slew to accommodate telecom ops because of the gimbaled Ka-band high gain antenna.

Three operational stellar reference units (SRU) are required to obtain the pointing knowledge requirement during operation with Sun sensors used to maintain pointing knowledge during safe mode. Fine spin rates are determined by both the SRU and one internally redundant inertial reference unit. All attitude control can be performed by 3 reaction wheels and desaturation utilizes 0.9 N hydrazine thrusters. In total CDIM has 6 Adcole 2-axis sun sensors, 4 Sodern SRUs, 1 internally redundant Northrop Grumman SIRU, and 4 Honeywell HR14 reaction wheels.

**Command and Data Handling**. CDIM's bus receives data from the payload at a maximum rate of 5 Mbit/s (up to 413 Gbit/day, §3.7). CDIM uses a dual-string cold RAD 750 chassis capable of recording data at up to 10 Mbit/s (100% margin) with 192 GByte (86% margin) flash memory.

The data storage system will store two days of data, the one currently being observed, and the previous day to ensure data can be retransmitted if necessary. If a single ground track is missed the data can be retransmitted over the following three days without loss. CDIM has the flexibility to alter its data storage timeline to allow transmission flexibility.

**Power**. CDIM's power subsystem is power positive in all modes at its worst-case 18° solar off-point and requires no solar panel deployments. Only the launch scenario and decontamination heaters will require battery operation. One 56-Ah Li-Ion battery is required for 3 hours of operations after LV umbilical separation until solar array is sun pointed. The maximum depth-of-discharge is 67%.

The solar panel is a 3.3-m outer-diameter, 2.4-m inner-diameter, toroidal lower panel with a total mechanical area of 4.17 m$^2$. 29.5% efficient GaAs cells provide 908 W BOL power and 820 W EOL power granting a 32% orbit average power margin, shown in **Table 4-2**.

**Propulsion**. CDIM's monopropellant blowdown hydrazine system is sized to perform ACS desaturations, station-keeping maneuvers, and two trajectory correction maneuvers (TCM). One TCM happens after LV separation and the other occurs for L2 insertion. Station-keeping will only occur once per orbit (once per six months) and ACS desaturations are scheduled weekly.

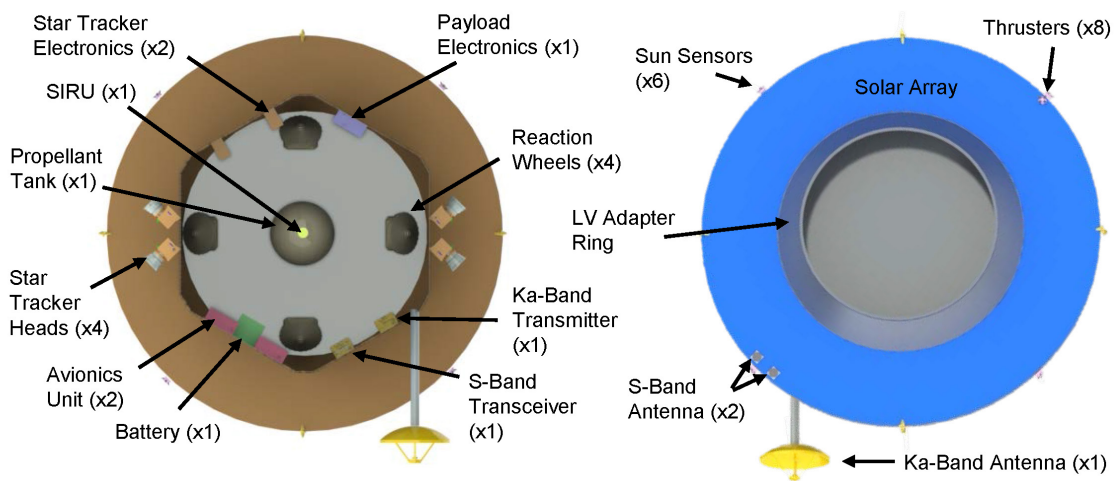

**Figure 4-4.** Ball Aerospace spacecraft bus, showing location of the major components and subsystems.

Eight MR-103C thrusters are positioned for ACS and station-keeping maneuvers and perform both TCMs. One ATK PSI P propellant tank has ample storage for the 3.7 kg propellant required for ACS and 14 kg for TCMs and station keeping yielding a total propellant mass margin of 38%.

**Structures**. CDIM's hexagonal aluminum bus structure is designed to support LV launch loads and is ideal for supporting the instrument's 3-bipod configuration, and with a single telecom gimbal for a 2-axis gimbaled High Gain Antenna. The spacecraft bus has no deployable structures since both the solar array and V-groove radiators are designed to fit within EELV 5-m fairings.

**Telecom**. CDIM transmits directly to Earth during all portions of its mission without interrupting science operations using a 0.65-m gimballed High Gain Antenna (HGA) and redundant Ka-band transmitter with amplification. Gimbaling the HGA enables a ~5% increase in observation time over the course of the mission. The system is designed to communicate with a 34-m DSN antenna at 150 Mbits/sec with 10.5 dB link margin capable of returning all science data in less than one hour per day.

Two S-band transponders with amplification are used with patch antennas to receive omnidirectional uplink and downlink at 2 kbps. S-band uplink has a link margin of 33 dB and downlink has 16.2 dB.

The primary telecom hardware consists of two Ka-Band transmitters, two 60-W Ka-Band TWTAs, one 0.65-m Ka-band High Gain Antenna, two S-band transponders with internal 8-W SSPAs, two S-band transmit patch antennas, and one waveguide transfer switch.

**Thermal**. CDIM operates in a stable thermal environment at L2 with only small variations in solar angle over the course of the mission. Additionally, the telescope is thermally decoupled from the bus by its bipod structure and V-groove radiators. The spacecraft bus does not provide any thermal control for the payload. Spacecraft thermal design uses flight-proven, low-risk passive temperature control including only PRT temperature sensors, mechanical thermostats, multi-layer insulation, flexible Kapton heaters, and white paint on structures.

## 4.7 Mission Operations and Ground Data System

CDIM has four primary mission phases: Launch, Early Operations, Cruise, and Science. Launch phase occurs during the first two weeks of the orbit and involves initial acquisition and systems checks. There are no mission-critical events that occur during this stage because solar arrays and V-groove radiators do not require deployments. CDIM will communicate with a DSN 34-m ground station twice per day with eight-hour ground tracks. The second phase occurs during the following two weeks to complete spacecraft checkout. Ground tracks in this phase decrease to once per day. Cruise phase consists of navigating to the desired L2 orbit, deploying the telescope cover, instrument decontamination, instrument calibration, and performing the L2 insertion burn. During the fourteen-week transit to Earth-Sun L2 orbit, CDIM will communicate with the ground four hours per day. The final mission phase, science operations, lasts 199 weeks and is designed for daily 1.5-hour passes. The spacecraft requires ~1 hour to transmit all science data each day (413 Gbit/day) with an additional 30 minutes added to account for retransmission. If one pass is missed, all data can be retransmitted without loss over the following three days. Over the mission lifetime 602 Tbit of science data will be transmitted to DSN, expanding to an estimated 1.5 PByte storage needed for the science data system. Number and duration of DSN 34-m BWG passes can be traded by increasing the size of the onboard storage card to handle passes every other day. The gimbaled HGA allows simultaneous data transmission and science operations, decoupling science observations and ground contact scheduling.

CDIM's science operations have a pre-planned science phase that allows for a high level of autonomy. Targets and durations do not require updating during the nominal four-year mission life. The instrument is also extremely simple to operate with only 'on' and 'off' states. These factors allow S-band uplink and RF navigation to occur only once per week.

The science data system uses standard NASA AMMOS capabilities and services with operations at JPL. Once data is received on the ground it enters the data pipeline at IPAC. The CDIM Science Team will perform data analysis in the IPAC environment, and IPAC will archive the data products.

# 5 TECHNOLOGY DEVELOPMENT PLAN

## 5.1 Detector Technology

**The CDIM instrument** is a simple high-throughput design that relies on established technologies for the telescope and most of the focal plane. Areas of required development are the 7.5 µm H2RG sensor and the linear variable filter elements (both currently assessed at TRL 4). Investments by NASA can readily advance the state of the art in these areas. Our plans leverage these investments and provide three-year programs to advance TRL to 6 in advance of flight development.

Five of the six CDIM bands for the focal plane (§3.6) are accommodated by existing, off-the-shelf detector products (TRL 9) from Teledyne that currently meet the mission requirements. The sixth band has a cutoff wavelength of 7.5 µm to reach redshifted H-alpha 656 nm at $z = 10$ (7.2 µm). In this section, we describe our plan to advance the TRL of the 7.5 µm detector from 4 to 5 in advance of flight development.

Our program addresses two developments required to enter a flight qualification program for the long wavelength detector:

1. Removal or qualification of the CdZnTe substrate layer to minimize the fluorescence from energetic particles in space (Waczynski et al., 2005); and
2. Improvement in "operability", or % pixels meeting CDIM requirements, in particular dark current.

A number of groups are already working with Teledyne to extend the cutoff wavelength further into the infrared (McMurtry et al., 2016a). It is possible that ongoing work funded by NEOCAM will improve performance in some or all of the areas that would be pursued by CDIM. Although the NEOCAM dark current requirement (200 e⁻s⁻¹) is relaxed by comparison to CDIM (0.17 e⁻s⁻¹), removal of most of the CdZnTe substrate upon which the HgCdTe is grown (using a molecular beam epitaxy process) is still required to minimize fluorescence (Dorn et al., 2016). H1RG 1k×1k test products are already available with cutoff wavelengths as long as 13 µm (McMurtry et al., 2016a). Devices with a 10 µm cutoff produced for the NEOCAM study have been substrate-thinned to 30 µm and exhibit dark current (>80% of pixels) of under 0.2 e⁻s⁻¹ at 35 K (McMurty et al., 2013), meeting the science requirements of CDIM. Complete removal of the substrate, which is not required for NEOCAM, becomes increasingly difficult as the cutoff wavelength increases due to variations in the properties of the alloy (Phillips et al., 2001). It is unclear whether complete removal of the CdZnTe substrate layer is required for CDIM. This will be investigated during radiation testing as part of a trade of fluorescence vs. operability. Although other missions have specified complete removal, it is not currently clear if a very thin remaining layer would be acceptable for CDIM, and such a layer would have the benefit of providing additional rejection of wavelengths shorter than 800 nm. It is reasonable to expect the achievable dark current of the 7.5 µm devices required by CDIM would be inherently lower than the existing 10-µm cutoff test sensors due to the reduced band gap. A goal of 95% operability is baselined for CDIM to match the rest of the focal plane. Reduced operability could be recovered with increased observing time. The planned observing sequence

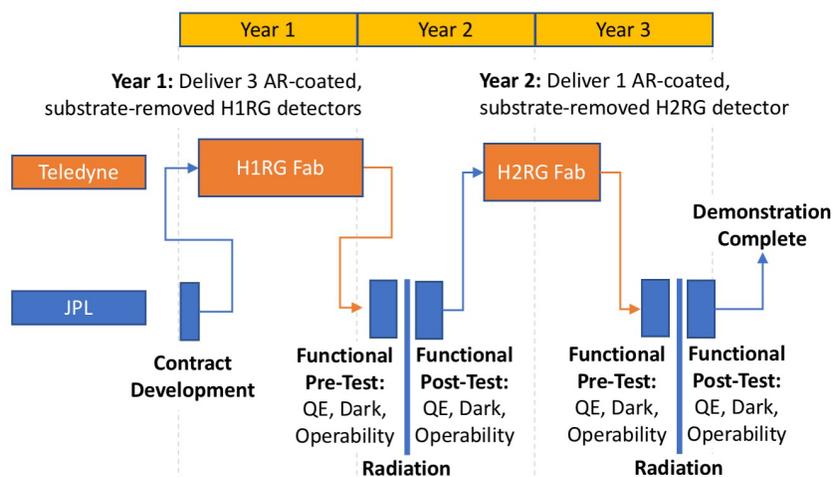

**Figure 5-1.** The planned technology development program for long wavelength detectors includes process development using the H1RG format and final delivery of a flight-like H2RG.

with two offset visits to each field to provide Nyquist sampling also mitigates this effect.

### 5.1.1 Development Plan

We propose a three-year plan to work with Teledyne to develop the required sensors for CDIM, summarized below. Remaining work to qualify 7.5-µm cutoff sensors for space flight would include:
- Removal of the CdZnTe substrate (or evaluation of the maximum acceptable thickness)
- Improvements to the MBE growth process to improve operability >95%
- Application of appropriate anti-reflection (AR) coatings to optimize quantum efficiency
- Radiation testing to verify operability and fluorescent properties

### 5.1.2 Schedule and Budget

In year 1, we contract Teledyne to produce three substrate-thinned (of different thickness), AR-coated 7.5-µm-cutoff H1RG sensors favoring quantity over size to maximize return. We would proceed to evaluate the key parameters of QE, dark current, and operability as a function of temperature. Radiation testing would verify that the sensors meet their background requirement in the space environment and also retain a high percentage operability. Performance feedback to Teledyne after six months of testing would inform a second round of device fabrication.

In the second round, we would produce a flight-format H2RG sensor with the optimal substrate thickness determined in the first round to demonstrate uniformity and achieve TRL5. A final set of results will establish the state of the art at the required cutoff wavelength and inform the project in Phase A. Further environmental testing would be required during normal flight development to achieve TRL6.

Our budget estimate for this work is $8M, representing a commitment of about 10% of the overall detector budget.

## 5.2 LVF Technology

The current state of the art of Linear Variable Filters (LVFs) has been demonstrated by industry: Omega, Viavi (OCLI), Materion (Barr Associates), and other research labs have fabricated filters between 0.4 µm to 4 µm (over several octaves) with 2 to 100 nm/mm spectral gradient. These LVF have in-band transmission > 85% and have out-of-band rejection 3 to 5 Optical Density (OD). State-of-the art LVF filters can be fabricated with area size 2 mm to 50 mm in the spectral direction and 15 mm to 50 mm in the cross-spectral direction.

Typical LVF filters fabricated for flight applications in the visible can survive: multiple -40°C to +80°C thermal vacuum cycles; 24 hr of 95% relative humidity; 10 krad radiation; and >25-g vibration. Omega performed extensive environmental tests of filters for the Indian Space Research Organisation.

The CDIM filter requirements are to fabricate 4 × 6 (LVF) filter matrix architecture, mounted in front of the CDIM detector focal plane assembly (**Figure 3-8**). These LVFs for the CDIM are required to have 6 segments covering 0.75 µm to 7.5 µm spectral range (**Table 3-3**) and have an out-of-band rejection of OD-5. The LVF will have R = $\lambda/\Delta\lambda \geq 300$ and will be illuminated by f/4.5 optics beam and angle of incidence (AOI) between 0° to 7°. The size of each filter segment will be 40 mm × 40 mm. These filters will be kept thermally stable at 35 K.

The main challenges with the LVF for the CDIM are extending the spectral range of the LVF > 4 µm to 7.5 µm (coating materials), achieving out-of-band rejection at ≥ OD-5 and controlling induced stress of the coating layers and substrates. These will be addressed in the LVF technology development plan, §5.2.1.

During this Probe study we invested in a short-term technology development task with Omega Optics (a potential vendor with an extensive flight hardware heritage) to develop the LVF filter models for the CDIM applications (**Figure 5-3**) and fabricate a 40 mm × 40 mm LVF prototype filter (**Figure 5-2**).

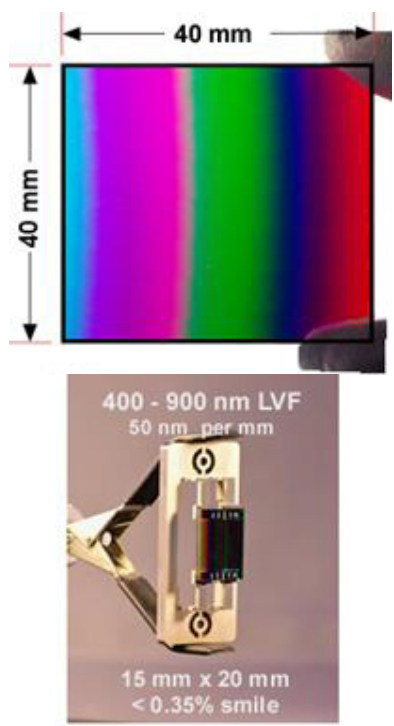

Figure 5-2. LVFs designed and fabricated by Omega Optical. *Lower panel:* Flight-qualified LVF filters delivered to the Indian Space Research Organization (ISRO). *Upper panel:* 40 mm × 40 mm LVF prototype (segment B) fabricated for the CDIM Probe study.

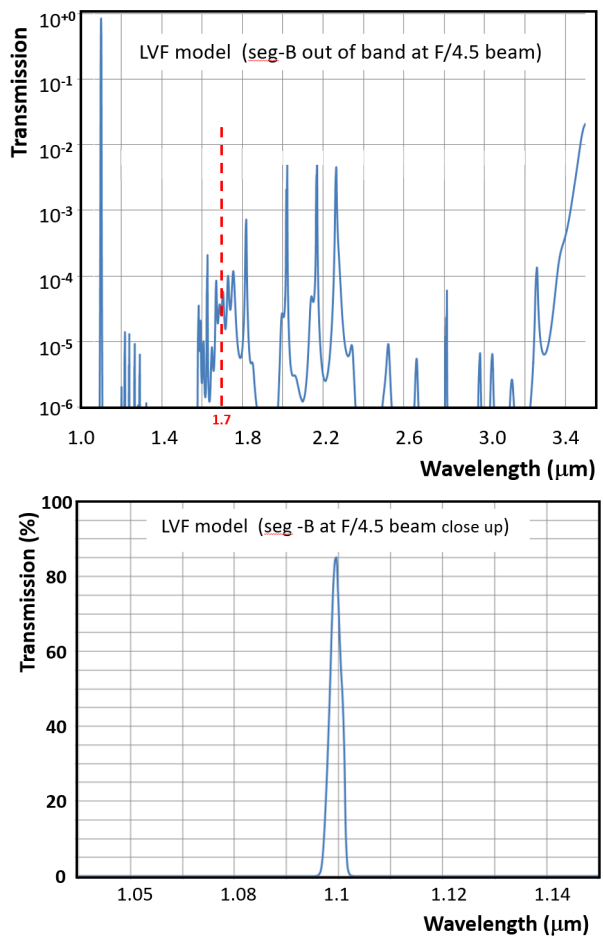

Figure 5-3. *Top panel:* Segment B model at 1.1 μm showing the out of band rejection at f/4.5 beam (note that detector has no response longward of 1.7 μm; **Table 3-3** and **Figure 3-7**). *Bottom panel:* Segment B model at an f/4.5 beam (close up).

Filter models were developed for all six LVF segments (A through F) operating at an f/4.5 beam. These models were constructed with the goal of meeting an optical density OD-5 (out-of-band rejection). Due to the extended rejection spectral range requirements, segments A, B, and C are found to be the most demanding segments with respect to out-of-band rejection.

We used the LVF segment-A model (**Figure 5-3**) as a demonstration model for the remaining segments. Segments B though F have been shown to have a similar performance but with an improved out-of-band rejection in segments D, E, and F. All models have been shown to meet the required f/4.5 beam as well the AOI between 0° and 7°.

### 5.2.1 Development Plan

We propose 4 tasks for the LVF technology development (**Table 5-1**). Some of the tasks may have an overlapping timeline.

**1. Develop Coating Materials and Processes**

We plan to work with Omega Optical to develop processes to use known IR coating materials in the 4 μm to 8 μm spectral range (Ge, Si, ZnS, and YF3), and to develop processes for the FeOx and ITO materials to improve blocking for all segments. This will include implementing Linear Variable Blocking Stacks designed and applied on each LVF filter segment.

OD-5 out-of-band rejection can also be achieved by adding a second air-spaced (gradient) blocking filter to the light path.

CDIM: Cosmic Dawn Intensity Mapper Final Report### 2. LVF coating stress

Develop processes to blend film layers to disrupt ordered film growth and use post-deposition annealing technology and the use of interface layers to mitigate the buildup of stress throughout the film stack.

### 3. Develop software code specific for LVF optimization

Modify existing optimization code developed by Omega to simultaneous optimize models at multiple wavelengths against multiple wavelength targets. This will improve the accuracy of LVF models for wide blocking spectral range requirements.

### 4. Fabricate prototype LVF segments for CDIM

Fabricate full-size prototype LVF segments. This will includes setting up a metrology testbed for testing the LVF (open spectrometer with different IR detectors covering the CDIM spectral range). To achieve TRL 6, we will follow flight environmental testing requirements and procedures on LVF prototypes – including multiple thermal cycles to the operating temperature of 35 K.

## 6  HERITAGE

The CDIM design takes advantage of heritage in the following hardware items:

**CDIM mirrors**. The CDIM beryllium mirrors are of a freeform figure, modest radius of curvature, and smaller in diameter than mirrors already fabricated for JWST (Lightsey, 2007).

**H2RG detectors**. Widely used in the infrared community and currently available from Teledyne. Flight versions have been developed for JWST (Rauscher et al., 2011) and Euclid (Waczynski et al., 2016). CDIM makes use of current performance measurements for its integrated model. The community is actively working with Teledyne to develop products that meet the requirements of the CDIM long wavelength band (e.g., NEOCam, McMurtry et al., 2016b).

**Linear Variable Filters.** LVFs in the near infrared (up to R~200) have been flown in the New Frontiers New Horizons/LEISA (with R=240 and 560; Reuter et al. 2008). An LVF full-color image of Jupiter using LEISA was featured on the cover of Science (Baker, 2007). More recently, OSIRIS-REx arrived at asteroid Bennu; its OVIRS instrument (Reuter et al., 2018) covers the wavelength range 0.4–4.3µm in 3 bands with R = 125–200.

**Thermal design**. The simple thermal design is based on the successful Planck mission (Leroy et al., 2006), exceeds CDIM requirements, and can be implemented with large margins on available payload mass. The design has substantial margins, possibly allowing one of the V-groove panels to be eliminated (future design trade).

**Spacecraft**. CDIM uses the Ball Configurable Platform (BCP) spacecraft that has successfully supported 12 missions on orbit. An additional three space vehicles are in System Integration and Test (I&T) and two are in development. While they have different structural forms, all spacecraft share a core architecture that has evolved over 20+ years. These programs span Earth sciences and scientific remote sensing, commercial imaging, and DoD operational remote sensing, including Kepler, World View 3 (WV-3), and Joint Polar Satellite System (JPSS-1). The CDIM spacecraft design draws directly from this extensive mission-performance history by using the same flight-qualified components and designs. This proven design has full redundancy and cross-strapping, ensuring it meets the four-year service life. BCP spacecraft consistently meet or exceed their mission lifetime and contract performance requirements, logging

| Table 5-1. LVF Technology Development budget | | | |
|---|---|---|---|
| **Task** | **LVF Technology development** | **Cost** | **Schedule** |
| 1 | Development of 4 µm to 8 µm materials and improve out-of-band blocking | $1.0M | 12 months |
| 2 | LVF coating stress | $1.0M | 8 months |
| 3 | Software optimization | $0.25M | 6 months |
| 4 | Fabricate prototype and performance/environmental testing to TRL 6 | $1.5M | 12 months |
| **TOTAL** | | **$3.75M** | **38 months** |

more than 70 years cumulative on-orbit time. The legacy BCP modular spacecraft design provides flexibility in I&T flow and enhances schedule assurance via parallel manufacturing and integration of the spacecraft.

# 7 COST AND RISK ASSESSMENT

The Cosmic Dawn Intensity Mapper (CDIM) fits within the cost guideline for a Probe-class mission. The total cost estimated by Team X is $929M (incl. launch vehicle, 30% development reserves, and 15% operations reserves).

The costs presented in this Report are ROM estimates from Team X – JPL's concurrent design facility; they are not point estimates or cost commitments. It is likely that each estimate could range from as much as 20% percent higher to 10% lower. The costs presented are based on Pre-Phase A design information, which is subject to change. All costs quoted in this Report are in $FY18.

In this Section we summarize the Team X cost; for some WBS cost elements we provide estimates from other sources, to aid in validation. The CDIM instrument is within family of similar instruments developed for NASA, and has modest technology development requirements at the subsystem level (see §5). The CDIM team partnered with Ball Aerospace to develop a Class-B spacecraft bus that meets the instrument and mission requirements that the CDIM science mission needs. This approach is low-risk in both implementation and in cost estimation. Mission operations at L2 were estimated by Team X; operations at L2 for missions such as Planck, and the planned JWST mission present no unusual challenges in observing, commanding, or data downlink.

The top two risks are both development risks, and both are technologies that can be addressed with pre-phase-A investment: (a) HgCdTe detectors with a 7.5-µm cutoff, and (b) linear variable filters (LVFs) that meet the CDIM requirements including environmental. These efforts can be carried in parallel with, and largely independent of, payload and mission development.

## 7.1 Cost Estimation by Team X and the CDIM Team

Total Mission costs are estimated primarily from the output of the Team X models. The CDIM team substituted costs where we felt that our estimates are higher fidelity than the corresponding model cost. These are discussed below.

### 7.1.1 Instrument Cost

Team X used two models to estimate the cost of the Instrument (Telescope, thermal control system, focal plane, and cold electronics) selected for which best captures the cost for its respective subsystem.
- Optical Telescope Assembly (OTA) – using multivariable parametric telescope cost model by Stahl & Henrichs (2016): $15.2M including reserves.
- Spectrometer – using NASA Instrument Cost Model (NICM VIII, July 16, 2018). For CDIM, there is no spectrometer, so the NICM tool models just the focal plane in this case.
  - System tool: B/C/D Development Cost = $142.7M at 50%-tile, with a caveat that total mass is outside of the range of the model, so the cost estimate is an extrapolation from the model's dataset.
  - Subsystem tool: B/C/D Development Cost = $152M at 50%-tile, with a caveat that thermal mass inputs were outside of the range of the model, so the cost estimate is an extrapolation from the model's dataset. As described in §3, most of the thermal mass comprises the conical panels of the V-groove radiator subsystem. These have a large surface area, but have loose mechanical tolerances and are functionally very simple; the model likely over-estimates the cost.

### 7.1.2 Detector Costs

The NICM System tool predicted a focal plane cost of $8.9M (as part of the 'Spectrometer Cost' in WBS 5.04.01). The CDIM Team replaced this model estimate with a much more conservative estimate of $70M based primarily on Euclid actuals; Team X adopted this estimate. We estimated the cost of developing and fabricating the 24-detector focal plane (§3.5) by scaling from the costs estimated and already incurred by the Euclid project. We also assumed that our 7.5-µm cutoff detectors cost the same as the NEOCam 10-µm detectors.

The specific assumptions in the CDIM cost estimation were (**Table 7-1**):

- Euclid (22 × 2.5 µm HgCdTe)
- NEOCam (4 × 5 µm plus 4 × 10 µm H2RG)
- Total cost scaled linearly according to detector variant
- Adjusted for inflation (2013–2017)
- 24 flight detectors plus a spare of each variant (Total: 30 units)
- 24 SIDECAR ASICs plus 6 spares

NEOCam requires 10-µm cutoff detectors, and is working with Teledyne Imaging Sensors Inc. (TIS) to develop long-wavelength cutoff detectors in their H2RG product line. CDIM assumes the NEOCam development will be complete, and that the needed 7.5-µm detectors are a product that can be purchased from Teledyne. 7.5-µm cutoff detectors (for lower dark current) have fewer development challenges than 10-µm detectors, but since these have not yet been demonstrated, we treat this as a technology development effort that would be undertaken in pre-Phase A or Phase A. Technology development costs are carried separately (see §5.1).

**Table 7-1.** CDIM cost estimate for the 24-detector focal plan is based conservatively on actuals for Euclid. See §5 for technology development plan for 7.5-µm cutoff detectors.

| Sensor | Total Cost | Unit | Quantity |
|---|---|---|---|
| 1.75 µm H2RG + SIDECAR | 4 | $M | 4 (+1 spare) |
| 2.5 µm H2RG + SIDECAR | 15 | $M | 12 (+3) |
| 5.0 µm H2RG + SIDECAR | 6 | $M | 4 (+1) |
| 7.5 µm H2RG + SIDECAR | 16 | $M | 4 (+1) |
| SIDECAR firmware | 4 | $M | – |
| Cable harness (CFCs) | 3 | $M | 30 |
| Teledyne management | 15 | $M | – |
| **Subtotal** | **63** | **$M** | **(30 ROICs + SIDECARs) (30 CFCs)** |
| **TOTAL (inflation adjustment)** | **70** | **$M** | **×1.10 (2013–2017)** |

The CDIM focal plane cost, based primarily on Euclid, is $70M. Team X adopted this higher-fidelity (and much higher) CDIM Team cost estimate. The cost for WBS 5 in **Table 7-2** reflects the Stahl telescope model and the NICM system tool for the spectrometer (but substituting the CDIM Team estimate for the focal plane).

### 7.1.3 Spacecraft Cost

The spacecraft cost (WBS 6) was estimated by Team X using JPL institutional cost models for an 'out of house' build, using design parameters provided by the CDIM team. The Team X estimate for WBS 6 and 10 was $243M.

Separately, the CDIM Team worked with our spacecraft partner Ball Aerospace to develop an independent estimate of the spacecraft costs. Ball estimated both WBS 6 and WBS 10 using SEER, based on the MEL and other design data in the Team X Report. Ball also estimated these costs using actual cost data from previous Ball spacecraft developed to Class B standards, and parameterized according to bus mass:

- WBS 6+10: $193M (SEER-H, at 50% confidence level) [WBS 6 = $159M; WBS 10 = $34M]
- WBS 6+10: $207M (fit to regression for historical Ball Class B missions)
- WBS 6+10: $243M (Team X) [WBS 6 = $220M; WBS 10 = $23M]

The Ball regression model for Class B spacecraft cost vs. mass is a tight relation, and is in reasonable agreement with SEER, while the Team X estimate appears out-of-family. Hence the CDIM team adopted SEER as the best estimate for WBS 6 and 10 (**Table 7-2**). This is reflected in §7.2 below.

### 7.1.4 Mission and Navigation Design (WBS 10)

WBS 10 was costed at $4.2M during an initial Team X session, but this cost was omitted in its Final Report. The CDIM team re-inserts this cost in its mission cost estimate (**Table 7-2**).

### 7.1.5 Science Team and Data Analysis Cost

Costs for WBS 4 were estimated by analogy with experience with WISE and Planck. CDIM has a closely analogous type of raw data product to WISE, and the first few analysis steps are very similar; the spectral data products are more complex, driven by science requirements. The SPHEREx Mission proposal did a very detailed cost estimate during NASA Phase A (SPHEREx would conduct a spectroscopic sky survey similar to CDIM, using a similar instrument architecture, observing scenario, and data pipeline). The CDIM team used SPHEREx for internal validation. CDIM Phase E costs are estimated at twice the SPHEREx cost, due to the longer prime mission (4 vs. 2 years).

Team X adopted the CDIM Team's estimates for WBS 4:

- Phase B-D: $25.0M
- Phases E-F: $40.0M

## 7.2 Total Mission Cost

Team X estimates the total mission cost at **$929M**. Reserves on Phases A-D are carried at 30% and Phases E-F Reserves are 15%, per JPL flight project guidelines (no reserves are carried on Launch Vehicle and spacecraft tracking). E&PO costs (WBS 11) are not included. Details of how the costs were estimated are explained in the Team X Final Report. A Team X cost summary by WBS is shown in **Table 7-2**.

For some WBS elements, the CDIM Team had better estimates than the Team X models (see §7.1), adopting overrides in those cases. The CDIM Team's best estimate of total mission cost is **$905.4M**.

## 7.3 Risk Approach

The CDIM instrument, spacecraft, and operations are designed to the standards of Mission Class B. We discuss below the major elements that could impact mission success. These are all development phase risks. No significant operational phase risks have been identified.

**Spacecraft.** The spacecraft (WBS 6) would be developed from a long line of successful Ball Aerospace Class B missions; the requirements on key system parameters such as pointing, data storage, and downlink, are within the family of spacecraft in the Ball's BCP-300 range (§4.6). This is considered by Ball to be low development risk.

| Table 7-2. CDIM mission costs estimated by Team X and CDIM Team | | |
|---|---|---|
| **Work Breakdown Structure (WBS) Elements** | **Team X Cost Estimate ($FY18)** | **CDIM Team Cost Estimate ($FY18)** |
| Development Cost (Phases A-D) | $690.1M | $666.4M |
| 1.0, 2.0, & 3.0 Management, Systems Engineering, and Mission Assurance | $51.5M | $50.8M |
| 4.0 Science | $25.0M | $25.0M |
| 5.0 Payload System | $170.5M | $170.5M |
| 5.01 Payload Management | $1.8M | $1.8M |
| 5.02 Payload Engineering | $1.4M | $1.4M |
| 5.04 Telescope (incl. II&T) | $167.3M | $167.3M |
| 5.04.01 Instrument (IR Spectrometer) | $152.0M | $152.0M |
| 5.04.02 Optical Telescope Assembly | $15.2M | $15.2M |
| 6.0 Flight System | $220.1M | $189.0M |
| 7.0 Mission Operations Preparation | $16.3M | $16.3M |
| 9.0 Ground Data Systems | $23.0M | $23.0M |
| 10.0 Assembly, Test, and Launch Operations (ATLO) | $24.6M | $34.0M |
| 12.0 Mission and Navigation Design | $0.0M | $4.2M |
| Development Reserves (30%) | $159.2M | $153.7M |
| Operations Cost (Phase E) | $89.0M | $89.0M |
| 1.0 Management | $4.5M | $4.5M |
| 4.0 Science | $40.0M | $40.0M |
| 7.0 Mission Operations | $23.6M | $23.6M |
| 9.0 Ground Data Systems | $9.7M | $9.7M |
| Operations Reserves (15%) | $11.1M | $11.1M |
| Launch Vehicle | $150.0M | $150.0M |
| **Total Cost (including Launch Vehicle)** | **$929.0M** | **$905.4M** |
| Note: See §7.1 for explanation of cost differences | | |

**Telescope.** The off-axis telescope features beryllium mirrors and metering structures, which greatly reduce the risk of optical alignment issues during cold testing. This was proven with Spitzer, which has a larger Be primary mirror; likewise, with the JWST mirror segments. CDIM can adopt the fabrication and test approach used by Spitzer (§3.2). This is considered low development risk.

**Thermal design.** CDIM uses the v-groove radiator approach designed at JPL, and implemented on Planck. Team X thermal models indicate substantial margins against requirements (perhaps enabling one v-groove to be eliminated). If higher-fidelity modeling indicated lower margins, an additional v-groove can be added, and the size of the focal plane radiator enlarged.

**Focal plane.** CDIM has a large focal plane, analogous to that of Kepler, and much smaller than the focal plane under development for WFIRST. Unlike WFIRST, CDIM does not have tight driving requirements on image quality, though it does require thermal stability, which is eased by operating in the benign L2 environment with an observing scenario that hardly changes the thermal load on the instrument (§4.6). The LVFs place demands on the optical design on telecentricity, in order to meet filter bandwidth requirements; the optical design, from JPL's Optical Systems Division, and described in the Team X Report, already meets these requirements. This is considered low development risk.

**Detectors.** CDIM requires four different types of Teledyne H2RG ($2048^2$ HgCdTe) detectors (§7.1.2). Devices with 2.5-µm long-wavelength cutoff have already been developed for Euclid; these can be used unmodified. Devices with 5.3-µm cutoff have already been developed for NEOCam; these can be used unmodified. Devices with 7.5-µm cutoff are not currently a commercial item; however, 10-µm devices are needed by NEOCam and that project is currently funding their development. The principal challenge is in CdZnTe substrate removal (to eliminate cosmic-ray fluorescence); this is harder for 10-µm than for 7.5-µm detectors. It is currently unclear whether CDIM in fact requires complete substrate removal. The current TRL for 7.5-µm cutoff detectors is approximately TRL 4.

Risk mitigation requires funding the technology development of these detectors at Teledyne, as discussed in §5.1. The risk may be partially mitigated by adopting the NEOCam 10-µm devices and accepting the science impact of higher dark current.

**Linear Variable Filters (LVFs).** LVFs are the enabling technology for CDIM, and as such are a high-consequence risk. These filters allow moderate-resolution spectroscopy of every astronomical object in a wide-field image, without the complexity of a slit spectrometer, or the limited field capability of an integral-field spectrograph (IFS). While LVFs have been successfully operated in space (including spectroscopic observations with the New Horizons LEISA instrument and OSIRIS-REx), the filters for CDIM are more demanding in their requirements. The alternative to LVFs – an integral-field spectrometer – is unlikely to fit in a Probe cost cap, and has its own major technology challenges.

Risk reduction requires technology funding early in development to design and simulate the filters, and to fabricate and test engineer model filters (including thermal cycling). Current TRL for the short-wavelength filters is TRL 4. A Pre-Phase-A technology effort is required to bring the LVFs to TRL 5, by fabricating and testing prototypes, as described in §5.2. The filter design work done during this Probe study by Omega Optical Inc. has already retired much of the design risk. To reach TRL 6, environmental testing (including thermal cycling) must be performed on engineering model filters.

# 8 MANAGEMENT AND SCHEDULE

## 8.1 Management

Management of the CDIM Mission would be governed primarily by the management guidelines outlined in a NASA Announcement of Opportunity (AO) for a Probe-class astrophysics mission. For this study, we assumed a specific management approach for the purposes of estimating mission costs. We assume that the mission would be PI-led, in a similar way to a New Frontiers mission in NASA's Planetary Science Division of the Science Directorate. In a future competition, the management structure would be proposed in response to the AO, and could different from the outline assumed in this Report.

Management of the mission would be the ultimate responsibility of the PI. Team X cost models (§7) assume JPL manages most elements, except the spacecraft and ATLO which would be subcontracted to a major partner, which we assume in this Report to be Ball Aerospace. JPL would be responsible for Management, System Engineering, Mission Assurance, Payload (including thermal design and I&T), Mission Operations, Ground Data System, and Mission Operations. Ball Aerospace would be responsible for the spacecraft bus, payload integration to the bus, and ATLO. NASA would be responsible for providing the launch vehicle and launch services. Elements of the payload would be sub-contracted: the telescope, focal plane design and fabrication, detectors, and LVFs.

The Science Data Center would be hosted at IPAC, where several NASA infrared missions (including WISE and Spitzer) have their Centers. The PI would be responsible for assembling a Science Team, and managing that Team's activities both pre-launch (including defining requirements for the data pipeline, and uplink commanding of the payload) and post launch (delivering data products to the community, §2.5).

## 8.2 Schedule

Team X costed the CDIM Mission assuming a 64-month development phase (A-D), with durations appropriate for a New Frontiers-class mission (**Figure 8-1**). The schedule makes the normal assumption about the technological readiness of the proposed mission: all elements must be at TRL 6 by the end of Phase B (PDR). For CDIM, there are two technology developments that must achieve TRL 6 by the date of this review (see §5); the technology development plan is designed to be consistent with this schedule.

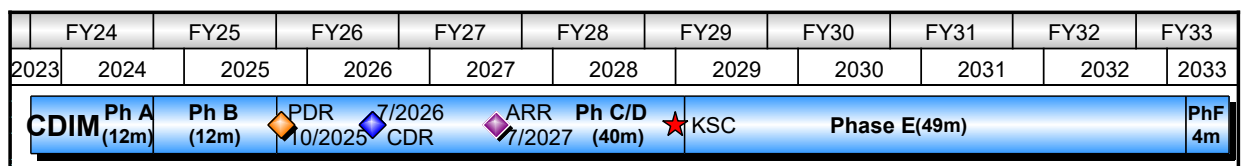

**Figure 8-1.** Development and operations schedule for CDIM. Phase E assumed to be 4 years, following a one-month in-orbit checkout. Any Extended Mission would be added to Phase E, prior to the closeout phase (Phase F).

## 8.3 Next steps

CDIM would need to complete a number of tasks to meet the requirements of a Probe Mission Announcement of Opportunity, prior to competitive selection to proceed into Phase A. Two of these tasks (7.5-µm cutoff H2RG detectors, and linear variable filters meeting CDIM requirements) are needed to reach TRL 5 prior to Step 1 selection and TRL 6 by PDR (see §5). Below we list the most important next steps.

- Continue simulations of science performance, including joint analysis of relevant space mission and ground telescope data
- Detailed flowdown of science requirements to instrument and mission design
- Simulations of observing scenarios for the three surveys
- Simulation of level 0 data (including telescope and detector performance, and spacecraft pointing)
- Develop detailed requirements for spacecraft (pointing, data storage and downlink, orbit maintenance, etc.)
- Continue thermal design for instrument passive cooling system, possibly leading to a simplification of the V-groove radiator design
- Technology maturation of the 7.5-µm cutoff H2RG detectors
- Technology maturation of the linear variable filters, including environmental testing of engineering prototypes

## 9 REFERENCES

Arendt, R. G., Kashlinsky, A., Moseley, S. H., Mather, J., 2016. Cosmic Infrared Background Fluctuations and Zodiacal Light. The Astrophysical Journal. 824, 26. http://stacks.iop.org/0004-637X/824/i=1/a=26

Azadi, M., et al., 2017. The MOSDEF Survey: AGN Multi-wavelength Identification, Selection Biases, and Host Galaxy Properties. The Astrophysical Journal. 835, 27. http://stacks.iop.org/0004-637X/835/i=1/a=27

Baker, J., 2007. Special Issue: New Horizons at Jupiter. Science. 318, 215-243. http://science.sciencemag.org/content/sci/318/5848/215.full.pdf

Baldwin, J. A., Phillips, M. M., Terlevich, R., 1981. Classification parameters for the emission-line spectra of extragalactic objects. Publications of the Astronomical Society of the Pacific. 93, 5.

Barkana, R., Loeb, A., 2001. In the beginning: the first sources of light and the reionization of the universe. Physics reports. 349, 125-238.

Becker, R. H., et al., 2001. Evidence for Reionization at z~ 6: Detection of a Gunn-Peterson Trough in az= 6.28 Quasar. The Astronomical Journal. 122, 2850.

Behroozi, P. S., Wechsler, R. H., Conroy, C., 2013. The average star formation histories of galaxies in dark matter halos from z= 0-8. The Astrophysical Journal. 770, 57.

Beletic, J. W., et al., 2008. Teledyne Imaging Sensors: infrared imaging technologies for astronomy and civil space. High Energy, Optical, and Infrared Detectors for Astronomy III, Vol. 7021. International Society for Optics and Photonics, pp. 70210H.

Benford, D. J., Lauer, T. R., Mott, D. B., 2008. Simulations of sample-up-the-ramp for space-based observations of faint sources. High Energy, Optical, and Infrared Detectors for Astronomy III, Vol. 7021. International Society for Optics and Photonics, pp. 70211V.

Blank, R., et al., 2012. H2RG focal plane array and camera performance update. High Energy, Optical, and Infrared Detectors for Astronomy V, Vol. 8453. International Society for Optics and Photonics, pp. 845310.

Böker, T., et al., 1999. The nicmos snapshot survey of nearby galaxies. The Astrophysical Journal Supplement Series. 124, 95.

Bouwens, R., et al., 2015. UV luminosity functions at redshifts z~ 4 to z~ 10: 10,000 galaxies from HST legacy fields. The Astrophysical Journal. 803, 34.

Bromm, V., Larson, R. B., 2004. The first stars. Annu. Rev. Astron. Astrophys. 42, 79-118.

Cerulo, P., et al., 2014. The morphological transformation of red sequence galaxies in the distant cluster XMMU J1229+ 0151. Monthly Notices of the Royal Astronomical Society. 439, 2790-2812.

Cerulo, P., et al., 2016. The accelerated build-up of the red sequence in high-redshift galaxy clusters. Monthly Notices of the Royal Astronomical Society. 457, 2209-2235.

Chang, T.-C., et al., 2015. Synergy of CO/[CII]/Ly α Line Intensity Mapping with the SKA. arXiv preprint arXiv:1501.04654.

Chary, R.-R., 2008. The stellar initial mass function at the epoch of reionization. The Astrophysical Journal. 680, 32.

Chary, R., Petitjean, P., Robertson, B., Trenti, M., Vangioni, E., 2016. Gamma-ray bursts and the early star-formation history. Space Science Reviews. 202, 181-194.

Coil, A. L., et al., 2015. The MOSDEF Survey: optical Active Galactic Nucleus Diagnostics at z~ 2.3. The Astrophysical Journal. 801, 35.

Comaschi, P., Ferrara, A., 2015. Probing high-redshift galaxies with Lyα intensity mapping. Monthly Notices of the Royal Astronomical Society. 455, 725-738.

Conroy, C., Gunn, J. E., 2010. The propagation of uncertainties in stellar population synthesis modeling. III. Model calibration, comparison, and evaluation. The Astrophysical Journal. 712, 833.

Cook, L., 1979. Three-mirror anastigmat used off-axis in aperture and field. Space Optics II, Vol. 183. International Society for Optics and Photonics, pp. 207-212.

DeBoer, D. R., et al., 2017. Hydrogen Epoch of Reionization Array (HERA). Publications of the Astronomical Society of the Pacific. 129, 045001.

Dijkstra, M., Wyithe, J. S. B., Haiman, Z., 2007. Luminosity functions of Lyα emitting galaxies and cosmic reionization of hydrogen. Monthly Notices of the Royal Astronomical Society. 379, 253-259.

Dorn-Wallenstein, T. Z., Levesque, E. M., 2018. Stellar Population Diagnostics of the Massive Star Binary Fraction. The Astrophysical Journal. 867, 125.

Dorn, M., et al., 2018. A monolithic 2k x 2k LWIR HgCdTe detector array for passively cooled space missions. High Energy, Optical, and Infrared Detectors for Astronomy VIII, Vol. 10709. International Society for Optics and Photonics, pp. 1070907.

Dorn, M. L., et al., 2016. Proton irradiation results for long-wave HgCdTe infrared detector arrays for Near-Earth Object Camera. Journal of Astronomical Telescopes, Instruments, and Systems. 2, 036002.

Eyles, L. P., Bunker, A. J., Stanway, E. R., Lacy, M., Ellis, R. S., Doherty, M., 2005. Spitzer Imaging of i′-drop Galaxies: Old Stars at z≈ 6. Monthly Notices of the Royal Astronomical Society. 364, 443-454.

Fan, X., Carilli, C., Keating, B., 2006a. Observational constraints on cosmic reionization. Annu. Rev. Astron. Astrophys. 44, 415-462.

Fan, X., et al., 2006b. Constraining the evolution of the ionizing background and the epoch of reionization with z∼ 6 quasars. II. A sample of 19 quasars. The Astronomical Journal. 132, 117.

Fan, X., et al., 2003. A survey of z> 5.7 quasars in the Sloan Digital Sky Survey. II. Discovery of three additional quasars at z> 6. The Astronomical Journal. 125, 1649.

Feng, C., Cooray, A., Keating, B., 2017. A Halo Model Approach to the 21 cm and Lyα Cross-correlation. The Astrophysical Journal. 846, 21.

Finkelstein, S. L., 2016. Observational Searches for Star-Forming Galaxies at z> 6. Publications of the Astronomical Society of Australia. 33.

Finkelstein, S. L., et al., 2015. The evolution of the galaxy rest-frame ultraviolet luminosity function over the first two billion years. The Astrophysical Journal. 810, 71.

Fixsen, D. J., Moseley, S. H., Arendt, R. G., 2000. Calibrating Array Detectors. The Astrophysical Journal Supplement Series. 128, 651. http://stacks.iop.org/0067-0049/128/i=2/a=651

Fosbury, R., et al., 2003. UV/optical properties of z∼ 2.5 radio galaxies. New Astronomy Reviews. 47, 299-302.

Furlanetto, S. R., Zaldarriaga, M., Hernquist, L., 2006. The effects of reionization on Lyα galaxy surveys. Monthly Notices of the Royal Astronomical Society. 365, 1012-1020.

Giallongo, E., et al., 2015. Faint AGNs at z> 4 in the CANDELS GOODS-S field: looking for contributors to the reionization of the Universe. Astronomy & Astrophysics. 578, A83.

Girard, J. J., Forrest, W. J., McMurtry, C. W., Pipher, J. L., Dorn, M., Mainzer, A., 2014. Cosmic ray response of megapixel LWIR arrays from TIS. High Energy, Optical, and Infrared

Detectors for Astronomy VI, Vol. 9154. International Society for Optics and Photonics, pp. 91542A.

Gong, Y., et al., 2011a. Intensity mapping of the [C II] fine structure line during the Epoch of Reionization. The Astrophysical Journal. 745, 49.

Gong, Y., Cooray, A., Silva, M. B., Santos, M. G., Lubin, P., 2011b. Probing reionization with intensity mapping of molecular and fine-structure lines. The Astrophysical Journal Letters. 728, L46.

Gong, Y., Silva, M., Cooray, A., Santos, M. G., 2014. Foreground contamination in Lyα intensity mapping during the epoch of reionization. The Astrophysical Journal. 785, 72.

Goulding, A., Alexander, D., 2009. Towards a complete census of AGN in nearby Galaxies: a large population of optically unidentified AGN. Monthly Notices of the Royal Astronomical Society. 398, 1165-1193.

Greene, J. E., Ho, L. C., 2007. A new sample of low-mass black holes in active galaxies. The Astrophysical Journal. 670, 92.

Greene, T. P., et al., 2017. λ= 2.4 to 5 μm spectroscopy with the James Webb Space Telescope NIRCam instrument. Journal of astronomical telescopes, instruments, and systems. 3, 035001.

Greig, B., Mesinger, A., 2016. The global history of reionization. Monthly Notices of the Royal Astronomical Society. 465, 4838-4852.

Greig, B., Mesinger, A., Haiman, Z., Simcoe, R. A., 2016. Are we witnessing the epoch of reionization at z= 7.1 from the spectrum of J1120+ 0641? Monthly Notices of the Royal Astronomical Society. 466, 4239-4249.

Hayes, M., Schaerer, D., Östlin, G., Mas-Hesse, J. M., Atek, H., Kunth, D., 2011. On the redshift evolution of the lyα escape fraction and the dust content of galaxies. The Astrophysical Journal. 730, 8.

Heneka, C., Cooray, A., Feng, C., 2017. Probing the Intergalactic Medium with Lyα and 21 cm Fluctuations. The Astrophysical Journal. 848, 52.

Kasliwal, M. M., et al., 2018. Spitzer Mid-Infrared Detections of Neutron Star Merger GW170817 Suggests Synthesis of the Heaviest Elements. Monthly Notices of the Royal Astronomical Society: Letters.

Kennicutt Jr, R. C., 1998. The global Schmidt law in star-forming galaxies. The Astrophysical Journal. 498, 541.

Kewley, L. J., et al., 2013. Theoretical evolution of optical strong lines across cosmic time. The Astrophysical Journal. 774, 100.

Konno, A., et al., 2014. Accelerated evolution of the Lyα luminosity function at z≳ 7 revealed by the Subaru ultra-deep survey for Lyα emitters at z= 7.3. The Astrophysical Journal. 797, 16.

Konno, A., et al., 2017. SILVERRUSH. IV. Lyα luminosity functions at z= 5.7 and 6.6 studied with∼ 1300 Lyα emitters on the 14–21 deg2 sky. Publications of the Astronomical Society of Japan. 70, S16.

Koopmans, L., et al., 2015. The cosmic dawn and epoch of reionization with the square kilometre array. arXiv preprint arXiv:1505.07568.

Kulkarni, G., Worseck, G., Hennawi, J. F., 2018. Evolution of the AGN UV luminosity function from redshift 7.5. arXiv preprint arXiv:1807.09774.

Latif, M. A., Ferrara, A., 2016. Formation of supermassive black hole seeds. Publications of the Astronomical Society of Australia. 33.

Leinert, C., et al., 1998. The 1997 reference of diffuse night sky brightness *. Astron. Astrophys. Suppl. Ser. 127, 1-99. https://doi.org/10.1051/aas:1998105

Lencioni, D. E., Digenis, C. J., Bicknell, W. E., Hearn, D. R., Mendenhall, J. A., 1999. Design and performance of the EO-1 Advanced Land Imager. Sensors, Systems, and Next-Generation Satellites III, Vol. 3870. International Society for Optics and Photonics, pp. 269-281.

Leroy, C., et al., 2006. Performances of the Planck-HFI cryogenic thermal control system. Space Telescopes and Instrumentation I: Optical, Infrared, and Millimeter, Vol. 6265. International Society for Optics and Photonics, pp. 62650H.

Levesque, E. M., Leitherer, C., Ekstrom, S., Meynet, G., Schaerer, D., 2012. The Effects of Stellar Rotation. I. Impact on the Ionizing Spectra and Integrated Properties of Stellar Populations. The Astrophysical Journal. 751, 67.

Lidz, A., et al., 2011. Intensity mapping with carbon monoxide emission lines and the redshifted 21 cm line. The Astrophysical Journal. 741, 70.

Lightsey, P. A., 2007. James Webb Space Telescope: a large deployable cryogenic telescope in space. Laser-Induced Damage in Optical Materials: 2007, Vol. 6720. International Society for Optics and Photonics, pp. 67200E.

Loose, M., Beletic, J., Garnett, J., Muradian, N., 2006. Space qualification and performance results of the SIDECAR ASIC. Space Telescopes and Instrumentation I: Optical, Infrared, and Millimeter, Vol. 6265. International Society for Optics and Photonics, pp. 62652J.

Madau, P., Dickinson, M., 2014. Cosmic star-formation history. Annual Review of Astronomy and Astrophysics. 52, 415-486.

Madau, P., Haardt, F., 2015. Cosmic reionization after planck: Could quasars do it all? The Astrophysical Journal Letters. 813, L8.

Maiolino, R., et al., 2008. AMAZE-I. The evolution of the mass–metallicity relation at $z > 3$. Astronomy & Astrophysics. 488, 463-479.

Matsuoka, Y., et al., 2018. SubaruHigh- z Exploration of Low-luminosity Quasars (SHELLQs). V. Quasar Luminosity Function and Contribution to Cosmic Reionization at $z = 6$. The Astrophysical Journal. 869, 150. http://stacks.iop.org/0004-637X/869/i=2/a=150

McGreer, I. D., Mesinger, A., D'Odorico, V., 2014. Model-independent evidence in favour of an end to reionization by $z \approx 6$. Monthly Notices of the Royal Astronomical Society. 447, 499-505.

McKelvey, M. E., et al., 2004. Radiation environment performance of JWST prototype FPAs. Focal Plane Arrays for Space Telescopes, Vol. 5167. International Society for Optics and Photonics, pp. 223-235.

McMurtry, C. W., Cabrera, M. S., Dorn, M. L., Pipher, J. L., Forrest, W. J., 2016a. 13 micron cutoff HgCdTe detector arrays for space and ground-based astronomy. SPIE Astronomical Telescopes + Instrumentation, Vol. 9915. SPIE, pp. 10.

McMurtry, C. W., et al., 2016b. Candidate 10 micron HgCdTe arrays for the NEOCam space mission. SPIE Astronomical Telescopes + Instrumentation, Vol. 9915. SPIE, pp. 8.

McMurty, C., et al., 2013. Development of sensitive long-wave infrared detector arrays for passively cooled space missions. Optical Engineering. 52, 091804.

Mesinger, A., Aykutalp, A., Vanzella, E., Pentericci, L., Ferrara, A., Dijkstra, M., 2014. Can the intergalactic medium cause a rapid drop in Lyα emission at $z > 6$? Monthly Notices of the Royal Astronomical Society. 446, 566-577.

Momose, R., et al., 2014. Diffuse Lyα haloes around galaxies at z= 2.2–6.6: implications for galaxy formation and cosmic reionization. Monthly Notices of the Royal Astronomical Society. 442, 110-120.

Momose, R., et al., 2016. Statistical properties of diffuse Lyα haloes around star-forming galaxies at z∼ 2. Monthly Notices of the Royal Astronomical Society. 457, 2318-2330.

Oesch, P., et al., 2016. A remarkably luminous galaxy at z= 11.1 measured with Hubble space telescope Grism spectroscopy. The Astrophysical Journal. 819, 129.

Oh, S. P., Haiman, Z., 2002. Second-generation objects in the Universe: Radiative cooling and collapse of halos with virial temperatures above 104 K. The Astrophysical Journal. 569, 558.

Omega Optical Inc., 2018. Design Study for Segmented Linear Variable Filters (LVF) from 0.75 to 7.5 microns. Technical Report.

Onoue, M., et al., 2017. Minor Contribution of Quasars to Ionizing Photon Budget at z∼ 6: Update on Quasar Luminosity Function at the Faint End with Subaru/Suprime-Cam. The Astrophysical Journal Letters. 847, L15.

Ouchi, M., et al., 2010. Statistics of 207 Lyα emitters at a redshift near 7: Constraints on reionization and galaxy formation models. The Astrophysical Journal. 723, 869.

Parsa, S., Dunlop, J. S., McLure, R. J., 2017. No evidence for a significant AGN contribution to cosmic hydrogen reionization. Monthly Notices of the Royal Astronomical Society. 474, 2904-2923.

Pettini, M., Pagel, B. E., 2004. [O iii]/[N ii] as an abundance indicator at high redshift. Monthly Notices of the Royal Astronomical Society. 348, L59-L63.

Phillips, J., Edwall, D., Lee, D., Arias, J., 2001. Growth of HgCdTe for long-wavelength infrared detectors using automated control from spectroscopic ellipsometry measurements. Journal of Vacuum Science & Technology B: Microelectronics and Nanometer Structures Processing, Measurement, and Phenomena. 19, 1580-1584.

Planck Collaboration, et al., 2018. Planck 2018 results. VI. Cosmological parameters. arXiv preprint arXiv:1807.06209.

Pontoppidan, K., Van Dishoeck, E., Dartois, E., 2004. Mapping ices in protostellar environments on 1000 au scales-methanol-rich ice in the envelope of serpens smm 4. Astronomy & Astrophysics. 426, 925-940.

Pullen, A. R., Doré, O., Bock, J., 2014. Intensity mapping across cosmic times with the Lyα line. The Astrophysical Journal. 786, 111.

Rauscher, B. J., Lindler, D. J., Mott, D. B., Wen, Y., Ferruit, P., Sirianni, M., 2011. The Dark Current and Hot Pixel Percentage of James Webb Space Telescope 5 μm Cutoff HgCdTe Detector Arrays as Functions of Temperature. Publications of the Astronomical Society of the Pacific. 123, 953.

Reuter, D., et al., 2018. The OSIRIS-REx visible and infrared spectrometer (OVIRS): spectral maps of the asteroid Bennu. Space Science Reviews. 214, 54.

Reuter, D. C., et al., 2008. Ralph: A visible/infrared imager for the New Horizons Pluto/Kuiper Belt mission. Space Science Reviews. 140, 129-154.

Robertson, B. E., Ellis, R. S., Furlanetto, S. R., Dunlop, J. S., 2015. Cosmic reionization and early star-forming galaxies: a joint analysis of new constraints from Planck and the Hubble Space Telescope. The Astrophysical Journal Letters. 802, L19.

Rodriguez-Ardila, A., Pastoriza, M. G., Viegas, S., Sigut, T., Pradhan, A. K., 2004. Molecular hydrogen and [Fe ii] in active galactic nuclei. Astronomy & Astrophysics. 425, 457-474.

Salvaterra, R., Ferrara, A., Dayal, P., 2011. Simulating high-redshift galaxies. Monthly Notices of the Royal Astronomical Society. 414, 847-859.

Shim, H., Chary, R.-R., 2013. Dissection of Hα Emitters: Low-z Analogs of z> 4 Star-Forming Galaxies. The Astrophysical Journal. 765, 26.

Shim, H., et al., 2011. z∼ 4 Hα emitters in the great observatories origins deep survey: tracing the dominant mode for growth of galaxies. The astrophysical journal. 738, 69.

Silva, M. B., Santos, M. G., Gong, Y., Cooray, A., Bock, J., 2013. Intensity Mapping of Lyα Emission during the Epoch of Reionization. The Astrophysical Journal. 763, 132.

Smit, R., Bouwens, R. J., Labbé, I., Franx, M., Wilkins, S. M., Oesch, P. A., 2016. Inferred Hα flux as a star formation rate indicator at z∼ 4–5: implications for dust properties, burstiness, and the z= 4–8 star formation rate functions. The Astrophysical Journal. 833, 254.

Sobacchi, E., Mesinger, A., 2014. Inhomogeneous recombinations during cosmic reionization. Monthly Notices of the Royal Astronomical Society. 440, 1662-1673.

Song, M., Finkelstein, S. L., Livermore, R. C., Capak, P. L., Dickinson, M., Fontana, A., 2016. Keck/MOSFIRE spectroscopy of z= 7–8 galaxies: Lyα emission from a galaxy at z= 7.66. The Astrophysical Journal. 826, 113.

Stahl, H. P., Henrichs, T., 2016. Multivariable parametric cost model for space and ground telescopes. Modeling, Systems Engineering, and Project Management for Astronomy VI, Vol. 9911. International Society for Optics and Photonics, pp. 99110L.

Stark, D. P., Schenker, M. A., Ellis, R., Robertson, B., McLure, R., Dunlop, J., 2013. Keck spectroscopy of 3< z< 7 faint Lyman break galaxies: the importance of nebular emission in understanding the specific star formation rate and stellar mass density. The Astrophysical Journal. 763, 129.

Trac, H., Cen, R., Mansfield, P., 2015. SCORCH. I. THE GALAXY–HALO CONNECTION IN THE FIRST BILLION YEARS. The Astrophysical Journal. 813, 54.

Tremonti, C. A., et al., 2004. The origin of the mass-metallicity relation: insights from 53,000 star-forming galaxies in the sloan digital sky survey. The Astrophysical Journal. 613, 898.

Tulloch, S., 2017. Persistence Characterisation of teledyne H2RG detectors. Scientific Detector Workshop, Baltimore, MD, USA.

Vanzella, E., et al., 2012. On the Detection of Ionizing Radiation Arising from Star-forming Galaxies at Redshift z ∼ 3-4: Looking for Analogs of "Stellar Re-ionizers". The Astrophysical Journal. 751, 70.

Venkatesan, A., Giroux, M. L., Shull, J. M., 2001. Heating and ionization of the intergalactic medium by an early X-ray background. The Astrophysical Journal. 563, 1.

Villar, V. A., et al., 2018. Spitzer Space Telescope Infrared Observations of the Binary Neutron Star Merger GW170817. The Astrophysical Journal. 862, L11. http://dx.doi.org/10.3847/2041-8213/aad281

Visbal, E., Loeb, A., 2010. Measuring the 3D clustering of undetected galaxies through cross correlation of their cumulative flux fluctuations from multiple spectral lines. Journal of Cosmology and Astroparticle Physics. 2010, 016.

Waczynski, A., et al., 2016. Performance overview of the Euclid infrared focal plane detector subsystems. High Energy, Optical, and Infrared Detectors for Astronomy VII, Vol. 9915. International Society for Optics and Photonics, pp. 991511.

Waczynski, A., et al., 2005. Radiation induced luminescence of the CdZnTe substrate in HgCdTe detectors for WFC3. Focal Plane Arrays for Space Telescopes II, Vol. 5902. International Society for Optics and Photonics, pp. 59020P.

Wang, F., et al., 2018. Exploring Reionization-Era Quasars III: Discovery of 16 Quasars at $6.4 \lesssim z \lesssim 6.9$ with DESI Legacy Imaging Surveys and UKIRT Hemisphere Survey and Quasar Luminosity Function at z~6. 7. arXiv preprint arXiv:1810.11926.

Willott, C. J., Bergeron, J., Omont, A., 2015. Star formation rate and dynamical mass of $10^8$ solar mass black hole host galaxies at redshift 6. The Astrophysical Journal. 801, 123.

Wu, X.-B., et al., 2015. An ultraluminous quasar with a twelve-billion-solar-mass black hole at redshift 6.30. Nature. 518, 512.

Yan, H., Dickinson, M., Giavalisco, M., Stern, D., Eisenhardt, P. R., Ferguson, H. C., 2006. The Stellar Masses and Star Formation Histories of Galaxies at z≈ 6: Constraints from Spitzer Observations in the Great Observatories Origins Deep Survey. The Astrophysical Journal. 651, 24.

Yu, Q., Tremaine, S., 2002. Observational constraints on growth of massive black holes. Monthly Notices of the Royal Astronomical Society. 335, 965-976.

Zemcov, M., et al., 2014. On the origin of near-infrared extragalactic background light anisotropy. Science. 346, 732-735. http://science.sciencemag.org/content/sci/346/6210/732.full.pdf

Zheng, Z.-Y., et al., 2017. First Results from the Lyman Alpha Galaxies in the Epoch of Reionization (LAGER) Survey: Cosmological Reionization at z∼ 7. The Astrophysical Journal Letters. 842, L22.

Zheng, Z., Cen, R., Trac, H., Miralda-Escudé, J., 2010. Radiative transfer modeling of Lyα emitters. I. Statistics of spectra and luminosity. The Astrophysical Journal. 716, 574.

Zheng, Z., Miralda-Escudé, J., 2002. Monte Carlo simulation of Lyα scattering and application to damped Lyα systems. The Astrophysical Journal. 578, 33.

# 2020 Astrophysical Decadal Survey - Probe Mission Preparatory Study
# Master Equipment List Based Parametric Total Lifecycle Cost Estimate

*Mission Name / Acronym:* **CDIM** — *The Cosmic Dawn Intensity Mapper*

*Cost Estimator:* Team X

*Date of Cost Estimate:* October 22, 2018

*Cost Estimate Based On:* Final Master Equipment List

| PROJECT PHASE | | COST [FY18 $M] |
|---|---|---|
| **Phase A** | | Note 1 |
| **Phases B-D** | Mgmt, SE, MA | $51.5 |
| | Science | $25.0 |
| | Telescope | $15.2 |
| | Instrument 1 | $155.3 |
| | Instrument 2 | n/a |
| | Instrument 3 | n/a |
| | Spacecraft, including ATLO | $244.7 |
| | MOS/GDS | $39.2 |
| | Launch Vehicle and Services | $150.0 |
| | Reserves | $159.2 |
| | ***Total Cost Phases B-D*** | ***$840.1*** |
| **Phase E-F** | Operations | $77.8 |
| | Reserves | $11.1 |
| | ***Total Cost Phases E-F*** | ***$88.9*** |
| | **TOTAL LIFECYCLE COST** | **$929.0** |

### Notes:

- This parametric cost estimate is based on the Probe's Master Equipment List derived from the Final Engineering Concept Definition Package that accurately reflects the mission described in the Probe's Final Report. This estimate is to be used only for non-binding rough order of magnitude planning purposes.

- Team X estimates are generally model-based, and were generated after a series of instrument and mission level studies. Their accuracy is commensurate with the level of understanding typical to Pre-Phase-A concept development. They do not constitute an implementation or cost commitment on the part of JPL or Caltech.

Note 1: Team X does not separate out Phase A costs from Phases B-D costs.

Note 2: CDIM Team estimates Total Lifecycle Cost to be $905.4M, using SEER-H to estimate costs for WBS 6 and 10.